\documentclass[10pt,preprint]{aastex}
\usepackage{graphicx,amssymb,natbib,longtable,lscape}
\usepackage{amsmath}

\newcommand{\kms}{km s$^{-1}$}
\newcommand{\etal}{et al.}

\newcommand{\sqd}{deg$^{2}$}
\newcommand{\ntot}{15855}
\newcommand{\ngal}{15041}
\newcommand{\ncone}{11941}
\newcommand{\nctwo}{3100}
\newcommand{\nhvc}{814}
\newcommand{\ndtot}{1013}
\newcommand{\ndxgal}{199}

\def\plotfiddle#1#2#3#4#5#6#7{{\centering \leavevmode \vbox to#2{\rule{0pt}{#2}}\includegraphics{#1}}}

\shorttitle{ALFALFA 40\% Catalog}
\shortauthors{Haynes \etal}

\begin{document}
\title{The Arecibo Legacy Fast ALFA Survey: The $\alpha.40$ 
HI Source Catalog, its Characteristics
and their Impact on the Derivation of the HI Mass Function}
\author{Martha P. Haynes, Riccardo Giovanelli, Ann M. Martin\altaffilmark{1},
        Kelley M. Hess\altaffilmark{2},
        Am\'elie Saintonge\altaffilmark{3},
        Elizabeth A. K. Adams, Gregory Hallenbeck\altaffilmark{1},
        G. Lyle Hoffman\altaffilmark{4},
        Shan Huang\altaffilmark{1},
        Brian R. Kent\altaffilmark{5},
        Rebecca A. Koopmann\altaffilmark{6},
        Emmanouil Papastergis\altaffilmark{1},
        Sabrina Stierwalt\altaffilmark{7},
        Thomas J. Balonek\altaffilmark{8},
        David W. Craig\altaffilmark{9},
        Sarah J.U. Higdon\altaffilmark{10},
        David A. Kornreich\altaffilmark{11},
        Jeffrey R. Miller\altaffilmark{12},
        Aileen A. O'Donoghue\altaffilmark{12},
        Ronald P. Olowin\altaffilmark{13},
        Jessica L. Rosenberg\altaffilmark{14},
        Kristine Spekkens\altaffilmark{15},
        Parker Troischt\altaffilmark{16},
        Eric M. Wilcots\altaffilmark{2}
}
\altaffiltext{1}{Center for Radiophysics and Space Research, Space Sciences Building, Cornell University, Ithaca, NY 14853.
        {\textit{e-mail:}} haynes@astro.cornell.edu, riccardo@astro.cornell.edu,amartin@astro.cornell.edu,
        betsey@astro.cornell.edu, ghallenbeck@astro.cornell.edu,shan@astro.cornell.edu,
        papastergis@astro.cornell.edu}
\altaffiltext{2}{Department of Astronomy, U. Wisconsin, Madison, WI. {\textit{e-mail:}} hess@astro.wisc.edu, 
        wilcots@astro.wisc.edu}
\altaffiltext{3}{Max Planck Institute for Extraterrestrial Physics and Max Planck Institute for Astrophysics, Garching bei
        Muenchen, Germany. {\textit{e-mail:}} amelie@mpe.mpg.de}
\altaffiltext{4}{Department of Physics, Lafayette College, Easton, PA 18042. {\textit{e-mail:}} hoffmang@lafayette.edu}
\altaffiltext{5}{National Radio Astronomy Observatory, 
        520 Edgemont Road, Charlottesville, VA 22903. {\textit{e-mail:}} bkent@nrao.edu}
\altaffiltext{6}{Department of Physics and Astronomy, Union College, Schenectady, NY. {\textit{e-mail:}} koopmanr@union.edu}
\altaffiltext{7}{Spitzer Science Center, California Institute of Technology, 1200 E. California Blvd.,
        Pasadena, CA 91125.{\textit{e-email:}} sabrina@ipac.caltech.edu}
\altaffiltext{8}{Department of Physics and Astronomy, Colgate University, Hamilton, NY. {\textit{e-mail:}} 
        tbalonek@mail.colgate.edu}
\altaffiltext{9}{Department of Math, Chemistry and Physics, West Texas A and M University, Box 60787, Canyon TX. 
        {\textit{e-mail:}} dcraig@wtamu.edu}
\altaffiltext{10}{Department of Physics, P.O. Box 8031, Georgia Southern University,
        Statesboro, GA 30460. {\textit{e-mail:}} shigdon@georgiasouthern.edu}
\altaffiltext{11}{Department of Physics and Astronomy, Humboldt State University, Arcata, CA 95521.  
        {\textit{e-mail:}} David.Kornreich@humboldt.edu}
\altaffiltext{12}{Physics Department, St. Lawrence University, 23 Romoda Drive, 
        Canton, NY 13617. {\textit{e-mail:}} jeff@stlawu.edu, aodonoghue@stlawu.edu}
\altaffiltext{13}{Department of Physics and Astronomy, St. Mary's College of California, 1928 St. Mary's Road, 
        Moraga, California 94575. {\textit{e-mail:}} rpolowin@stmarys-ca.edu}
\altaffiltext{14}{School of Physics, Astronomy and Computational Science, George Mason University, 
        MS 3F3, Fairfax, VA 22030. {\textit{e-mail:}} jrosenb4@gmu.edu}
\altaffiltext{15}{Department of Physics, Royal Military College of Canada, PO Box 17000, Station Forces,
        Kingston, Ontario K7K 7B4, Canada. {\textit{e-mail:}} Kristine.Spekkens@rmc.ca}
\altaffiltext{16}{Department of Physics, Hartwick College, Johnstone Science Building, 1 Hartwick Drive, 
        Oneonta, NY 13820. {\textit{e-mail:}} TroischtP@hartwick.edu}

\begin{abstract}
We present a current catalog of 21 cm HI line sources extracted from the Arecibo Legacy
Fast Arecibo L-band Feed Array (ALFALFA) survey over $\sim$2800 \sqd~
of sky: the $\alpha.40$ catalog. Covering 40\% of the final survey area, the 
$\alpha.40$ catalog contains \ntot~ sources in the regions $07^h30^m <$ R.A. $< 16^h30^m$,
$+04^\circ <$ Dec. $< +16^\circ$ and $+24^\circ <$ Dec. $< +28^\circ$ and
$22^h <$ R.A. $< 03^h$, $+14^\circ <$ Dec. $< +16^\circ$ 
and $+24^\circ <$ Dec. $< +32^\circ$. Of those, \ngal ~are certainly extragalactic,
yielding a source density of 5.3 galaxies per \sqd, a factor of 29 improvement
over the catalog extracted from the HI Parkes All Sky Survey. 
In addition to the source centroid positions, HI line flux densities,
recessional velocities and line widths, the catalog includes the coordinates
of the most probable optical counterpart of each HI line detection, and a separate
compilation provides a crossmatch to identifications given in the photometric
and spectroscopic catalogs associated with the Sloan Digital Sky Survey Data 
Release 7. Fewer than 2\% of the extragalactic HI line sources cannot be 
identified with a feasible optical counterpart; some of those may be rare 
OH megamasers at 0.16 $< z <$0.25. A detailed analysis is presented of the 
completeness, width dependent sensitivity function and bias
inherent of the $\alpha.40$ catalog. The impact of survey selection, 
distance errors, current volume coverage and local large scale structure on 
the derivation of the HI mass function is assessed. While $\alpha.40$ does not yet 
provide a completely representative sampling of cosmological volume, derivations 
of the HI mass function using future data releases from ALFALFA 
will further improve both statistical and systematic uncertainties.

\end{abstract}

\section{Introduction}

The evolution of baryons within their dark matter halos and the morphologies of the 
resulting systems depend on the merger and accretion history of the parent halos. 
Major efforts of galaxy evolution studies today focus on how galaxies 
acquire the gas which fuels their star formation and what processes drive the 
distinctions between the red sequence and the blue cloud. Still, 
our view of the extragalactic universe is only as complete as our methods 
for cataloging the galaxies that populate it. 
While the public wide area optical/IR and associated spectroscopic surveys are good at 
detecting luminous ellipticals, bright spirals and bursting or active galaxies, 
they are substantially less complete in tracing the low surface brightness,
dwarf and gas-rich galaxy populations that actually dominate the local population. 
Each catalog derived from an individual survey has its own built-in limitations and biases
which affect our ability to construct a true census of the present day universe.

Because of its relatively simple physics, the HI line provides a useful tracer of the cool
gas mass and of the star formation {\it potential} in nearby galaxies and probes the
very population of modest luminosity, gas rich objects which are often underrepresented
in surveys selected by optical/IR properties. While it is clear that most stars 
form out of molecular rather than atomic hydrogen, the molecular clouds themselves develop 
through the collapse of overdensities in the more diffuse, neutral medium. Thus,
while the connection of HI to star formation is on small scales indirect, 
the global HI content serves as a tracer of relative SF potential.
However, at present, HI line measurements yield HI masses $M_{HI}$ for
far fewer galaxies than those for which stellar masses $M_*$ are available from optical/IR
wide area surveys. In fact, only now are HI surveys 
adequate in terms of volume sensitivity to sample a cosmologically
significant volume \citep{Martin10}.

After the pioneering results delivered by small-scale surveys such as the
Arecibo HI Strip Survey \citep[AHISS:][]{Zwaan97} and the Arecibo Dual Beam
Survey\citep[ADBS:][]{RS02},
the advent of multi-feed array receivers on large 
single dish telescopes made possible wide-area 21~cm HI line surveys, such as the 
HI Parkes All-Sky Survey, \citep[HIPASS:][]{Barnes01, Meyer04, Wong06}
and the companion HI Jodrell Bank All-Sky Survey \citep[HIJASS:][]{Lang03}.
While covering a large fraction of the sky, these
surveys failed to sample a cosmologically fair volume  because their mean depth was
too shallow, typically $<40$ Mpc, and they were limited in both 
angular and spectral resolution and in sensitivity. As a result,
HIPASS sampled only sparsely both the most HI-rich --- but rare --- objects 
and the lowest halo mass systems --- detectable only if very nearby and with very
narrow HI line widths--- and,
because of the large Parkes antenna beam (15.5\arcmin), suffered 
from confusion in the identification of optical counterparts (OCs). 

The advent of a similar seven feed array at Arecibo (``ALFA'', the Arecibo L-band Feed Array) 
has enabled a second-generation wide area extragalactic HI line 
survey, ALFALFA, the {\bf A}recibo {\bf L}egacy {\bf F}ast {\bf ALFA} survey 
\citep{Gio05a, Gio05b, Gio08, Haynes08}. Initiated in February 2005,
survey observations are now more than 90\% complete. In this paper, we present the
catalog of HI detections covering about 40\% of the planned survey sky area,
referred to hereafter as the $\alpha.40$ catalog.
Both by design and because of improvements made possible by
the accumulation and analysis of more survey data, the catalog presented here 
both extends and supercedes earlier ones presented by 
\citet{Gio07, Saintonge08, Kent08, Martin09, Stierwalt09}. In addition, the ALFALFA data release
presented here includes, where applicable, a cross reference to the optical survey 
dataset corresponding to Data Release 7 (DR7) of the Sloan Digital Sky Survey 
\citep[SDSS:][]{Abazajian09}. 

The availability now of a large body of ALFALFA data, constituting 40\% of the
expected final survey, allows us to undertake an
examination of the characteristics of its catalog of HI sources. \citet{Martin10}
and \citet{Toribio11a} have presented earlier considerations of survey characteristics
for subsets of the $\alpha.40$ catalog
specifically in the context of using the ALFALFA survey to derive the HI
mass function (HIMF) and to establish a standard of normal HI content for galaxies
in low density environments, respectively. Here, we examine the full
$\alpha.40$ catalog, discuss its identification of optical counterparts, 
and compare parameters derived from its measurements with 
those available in the previous compilation of targeted HI line observations 
presented by \citet{Springob05a}. We also present a more detailed look at the
completeness of $\alpha.40$ and how HI source catalog limitations in general can 
affect measurements of the HIMF.

This paper is organized as follows: In \S\ref{sec:observations}, we discuss the 
observational strategy, sky coverage, and data processing associated with the production of the
ALFALFA dataset and its final data products. \S\ref{sec:datapres} presents the
$\alpha.40$ catalog of HI sources. The identification of the optical counterparts (OCs)
of the HI sources is discussed in \S\ref{sec:oc}. In that section, we present
the crossmatch of the $\alpha.40$ catalog to the SDSS DR7 database and discuss those 
circumstances under which the ALFALFA detection is not associated with an OC.
A comparison of the HI line parameters
derived from the ALFALFA survey with those extracted from the large targeted HI
dataset presented in \citet{Springob05a} is used in \S\ref{sec:validate}
to validate the photometric and spectral calibration underlying the
ALFALFA source parameters. An analysis of the survey completeness and reliability
is presented in \S\ref{sec:completeness} followed in \S\ref{sec:impact} with
a discussion of how the $\alpha.40$ survey characteristics
impact its cosmological applications, in particular, the derivation of the HIMF.
A brief summary of the main points of this paper is given in \S\ref{sec:conclusion}. 

\section{Data}\label{sec:observations}
The ALFALFA observing strategy has been discussed in detail in \citet{Gio05a} and \citet{Kent11}. 
Of particular note to this data release, observations during a given observing session
use the ALFA seven-beam receiver parked on the meridian with data acquired in ``almost
fixed'' drift-scan mode; minor motion of the telescope is permitted so that the position 
of the central beam tracks in constant J2000 Declination. With the feed arm positioned
along the meridian at azimuths near 180$^\circ$ (for declinations north of the Arecibo
zenith at Dec. = 18$^\circ$21\arcmin) ~or 360$^\circ$ (for declinations south of zenith), 
the feed array is rotated by 19$^\circ$ so that the seven beams sweep out tracks 
equally spaced in declination by about 2.1\arcmin. In nearly all circumstances, a given
observing run is dedicated to a single declination track. The 2-D (time versus frequency)
drift scan datasets are converted
from FITS to IDL format and run through an initial bandpass calibration and subtraction,
normally within 24 hours of acquisition. 

In contrast to traditional total power, position switched pointed observations, 
a drift-scan survey (of which ALFALFA is certainly not the first example)
collects spectra continuously (almost) without moving the telescope. In the case of
the ALFALFA survey, the sampling rate is 1 Hz, i.e. a spectrum of 4096 spectral
channels (a ``record'') is recorded every second for each polarization of
every beam of the feed array. The slowly-changing characteristics of the
bandpass with time can thus be monitored effectively. The ALFALFA pipeline does so 
by separately monitoring the behavior of each spectral channel across the time domain,
through a robust, low-order polynomial fit (which skips over sources), outside
of the spectral region dominated by Galactic emission. For each 600 record
unit (a 10 minute drift ``scan''), we thus obtain a two-dimensional map of the
bandpass which can be ``subtracted'' from each spectral record. Such ``sky
subtraction'' is thus conceptually similar to that of the traditional
position-switching mode, although the duration of the ``off'' is much larger
than that ``on'' source, gaining $\sqrt{2}$ in sensitivity with respect to
standard position-switching observations. During the same processing step, 
continuum subtraction is also performed, and a separate continuum map is recorded.

For spectral channels affected by Galactic HI emission, such ``sky subtraction''
is not an option, and the bandpass subtraction cannot be applied in the same
manner as for spectral channels away from the Galactic signal. In this
case, the spectral shape of the bandpass across the Galactic emission region
is adopted as a linear interpolation between the two Galactic emission-free sides of
the spectrum. Thus, the flux calibration of Galactic features processed by the
standard ALFALFA pipeline is not accurate.

Each 2-D bandpass-subtracted dataset for each beam and each polarization
is examined interactively and flagged for radio frequency interference (RFI); regions
characterized by lowered quality  
(due to standing waves, gain instabilities etc) are assigned a lower weight. While this 
step (known as {\it ``flagbb''}) is laborious, the facts that the continuum information is retained
and the RFI is not median filtered away enables the further use of the dataset to look
for HI absorption, for the derivation of upper limits at arbitrary positions in 
3-D, and for stacking analysis \citep{Fabello11}. The flattened and flagged 2-D line and continuum maps
are archived as Level I datasets. 

Once the set of drift scans providing full coverage for a complete strip in declination is
flagged in this manner, the set of evenly gridded data cubes is generated. Details of
the gridding process are given in \citet{Kent11} and summarized here.
The grids are square in the angular dimension, 2.4$^\circ$ on a side,  
evenly sampled at 1\arcmin~ spacing. Their center positions on the sky are spaced 8$^{min}$ apart
in R.A. and centered on odd integer declinations; the spatial dimensions of a grid
are 144 by 144 pixels. For convenient access using modest data
processors, each spatial grid is split into four, partially overlapping subgrids,
each covering 1024 frequency channels. The grid generation algorithm also converts
the spectral intensities from units of antenna temperature to mJy/beam in flux density,
correcting for zenith angle variations in the gain of the telescope. A first step in the 
examination of the grids performs an astrometric fit to the continuum sources within them;
this fit is then used to subtract off the residual telescope pointing errors \citep{Gio07,Kent08,
Kent11}. Grids are then flatfielded and rebaselined in both the angular and spectral dimensions to 
improve their quality by accounting for variations in gain, calibration and other systematic
blemishes. ``Flatfielding'' here corresponds to the process by which pixel-to-pixel
variations within each channel map, caused mainly by continuum fluctuations, are accounted for.
For spectral channels away from Galactic emission, extragalactic HI sources are typically
small in comparison with the angular size of ALFALFA data cubes (``grids'' of $2.4^\circ 
\times 2.4^\circ$). Large--scale variations in the continuum level which may not have been
effectively removed by the bandpass subtraction procedure can be identified by 
robust-fitting a two-dimensional surface (in the angular domain) from the channel map. 
In the absence of very strong continuum sources, this correction is generally small 
and it does not affect noise statistics in any significant way. 

After the angular flat fielding is performed, residual, localized spectral baseline 
features are also removed by subtracting low order polynomial fits to 
the signal free portions of the spectral domain around emission features. These arise,
for example, from standing waves produced by multiple reflections of continuum source
emission within the optical path.
  
Signal extraction is applied following \citet{Saintonge07a}, and once a catalog
of candidate detections has been obtained, the grid is interactively examined, the global
profiles are extracted, fluxes are measured, OCs are identified and remarks are recorded.
It should be noted that this interactive process improves the definition of source
parameters beyond the model fitting used by the automatic signal extractor; this point,
and the resultant reliability and completeness of the catalog,
is discussed more fully in \S\ref{sec:impact}.
The final catalog of sources is constructed following a process of culling poorer quality
detections where a source is contained in adjacent overlapping grids and running a series of
data quality checks.

The catalog presented here supercedes previous ALFALFA data releases for several reasons mainly having
to do with (1) the increased size of the available dataset which yields better understanding of pointing errors,
gain variations and other instrumental artifacts, (2) improved SDSS coverage since the first catalogs
were produced, (3) improvements in the algorithm used to make global profile measurements and (4)
increased contiguous coverage. Some earlier measurements tended to underestimate fluxes
for the brightest and more extended sources, a systematic effect for which a correction
is now applied (see \S\ref{sec:validate} for the comparison of flux density measurements
with published values). In most cases, changes to the flux density measurements included in earlier
data releases are minor, but the current catalog is intended to replace the earlier ones entirely.
It should be noted that further revisions of parameters for sources
located near edges of the current grid coverage will come in the future in those cases
when a newer grid in an adjacent strip better encompasses the source or contributes a higher 
quality dataset. By its nature as a cumulative drift scan survey, the harvest of ALFALFA
will both grow and improve over time.

\begin{figure}
\begin{center}
\plotfiddle{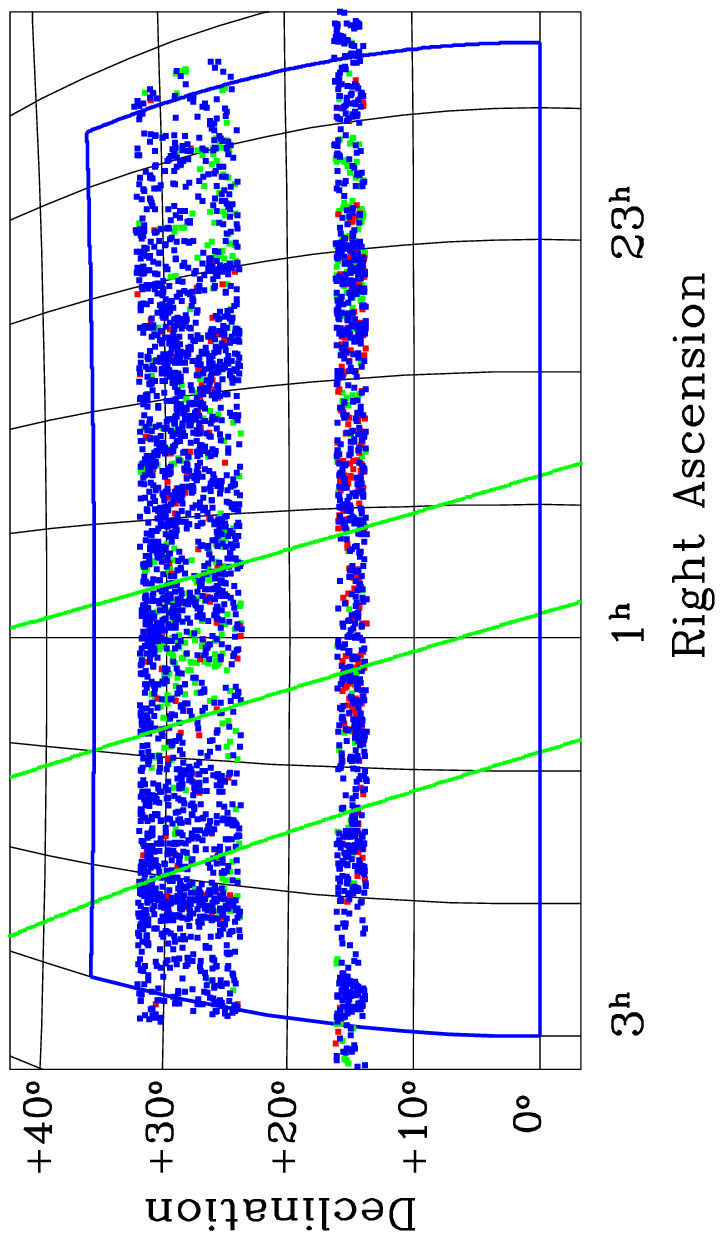}{2.0in}{0}{87}{87}{-580}{-335}
\plotfiddle{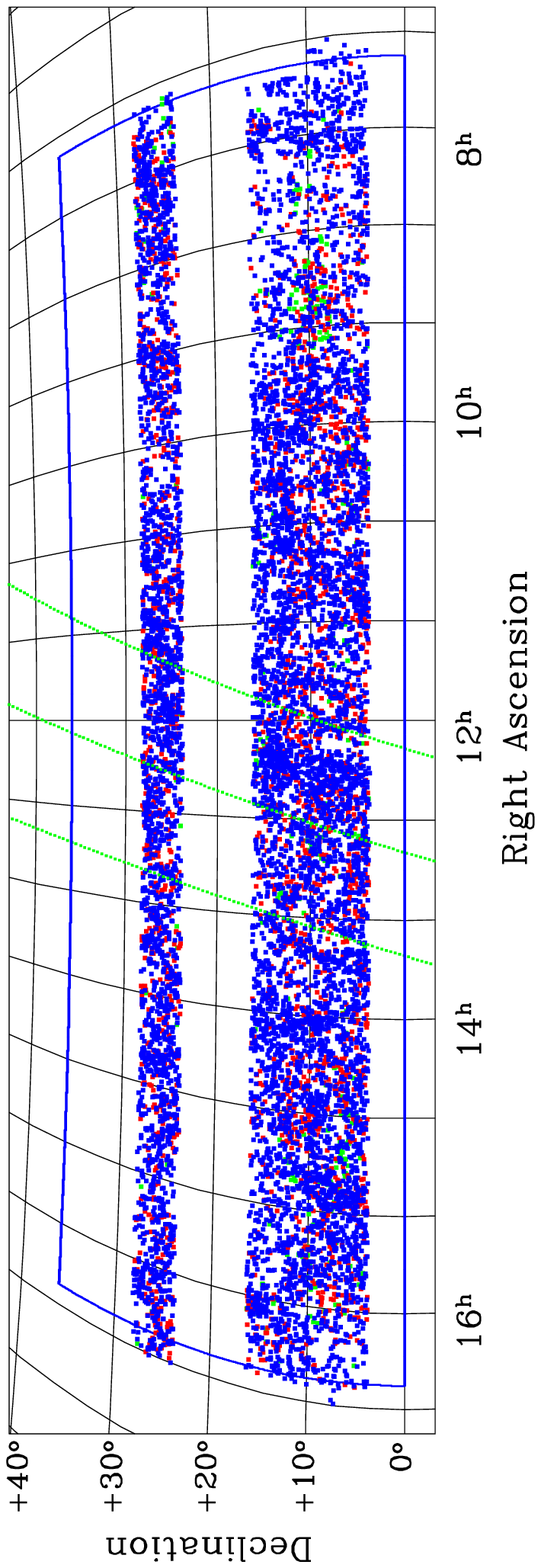}{2.0in}{0}{87}{87}{-360}{-280}
\end{center}
\vskip 6.5cm
\caption{Sky distribution, in equatorial coordinates on 
an Aitoff grid projection, of the current $\alpha.40$ catalog detections.
Upper panel: the ``fall ALFALFA sky'' (anti-Virgo direction) region; 
lower panel: the ``spring ALFALFA sky'' (Virgo direction) region.
Blue, red and green symbols identify the Code 1 (best quality), 2 
(priors) and 9 (HVC) sources respectively. The green diagonal lines in
each panel trace
the supergalactic plane and SGL $\pm$ 10$^\circ$.}\label{fig:aitoff}
\end{figure}

The full ALFALFA survey is intended to cover 7000 \sqd~ of sky in two regions of high
Galactic latitude within $18^\circ$ of the Arecibo zenith. All declinations will
be covered $0^\circ <$ Dec. $< +36^\circ$. Since all observations are conducted
during nighttime hours, the two regions are referred to as ``spring'' and ``fall'.
The ``spring'' region extends from 
$07^h30^m <$ R.A. $< 16^h30^m$ while the ``fall'' ALFALFA region encompasses from
$22^h <$ R.A. $< 03^h$.  Some sources are found outside the stated R.A. boundaries
where the actual drift scan observations extended beyond the nominal map area.
Some priority has been given to completing areas within the SDSS spectroscopic survey 
footprint, and the pace of observing has been dictated by the availability of telescope
time. Figure \ref{fig:aitoff} illustrates the area of the sky contained in the 
$\alpha.40$ catalog presented here: regions $07^h30^m <$ R.A. $< 16^h30^m$,
$+04^\circ <$ Dec. $< +16^\circ$ and $+24^\circ <$ Dec. $< +28^\circ$  (the ``spring'' region)
and $22^h <$ R.A. $< 03^h$, $+14^\circ <$ Dec. $< +16^\circ$ 
and $+24^\circ <$ Dec. $< +32^\circ$ (the ``fall'' region).

\section{Catalog Presentation}\label{sec:datapres}

We present in Table \ref{tab:catalog} the measured parameters for \ntot \ detections, \ngal \ of which are 
certainly
associated with extragalactic objects. An additional \nhvc \ are detected
at velocities which suggest they may not be
extragalactic but are more likely to be Galactic high velocity cloud (HVC) features. 
The contents of Table \ref{tab:catalog} are 
as follows:

\begin{itemize}

\item Col. 1: Entry number in the Arecibo General Catalog (AGC), a private database
   of extragalactic objects maintained by M.P.H. and R.G. The AGC entry normally 
   corresponds both to the OC and the HI line source except in the cases of 
   HVCs and other HI sources 
   which cannot be associated with an optical object with any high
   degree of probability. In those cases, the AGC number corresponds only to the
   HI detection. An AGC number is assigned to all ALFALFA sources; it is intended to be used as
   the basic cross reference for identifying and tracking ALFALFA sources as new data acquired in overlapping
   regions supercedes older results. Note that in previous ALFALFA catalogs, an index number was used, 
   a practice no longer employed; a cross-reference to these older identifications is provided in 
   Table \ref{tab:comments}. The designation of an 
   ALFALFA source referring only to its HI emission (without regard to its OC)
   should be given using the prefix ``HI'' followed by
   the position of the HI centroid as given in Col. 3 of Table \ref{tab:catalog}.

\item Col. 2: Common name of the associated OC, where applicable. Further discussion of the process of
   assigning optical counterparts is presented in \S\ref{sec:ocfind}.

\item Col. 3: Centroid (J2000) of the HI line source, in hhmmss.sSddmmss, after correction for systematic 
   telescope pointing errors, which are on the order of 20\arcsec \ and depend on declination. The systematic
   pointing corrections are derived from an astrometric solution for the NRAO Very Large Array Sky Survey (NVSS)
   radio continuum sources  \citep{Condon98} found in the grids.  As discussed in \citet{Gio07} and 
   \citet{Kent08}, the assessment of centroiding errors is complicated
   by the nature of 3-D grid construction from the 2-D drift scans, those often acquired in widely
   separated observing runs, and, for resolved/confused sources, unknown source structure. As those authors suggest
   the best assessment of HI centroid error is 
   accomplished by comparison of the HI centroids with the positions of the adopted OCs. An analysis of the positional
   offsets of the HI centroids from the positions of the OCs yields a relation for
   the median error in the HI position err$_{med, HI}$ as a function of the signal-to-noise ratio, 
   S/N (see Col. 7), for the $\alpha.40$ sample:
    \begin{equation}
     err_{med,HI} (arcsec) = \left\{
     \begin{array}{lr}
      71.\, - 79. ~log S/N  \, + 26.~log (S/N)^2 \,\,\, &   log S/N < 1.6 \\
       11 \,\,\,                         &   log S/N \geq 1.6
     \end{array}
   \right.
    \label{eq:poserr}
    \end{equation}
    On average, the positional offset is about 18\arcsec, but it can, in rare instances exceed 
    1\arcmin; those cases
    are noted in the comments included in Table \ref{tab:comments}.

\item Col. 4: Centroid (J2000) of the most probable OC, in hhmmss.sSddmmss, associated with 
        the HI line source, where applicable.  The OC has been identified and its likelihood
	has been assessed interactively using tools provided
	through the {\it SkyView} website or the SDSS Explore Tool, in addition to to the NASA Extragalactic
        Database (NED) and the AGC and make
        use of judgmental criteria including redshift (when known), size, morphology and optical color. 
        The optical positions are normally estimated to be 3\arcsec~ or better but may be larger in exceptional 
        cases (very low surface brightness or peculiar, disturbed objects). The process of assignment of the most
        probable OC is discussed in \S\ref{sec:ocfind}. It should be noted that only one OC is assigned per
        HI source although in reality confusion within the telescope beam is a possibility. Suspected cases
        of confusion or ambigous assignment of the OC are noted in the comments included in 
        Table \ref{tab:comments}.

\item Col. 5: Heliocentric velocity of the HI source, $cz_{\odot}$ in \kms, measured as the midpoint 
	between the channels at which the flux density drops to 50\% of 
	each of the two peaks (or of one, if only one is present) at each
	side of the spectral feature; see also \citet{Springob05a}.
        The error on $cz_\odot$
	to be adopted is half the error on the width, tabulated in Col. 6.

\item Col. 6: Velocity width of the HI line profile, $W_{50}$ in \kms, measured at the 50\%
	level of each of the two peaks, as described in Col. 5 and corrected
	for instrumental broadening. No corrections due to
	turbulent motions, disk inclination or cosmological effects are
	applied. The estimated error on the velocity width, $\epsilon_w$, in \kms, follows,
        in parentheses. This error is the sum in quadrature of two components: a
	statistical error and a systematic error associated with the subjective guess with 
        which the person performing parameter extraction estimates the spectral boundaries of 
        the feature, flagged 
        during the interactive assessment of candidate detections. In the majority of cases,
	the systematic error is significantly smaller than the statistical
	error; thus the former is ignored.

\item Col. 7: Integrated HI line flux density of the source, $S_{21}$, in Jy~\kms. This value
	corresponds to the total HI line flux measured on the integrated spectrum obtained by
	spatially integrating the source image over a solid angle of at
        least $7$\arcmin $\times 7$\arcmin ~and dividing by the sum of the survey beam
	values over the same set of image pixels \citep[see][]{Shostak80, Kent11}. 
	Estimates of integrated flux densities for very extended sources with
	significant angular asymmetries can be misestimated by our 
	algorithm, which is optimized for measuring sources comparable with
	or smaller than the survey beam. A special catalog with parameters
	of extended sources will be produced after completion of the survey. 
	The issue is especially severe for extended HVCs
	that exceed in size that of the ALFALFA data cubes. In these specific 
	cases, only the flux in the knots of emission is measured. In general, 
        the HVCs have been catalogued here applying the same 
	kind of S/N selection threshold as for the extragalactic signals, with 
	the exception of the southern extension of Wright's cloud, where, in addition
	to a bulk measurement of the portion of the cloud lying within this region, a 
	selection of the brightest knots was measured to trace the structure.
	See Column 12
	and the corresponding comments for individual objects. 
	The estimated uncertainty of the integrated flux density, in Jy \kms, 
	is given in parentheses. 

\item Col. 8: Signal--to--noise ratio S/N of the detection, estimated as 
   \begin{equation}
	S/N=\left (~\frac{1000S_{21}}{W_{50}} \right ) \frac{w_{smo}^{1/2}}{\sigma_{rms}}
	\label{eq:eqsn}
	\end{equation}
   where $S_{21}$ is the integrated flux density in Jy \kms, as listed in Col. 7;
   the ratio $1000 S_{21}/W{50}$ is the mean flux density across the feature in mJy;
   $w_{smo}$ is either $W_{50}/(2\times 10)$ for $W_{50}<400$ \kms \ or
   $400/(2\times 10)=20$ for $W_{50} \geq 400$ \kms ($w_{smo}$ is a
   smoothing width expressed as the number of spectral resolution
   bins of 10 \kms \ bridging half of the signal width; the raw spectra are
   sampled at 24.4 kHz $\sim$ 5.5 \kms ~at $z\sim0$); and $\sigma_{rms}$
   is the r.m.s noise figure across the spectrum measured in mJy at 10
	\kms \ resolution, as tabulated in Col. 9.

\item Col. 9: Noise figure of the spatially integrated spectral profile, $\sigma_{rms}$,
	in mJy. The noise figure as tabulated is the r.m.s. as measured over the signal-- and
	RFI--free portions of the spectrum, after Hanning smoothing to a spectral
	resolution of 10 \kms.

\item Col. 10: Adopted distance in Mpc, $D_{Mpc}$. For objects with $cz_{\odot} > 6000$ \kms, 
	the distance is simply estimated as $cz_{cmb}/H_\circ$ where $cz_{cmb}$ is the recessional velocity
	measured in the Cosmic Microwave Background reference frame \citep{Lineweaver96} and $H_\circ$ is
	the Hubble constant, adopted to be 70 \kms Mpc$^{-1}$. For objects with
	$cz_{cmb} < 6000$ \kms, we use the local universe peculiar velocity model of \citet{Masters05}, 
        which is based on data
	from the SFI++ catalog of galaxies \citep{Springob07} and results
	from analysis of the peculiar motions of galaxies, groups, and clusters, using a combination of primary
	distances from the literature and secondary distances from the Tully-Fisher relation. The resulting model
	includes two attractors, with infall onto the Virgo Cluster and the Hydra-Centaurus Supercluster, as well as
	a quadrupole and a dipole component. The transition from one distance estimation method to the other
	is selected to be at $cz_{\odot}=6000$ \kms \ because the uncertainties in each method become 
        comparable at that distance.
	Where available, primary distances as available in the published literature are adopted. When the galaxy
	is a known member of a group \citep{Springob07}, the group systemic recessional velocity $cz_{cmb}$ is 
        used to determine the distance
	estimate according to the general prescription just described.
	
\item Col. 11: Logarithm of the HI mass $M_{HI}$, in solar units, computed via the standard
	formula $M_{HI}=2.356\times 10^5 D_{Mpc}^2 S_{21}$ and assuming the distance given in Col. 10.
        No correction for HI self-absorption has been applied.

\item Col. 12: This column contains three relevant coded flags:

        The first code, assigned as an integer value of 1, 2 or 9, refers to the category of 
        the HI detection defined as follows: 
	
	\hskip 8pt Code 1 refers to sources 
	of S/N and general qualities that make it a reliable detection. These signals exhibit
	a good match between the two independent polarizations observed by ALFALFA,
	a spatial extent consistent with the telescope beam (or larger), an RFI-free spectral profile,
	and an approximate minimum S/N threshold of 6.5 \citep{Saintonge07a}. These
	criteria lead to the exclusion of some candidate detections with S/N $>6.5$; likewise, some
	features with S/N slightly below this soft threshold are included, due to optimal overall characteristics
	of the feature, such as well-defined spatial extent, broad velocity width, and obvious association with an OC. 
        We estimate that the detections with code 1 in Table \ref{tab:catalog} are nearly 100\% reliable; 
        the completeness and reliability of
        the $\alpha.40$ catalog are discussed in \S\ref{sec:impact}.

	\hskip 8pt Code 2 refers to sources categorized as ``priors''. They are sources of low S/N ($\lesssim$ 6.5),  
        which would ordinarily not be considered
	reliable detections by the criteria set for code 1, but which have been matched with OCs with
	known optical redshifts coincident (to within their errors) with those measured in the HI line. We include 
        them in our catalog because they are very likely to be real.  In general, however, they should not be used in 
        statistical studies which require well-defined completeness limits; this point is further discussed in \S\ref{sec:impact}.

	\hskip 8pt Code 9 refers to objects assumed to be HVCs; no
	estimate of their distances is made.

        Of the \ntot\ sources included in this data release, \ncone\ are classified as source code 1, \nctwo\
        are code 2, and \nhvc\ are code 9.

        The second code, assigned as an alphabetic character, refers to a category reflecting the status of the
        cross identification of the ALFALFA detection with an entry in the SDSS DR7 database, as judged by the 
        ALFALFA team. This code is used to identify galaxies which lie outside the SDSS DR7 sky footprint or for which 
        there are clearly issues with the identification. It should be noted that this code refers only to the cross
        match with SDSS DR7. The cross-reference and basic parameters of the OCs is given in Table \ref{tab:sdsscross}.
        This code and its interpretation are as follows:

        \hskip 8pt  I: \hskip 3pt ``identified'': The PhotoObjID is set but no other indicative flags have been
                  applied; this code applies whether or not there is a SDSS spectroscopic counterpart.

        \hskip 8pt   O: \hskip 3pt ``outside DR7'': The SDSS OC lies outside of the SDSS DR7 footprint and thus
                                   no DR7 crossmatch can be performed. 

        \hskip 8pt   U: \hskip 3pt ``unidentified'': No SDSS OC has been identified, but the object lies within
                                   the SDSS DR7 footprint.

        \hskip 8pt   N: \hskip 3pt ``no DR7 photometric ID'': No SDSS DR7 photometric source has been identified; 
                 assignment of this code can result from proximity to
                 bright star, satellite trails, incomplete coverage or for other reasons.
  
        \hskip 8pt   M: \hskip 3pt ``missing'': The OC is in the SDSS DR7 footprint region but
         neither a PhotoObjID or a SpecObjID are returned to queries of the SDSS DR7 database.

        \hskip 8pt   P: \hskip 3pt ``photometry suspect'': The SDSS DR7 photometry for the associated 
         PhotoObjID are suspect for some reason 
         as judged by the ALFALFA team. Assignment of this code often is associated with the identification
         of multiple near-equal-flux photometric objects within an obviously single OC.
         Such cases apply often to very large optical objects or to faint, low surface brightness and/or patchy 
         systems. The optical photometry associated with the SDSS ``parent'' object may be adequate but caution
         should be exercised.

         \hskip 8pt   D: \hskip 3pt ``displaced SDSS object'': The SDSS Photo/SpectID is displaced from 
         the optical galaxy center, as identified
         by ALFALFA team. The PhotoObjID may be legitimate; often this is brightest photometric ``child''. Because of the
         displacement, the SDSS redshift may not reflect the systemic recessional velocity of the galaxy.

         \hskip 8pt  T: \hskip 3pt ``two SDSS objects'': The SDSS PhotoObjID associated with the 
         galaxy center is displaced from the
         target associated with the SDSS SpectObjID, as judged by the ALFALFA team, i.e., 
         the best PhotoObjID does not coincide with the SpectObjID. Usually, 
         the SpectObjID is an offcenter HII region or other bright knot
         within the target galaxy.

         \hskip 8pt   S: \hskip 3pt ``superposed SDSS object'': The SDSS redshift corresponds to a superposed 
         foreground star or background QSO.

        \hskip 8pt    B: \hskip 3pt ``bad SDSS solution'' The SDSS redshift is unreliable or rejected for some 
         unspecified reason.

 	The third code, given as an asterisk where applicable, indicates that a comment regarding the HI detection
        and/or the assignment of the OC is included for this source in Table \ref{tab:comments}.

\end{itemize}

Only the first few entries of Table \ref{tab:catalog} are listed in the printed version of this
paper. The full content of Table \ref{tab:catalog} is accessible through the electronic version
of the paper and will be made available also through our public digital 
archive site\footnotemark \footnotetext{http://arecibo.tc.cornell.edu/hiarchive/alfalfa/} and
the ALFALFA project data site\footnotemark \footnotetext{http://egg.astro.cornell.edu/alfalfa/data}.

In addition to the HI emission sources presented in Table \ref{tab:catalog}, it is expected that the ALFALFA
spectral cubes will also contain evidence for HI in absorption. \citet{Darling11} discuss a pilot program
which uses an adaptation of the ALFALFA pipeline to search for HI absorption along the line of sight to NVSS
sources in a small number of the ALFALFA cubes. The known HI absorber in the interacting system UGC 6081 was
recovered. Because the standard ALFALFA reduction is not designed to look for such phenomena, the HI 
absorption detection is not included in Table \ref{tab:catalog}, and the reader is referred to 
\citet{Darling11} for its parameters.

Table \ref{tab:comments} contains comments about entries in Table \ref{tab:catalog} which have been
recorded in the course of extracting source parameters and identifying the OCs.
The second column contains a cross-reference to the catalog identification used in earlier papers,
which is no longer used. We repeat in Column 3 the HI detection code assigned for each source (the
first code in Column 12 of Table \ref{tab:catalog} described above). It should be noted that
angular separations given in these comments are referenced to the centroid of the HI source, not the
position of the OC. These notes are somewhat heterogeneous in nature, having been incorporated
during the process of data reduction by the individual responsible for source extraction. Since the
extraction has been performed over a period of several years, the databases available to the
person making the comments have evolved; thus the mention of nearby neighbors is not intended
to be complete and should not be used in any derivation of local density. In some
cases these notes identify issues with data quality, certainty of the OC 
or parameter extraction. The presence of a note does not mean necessarily that parameters
are less certain than their errors indicate, as we have a tendency to err on the conservative
side of casting doubt. They are included here because they provide an additional contribution to
the legacy value of the dataset.

\begin{figure}
\begin{center}
\epsscale{0.7}
\plotone{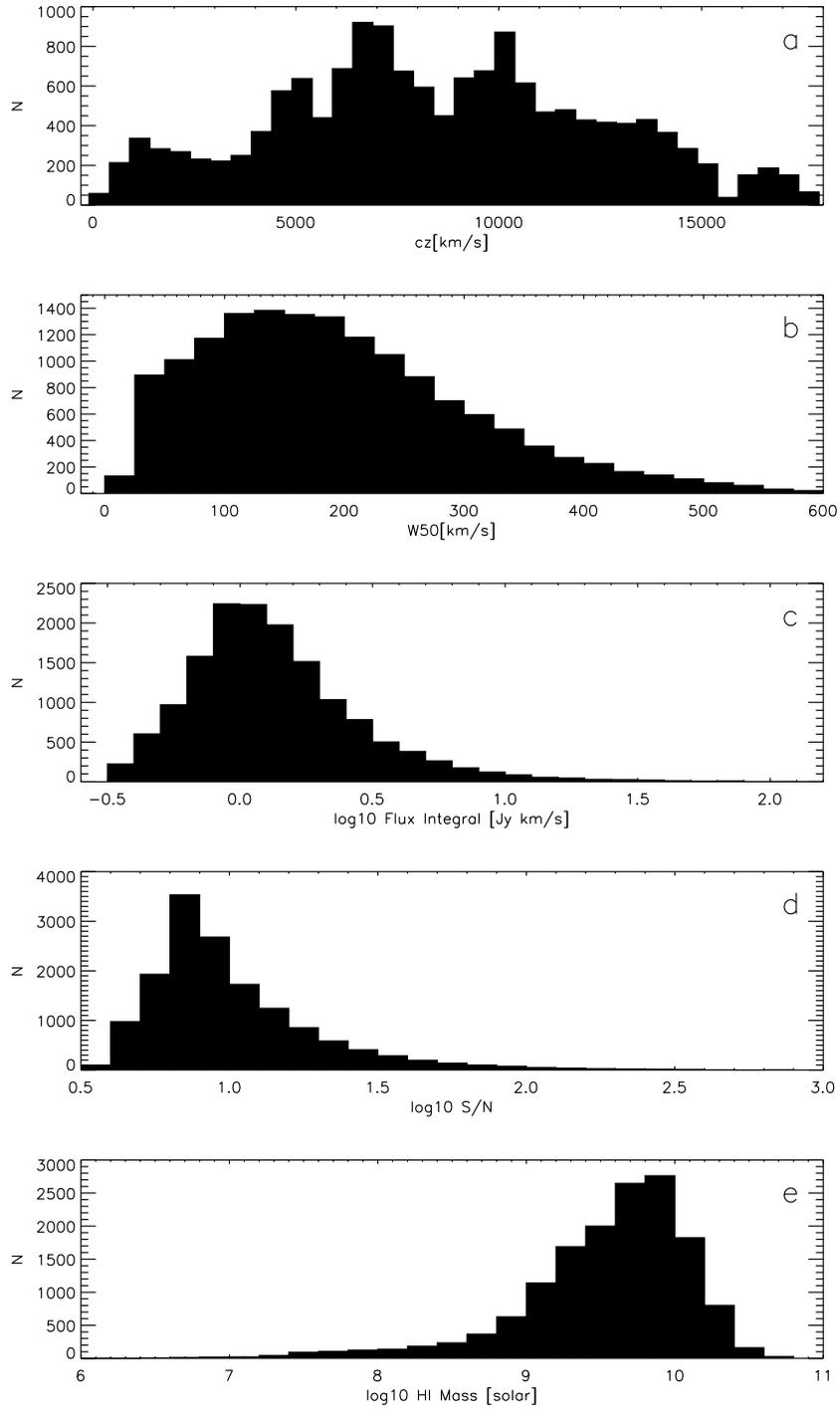}
\end{center}
\caption{Histograms of the distributions of redshift $cz$, $W_{50}$, log $S_{21}$, log S/N and log $M_{HI}$
(top to bottom) for the $\alpha.40$ catalog sample presented in Table \ref{tab:catalog}.}\label{fig:histograms}
\end{figure}

\begin{figure}
\begin{center}
\epsscale{1.0}
\plotone{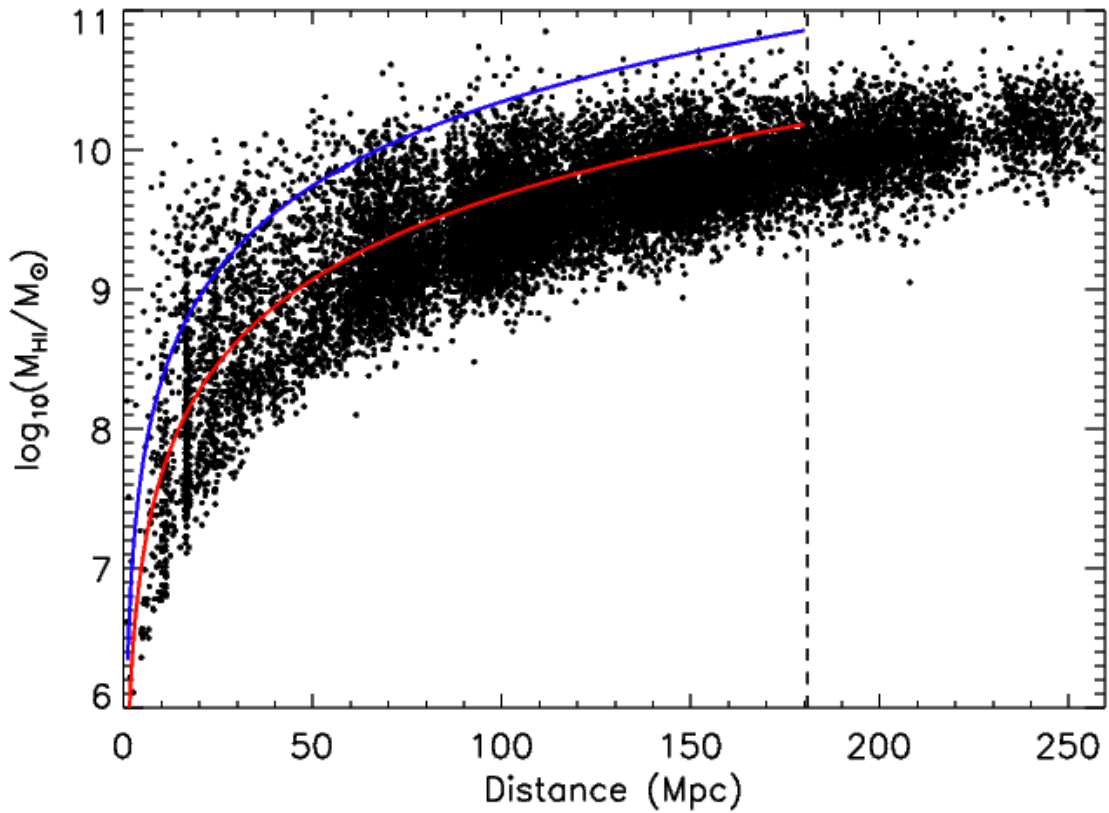}
\end{center}
\caption{Spaenhauer diagram
for the $\alpha.40$ catalog sample presented in Table \ref{tab:catalog}.
The superposed blue (upper) curve traces the HIPASS completeness limit, while the
red (lower) curve traces that survey's detection limit. The vertical dashed line
indicates the outer limit in distance corresponding to the HIPASS bandpass
edge; HIPASS did not sample any volume at larger distances. The vertical
overdensity points
evident at 17 Mpc is the Virgo cluster; the paucity of points at $\sim$225 Mpc
arises because many nights of ALFALFA observations are contaminated by strong
RFI generated by the FAA radar at the San Juan airport. A less pronounced
gap evident at $\sim$85 Mpc arises from occasional much milder contamination 
from a harmonic of the radar at 1380 MHz and from rare burst events associated with 
the US Air Force NUclear DETonation detection (NUDET) system aboard the 
Global Positioning System (GPS) which transmits at 1381 MHz.}\label{fig:spanplot}
\end{figure}

Subsets of the $\alpha.40$ catalog have been included in the derivation of the HI mass function
\citep{Martin10} and the HI width function \citep{Papastergis11}; both papers include discussion
of the sample characteristics, limitations and biases. Similar to figures shown in those
papers, Figure \ref{fig:histograms} illustrates
the distributions of (top to bottom) redshift $cz$, $W_{50}$, log $S_{21}$, log S/N and log $M_{HI}$
for the full ALFALFA $\alpha.40$ sample presented in Table \ref{tab:catalog}, while
Figure \ref{fig:spanplot} shows the corresponding Spaenhauer plot. Further discussion of the
impact of survey characteristics on cosmological issues and specifically on the derivation of
the HI mass function is given in \S\ref{sec:impact}.

\section{Optical Counterparts of ALFALFA Sources}\label{sec:oc}

The principal aim of ALFALFA is to catalog all gas-bearing extragalactic objects in the local universe.
An integral part of understanding this HI census is similarly identifying the stellar counterpart associated
with each HI source, or even more importantly, rejecting that such a counterpart exists. During the
ALFALFA data reduction process, optical images from the Palomar Digital Sky Survey (DSS2)
and, where available, the SDSS are interactively examined alongside the ALFALFA HI
dataset and the most probable OC is identified and recorded. While this assignment may not 
be correct in individual cases, it provides a first approach to understanding the relationship
between the HI source and its stellar counterpart. The notes included in Table \ref{tab:comments} record
comments on this process made by the ALFALFA team member performing this interactive
stage of the data analysis. In this section, we describe 
the process by which OCs are identified and discuss unresolved issues, provide a cross reference
of sources to the SDSS DR7 database and summarize general results on the evidence
for ``optically-dark'' galaxies.

\subsection{Identifying Optical Counterparts}\label{sec:ocfind}

We make use of Virtual Observatory tools embedded
in the IDL-based ALFALFA reduction package \citep{Kent11} to access several public imaging and
catalog databases at several stages in the data reduction process. During the process
of HI parameter measurement (the routine called {\it ``galflux''}), both DSS2(B) and SDSS
images are examined to identify interactively the most probable OC of each ALFALFA source.
Because of their generally superior quality and ancillary information, preference is given 
to the SDSS images where they are available. 
Entries in our internal AGC database as well as those listed in the NASA Extragalactic Database (NED) 
can be retrieved and examined. The ALFALFA team member processing each source uses the
available public information as to color, morphology, redshift, separation from the HI centroid
in combination with his/her scientific judgement in assigning an optical counterpart.
It should be noted however that because the catalogued data presented here were 
reduced over a three year time period, not all current information/data were available
at the time this assignment was made. Consistency checks are made later to look for 
redshift discrepancies or cases of large positional offset. 

\begin{figure}
\begin{center}
\epsscale{1.0}
\plotone{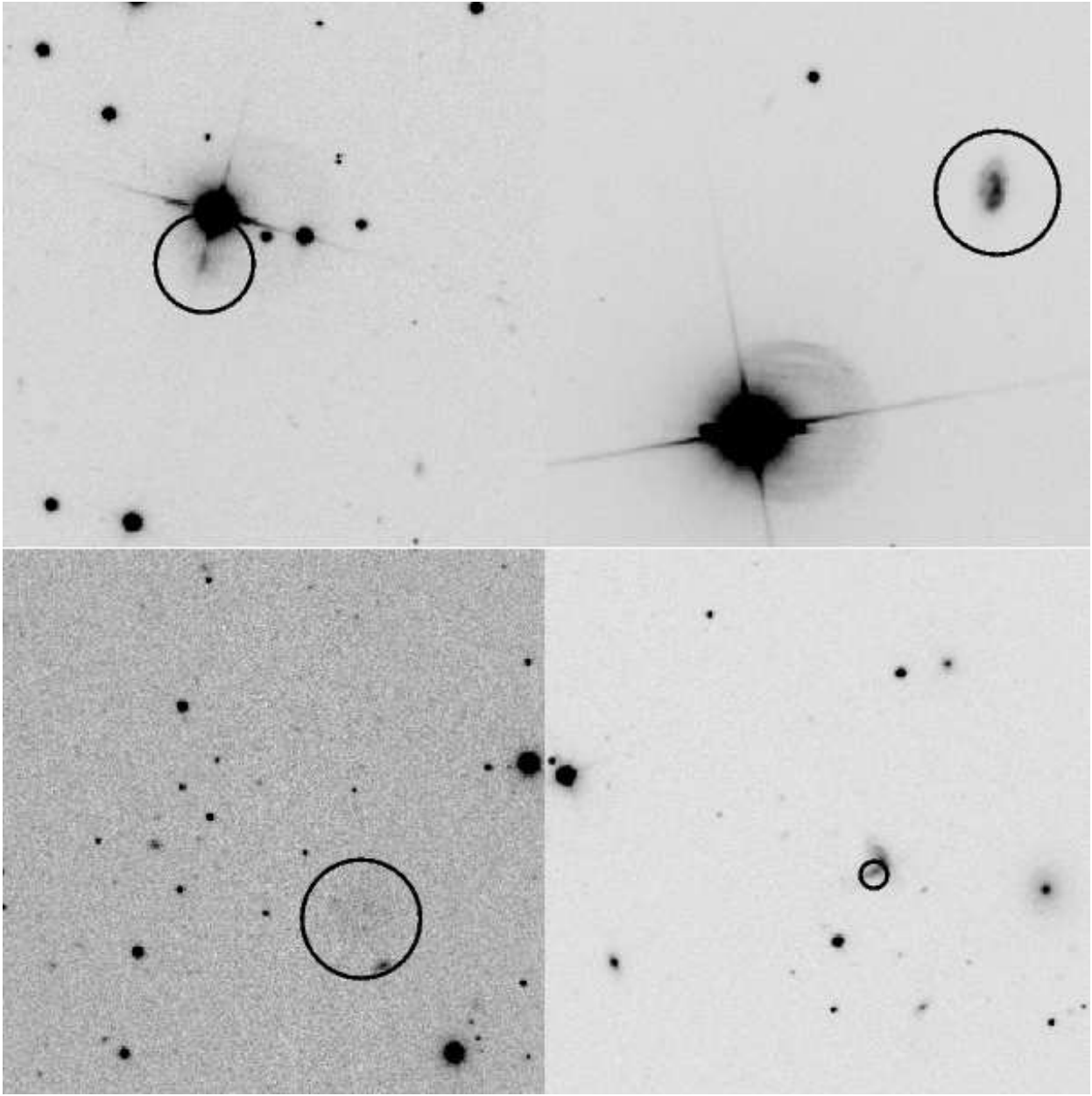}
\caption{Illustrative examples of issues related to the identification of
the OCs of ALFALFA HI sources. Each panel is a 3\arcmin ~by 3\arcmin ~frame
extracted from the Montage data product of SDSS g-band
images centered on the position of the ALFALFA HI source. 
In each frame, the superposed circle, of arbitrary
size, identifies the adopted OC. See text for details of individual cases.}
\label{fig:ocfind}
\end{center}
\end{figure}

With that caveat in mind, Figure \ref{fig:ocfind} shows several examples which illustrate the process
of identification of OCs and the uncertainties inherent in it. Each panel shows a 3\arcmin ~by
3\arcmin~ SDSS g-band image centered on the HI centroid. The superposed circle marks the
OC identified in Table \ref{tab:catalog}; the size of each circle is arbitrarily chosen for
best illustration of the target. The panels are intended to illustrate some of 
the challenges of assigning the OC by highlighting four specific cases as follows:

\begin{itemize}

\item The upper left image is centered on the best-fit position of the HI source
detected at HI095452.2,+142907, a weak
source of S/N = 7.3. The corresponding OC AGC 193821 is identified as the small galaxy SDSS 
J095453.79+142910.0
22\arcsec ~from the HI centroid and partly contaminated by the diffraction spike of 
the bright foreground star; the galaxy is more evident in the DSS2(B) image. There is no
further optical information.

\item The upper right image is centered on the position of HI123120.9+050402, a marginal
ALFALFA detection with a S/N = 4.9. The OC AGC 220720 is identified as 
VCC 1347 = CGCG 042-143 = J123117.00+050429.3, a small
spiral galaxy offset from the HI centroid by about 64\arcsec; the large offset is not
surprising given the low S/N of the HI detection. The SDSS optical redshift is 9830 $\pm$ 30 \kms,
slightly off the HI $cz$ of 9873 $\pm$ 4 \kms. Because of the low S/N of the HI emission profile
but the coincidence with an optical galaxy with an adequately close redshift match, the
optical identification is made and the source is designated as a ``prior'' and
assigned an ALFALFA detection category code of 2.

\item The lower left image is centered on the position of HI 152240.3+055017, a very narrow
($W_{50} = 24$ \kms) feature at $cz$ = 1796 \kms. The OC is identified as a dwarf galaxy
AGC 258471
better evident in the DSS2(B) image at J152238.7+054945; the SDSS pipeline identifies at least
five photometric objects within the LSB emission associated with the dwarf so that its magnitude
is poorly measured. The offset of the HI centroid from the optical object is about 38\arcsec.

\item The lower right image is centered on the position of HI 160743.9+272201, a source of
S/N = 10.9. As evident in the SDSS g-band images, there are several objects in the field,
including a close pair associated with SDSS spectroscopic target 
J160744.75+272140.2 = KUG 1605+275 NED01 with 
a redshift from the SDSS of 23676 $\pm$ 31 \kms. The redshift is too high to be
associated with the ALFALFA HI source; several other galaxies in the vicinity
of this system have similar redshifts. Careful
examination of the SDSS image shows a second object, which is not identified in the
SDSS photometric database and which appears to be partly overlapping with J160744.75+272140.2
but in its foreground at J160743.9+272201. We identify the HI source with this
foreground blue galaxy which becomes AGC 749361.

\end{itemize}

We emphasize again that because of the ALFALFA centroid position uncertainty and its relatively 
large beam size, assignment of the most probable OC is a reasonable but not a perfect process.
Furthermore, it will continue to be a dynamic one, striving for improvement when new data provide
improved detail. For example, the current dataset does not include yet a systematic incorporation
of data from the SDSS III survey or its DR8.

\subsection{Cross Reference with the SDSS}\label{sec:sdsscross}

Although not available at the time of earlier ALFALFA data releases, the completion of the SDSS
legacy survey has afforded the opportunity to cross reference the ALFALFA and SDSS datasets where the two
share footprints. As a new feature of this and future ALFALFA catalog releases, here we provide, 
in Table \ref{tab:sdsscross}, the cross identifications of ALFALFA sources with the photometric 
and spectroscopic catalogs associated with the SDSS,
in this instance, with the data release DR7 \citep{Abazajian09}. 
Entries in Table \ref{tab:sdsscross} are as follows:

\begin{itemize}

\item Col. 1: the source AGC number, identical to Col. 1 of Table \ref{tab:catalog}

\item Col. 2: the HI detection category code, identical to the 1st (integer) code in Col. 12 of Table \ref{tab:catalog}

\item Col. 3: The SDSS cross reference catagory, identical to the 2nd code in Col. 12 of Table \ref{tab:catalog}

\item Col. 4: The SDSS DR7 photometric catalog object identification number (PhotoObjID), where applicable. 

\item Col. 5: The SDSS DR7 spectroscopic catalog object identification number (SpecObjID), where applicable.

\item Col. 6: The r-band model magnitude corresponding to the photometric object or its SDSS parent.

\item Col. 7: The (u-r) color associated with the OC from the SDSS as reported in the DR7.
              This value is used in Figure \ref{fig:baldryplot}, and in order to allow direct comparison with Figure
              9 of \citet{Baldry04}, it has not been corrected for extinction or redshift.

\item Col. 8: The redshift corresponding to the SDSS DR7 spectroscopic catalog object, extracted from the SDSS 
              DR7 database, where applicable.

\item Col. 9: The error on the redshift given in Col. 8, extracted from the SDSS DR7 database, where applicable.

\end{itemize}

It is important that potential users understand the limitations associated with this ALFALFA--SDSS cross-reference.
As noted by \citet{Gio07} and discussed in \S\ref{sec:datapres}, the ALFALFA HI centroid accuracy is of order 20\arcsec, but 
increases as the S/N decreases, as given in Equation \ref{eq:poserr}. Furthermore, as is well known, the standard SDSS 
image reduction pipeline suffers from source blending, and more importantly, shredding, particularly
in the sources whose light distributions are patchy or of low surface brightness. The current ALFALFA
reduction process includes an interactive step of direct examination of the SDSS imagery and issues
associated with blending/shredding are noted immediately. However, earlier ALFALFA datasets which
predated release of DR7 were not subject to such individual cross examination. While attempts
have been made to flag and check suspicious cases, it is likely that some misidentifications remain.

The DR7 photometric catalog object identification number given in Column (4)
is the PhotoObjID whose magnitude and position given in SDSS DR7 corresponds most likely to the OC; the
actual ``best magnitude'' may be associated with the SDSS pipeline ``parent''. Users are cautioned
to understand fully issues associated with blending, shredding and poor sky subtraction and to make
use of warning flags and other quality indicators when using the photometry associated with
the photometric object given here. Particularly relevant discussions of background subtraction
issues are given in \citet{West10} and \citet{Blanton11}.
Similarly, the spectroscopic identification refers to the
most probable and most closely related SDSS spectroscopic target. This cross match likewise 
can suffer from issues of positional offset, signal-to-noise etc. and should be treated with
similar caution. The SDSS DR7 cross reference category given in Col. 3. of Table \ref{tab:sdsscross}
(and also as one of the two codes given in Col. 12 of Table \ref{tab:catalog}) provides
further comment on quality issues as identified by members of the ALFALFA team. However,
because some of the processing of ALFALFA data predates the release of SDSS DR7, this
code assignment should not be considered complete: many but not all sources have been revisited
after the release of DR7. The intent of providing the cross reference is to make statistical studies
more convenient and potentially homogeneous. But again, we emphasize the importance of
visual examination of individual cases where such attention is critical to the 
drawing of scientific conclusions.

Of the \ngal ~extragalactic (i.e., non-HVC) objects listed in Table \ref{tab:catalog}, 2312
lie outside the SDSS DR7 footprint and \ndxgal ~are classified as ``dark'' 
(see \S\ref{sec:darkgals}).
Of the ones with identified OCs and included in DR7, 11740 are
assigned SDSS code ``I''  (meaning the SDSS photometric identification is acceptable and there
is no issue with the spectroscopic identification where such exists)
while the others are given a code in Table \ref{tab:sdsscross}
indicating a recognized issue with either the SDSS photometry or spectroscopy. The ALFALFA fall portion
of the sky contains some regions for which only photometry is available; in the spring region,
the photometric and spectroscopic footprints overlap more completely. Of the
11240 ALFALFA spring sky galaxies with a corresponding SDSS photometric ID (of any code), 9377 (83\%)
have an associated entry in the SDSS spectroscopic catalog, and 1863 (17\%) do not.

\begin{figure}
\begin{center}
\epsscale{1.1}
\plotone{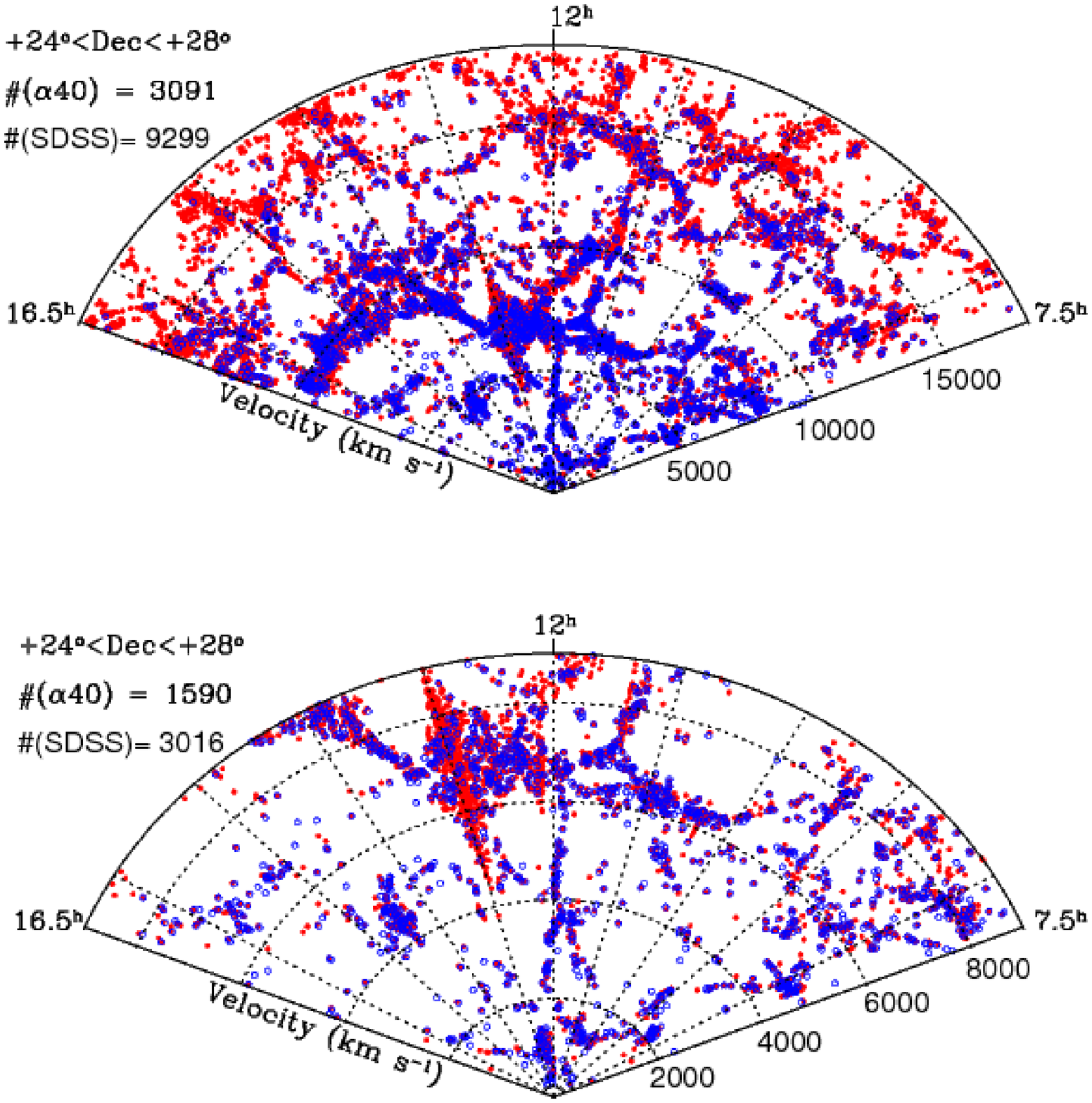}
\end{center}
\caption{Cone diagrams showing the distribution of $\alpha.40$ HI sources
(blue open circles) and those with optical redshifts from the SDSS 
(filled red circles) within the spring sky strip covering $24^\circ < Dec.
< +28^\circ$. 
The upper diagram shows the volume extending over the
full ALFALFA bandwidth to 18000 \kms ~(including regions impacted by
terrestrial interference). The bottom diagram contains only the
volume to 9000 \kms.}\label{fig:conespr26}
\end{figure}

\begin{figure}
\begin{center}
\epsscale{1.0}
\plotone{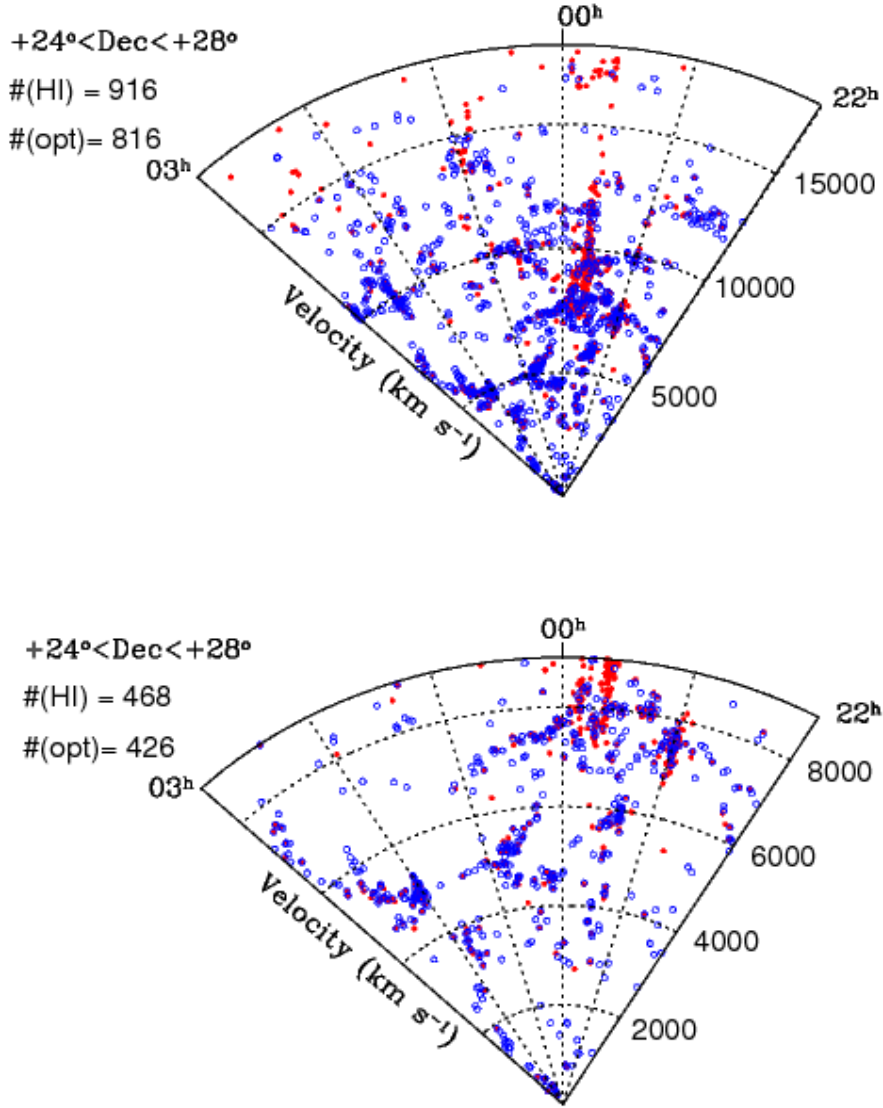}
\end{center}
\caption{Cone diagrams showing the distribution of $\alpha.40$ HI sources
(blue open circles) and those with reported optical redshifts 
(filled red circles) within the fall sky strip covering $24^\circ < Dec.
< +28^\circ$. 
The upper diagram shows the volume extending over the
full ALFALFA bandwidth to 18000 \kms ~(including regions impacted by
terrestrial interference). The bottom diagram contains only the
volume to 9000 \kms. The lack of coverage by the SDSS is evident 
in the paucity of optical redshifts in comparison with Figure 
\ref{fig:conespr26}.
}\label{fig:conefall26}
\end{figure}

\begin{figure}
\begin{center}
\epsscale{1.0}
\plotone{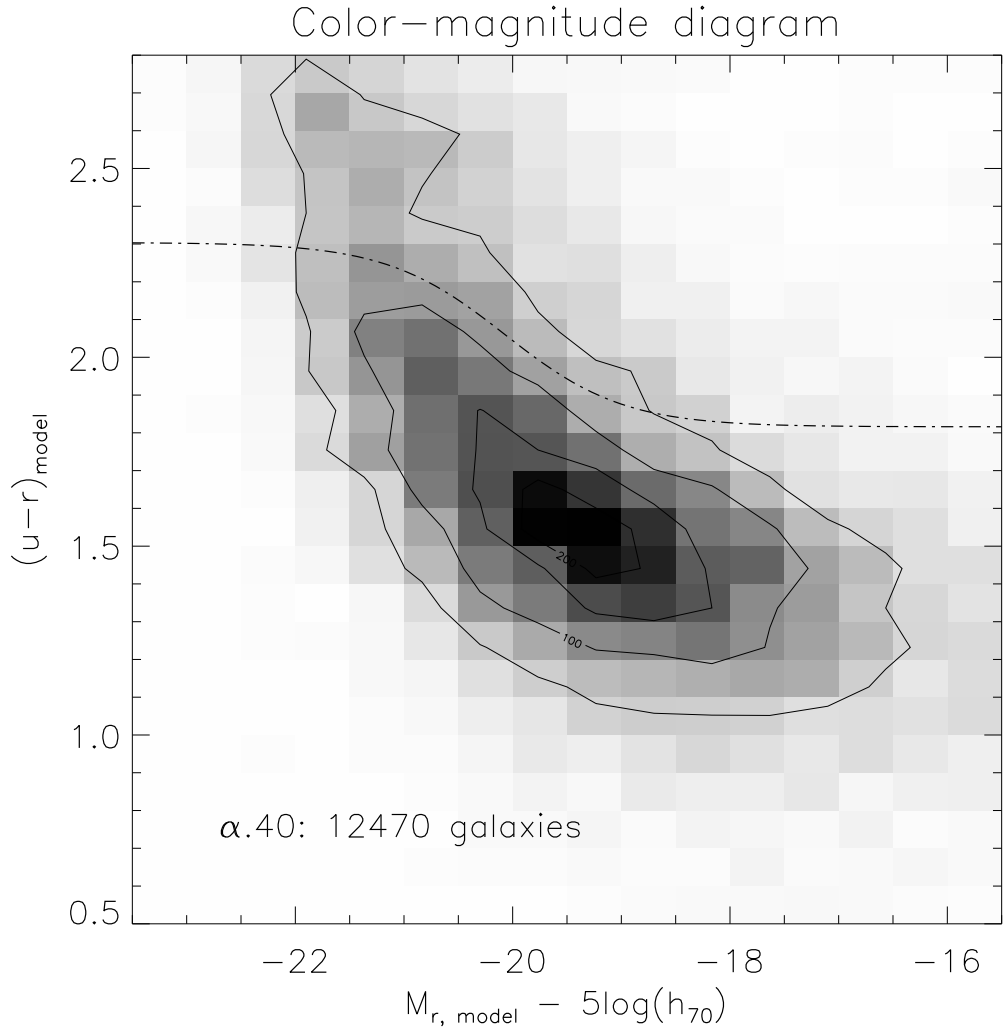}
\caption{Gray scale color magnitude diagram, based on SDSS DR7 photometry, for
the ALFALFA-SDSS overlap sample using the model magnitudes and
colors as given in Table \ref{tab:sdsscross}. The x and y ranges are matched to
Figure 2 of \citet{Baldry04} for comparative purposes. The superposed dashed line is the 
optimum divider given as Equation 11 of that paper which separates the
red sequence from the blue cloud.}
\label{fig:baldryplot}
\end{center}
\end{figure}

Figures \ref{fig:conespr26} and \ref{fig:conefall26} provide graphical illustrations of the relative
strengths and weakness of the ALFALFA and the SDSS surveys as tracers of
the large scale structure in the local universe. Figure \ref{fig:conespr26} shows a cone
diagram of a four degree wide slice of the ALFALFA spring sky centered on Dec. = +26$^\circ$ and
including the full ALFALFA bandpass redshift range $cz <$ 18000 \kms.
The upper cone extends over the full $cz$ range covered by ALFALFA and the lower one, only
the inner $cz <$ 9000 \kms. Blue open circles mark the locations of galaxies detected by ALFALFA,
while red filled ones denote objects with redshifts from the SDSS DR7. The fall-off in the 
density of blue points follows the distribution seen in Figure \ref{fig:spanplot}. The ``finger of God''
radial line-up of optical--$cz$ (red) points so prominent in the lower diagram is the Coma
cluster Abell 1656. Galaxies in that cluster are well known to be strongly HI deficient
\citep{GH85,Magri88} so that ALFALFA detects very few of them. As indicated by the numbers superposed
on the diagram, the number of SDSS spectroscopic targets in the full ALFALFA volume is about
three times the number of ALFALFA HI sources; in the inner volume illustrated in the bottom
diagram, that ratio drops to two and to $\sim 1$ for field galaxies at $cz <$ 5000 \kms. 
While strong bias against finding HI sources in the
regions of rich clusters is clearly evident, the HI-bearing galaxies trace well the large-scale
supercluster structures and include some of the most isolated objects found in this
nearby volume.

For comparison, Figure \ref{fig:conefall26} shows a similar cone plot covering
a four degree wide slice of the ALFALFA fall sky centered on Dec. = +26$^\circ$.
The SDSS spectroscopic survey did not cover this region; the red filled circles mark
objects with optical redshifts available from the literature. In this part of the sky,
ALFALFA sources contribute the majority of redshifts even at its outer boundary.
It should be noted that the slice of the sky sampled in Figure \ref{fig:conespr26} 
covers a strongly overdense region of the local universe, the Coma-Abell 1367 supercluster,
whereas the fall region lies to the south of the main filament of the Pisces-Perseus
supercluster and includes a portion of the void in front of it. 
As in all studies of the local universe, the actual
large scale structure contained in the survey volume
can leave a strong imprint on the observed distribution of
galaxies and their properties in limited samples. Further discussion of the impact
of large scale structure on cosmological inference is included in \S\ref{sec:impact}.

Making use of the SDSS cross reference tabulation, Figure \ref{fig:baldryplot} presents a 
color magnitude diagram (CMD) for the $\alpha.40$--SDSS overlap
sample for comparison with similar diagrams extracted from the SDSS photometric survey alone. 
A similar CMD was independently constructed by \citet{Toribio11a} for a sample of ALFALFA
galaxies found in low density environments. In Figure \ref{fig:baldryplot},
grayscale and contours show the distribution of the HI-selected sample and the
axes correspond to the range illustrated in Figure 9 of \citet{Baldry04}. The superposed
dashed line shows the optimum divider used by \citet{Baldry04} to separate galaxies
on the red sequence (above the curve) from those in the blue cloud 
(below it) and given by their equation 11.
For the purpose of comparison with their Figure 9, no corrections for redshift or internal
extinction have been applied to the magnitudes used to construct Figure \ref{fig:baldryplot}.
Figure \ref{fig:baldryplot} can also
be compared with Figure 4 of \citet{Tempel11} who used a large sample of galaxies from
SDSS DR7 and did apply a K-correction; in their figure, those authors also categorize 
separately elliptical and spiral galaxies according to the SDSS catalog parameter
$f_{deV}$, the fraction of the galaxy's luminosity contributed by the deVaucouleurs
profile. Clearly, the $\alpha.40$ catalog is dominated by blue spiral galaxies and
is strongly biased against the red sequence. As discussed by \citet{Tempel11}, some of the
luminous, red objects are truly red, luminous and gas-bearing objects; other
luminous objects appear red because they are edge-on disks 
for which the internal extinction correction is
significant. Still, Figure \ref{fig:baldryplot} confirms the conclusion of
\citet{Masters10} that the most luminous gas-rich population 
includes a significant fraction of red galaxies.
Further discussion of the stellar and star forming properties as derived 
from SED fitting the photometry provided by the SDSS in the optical and
the FUV/NUV by the Galaxy Evolution Explorer (GALEX)
satellite for the ALFALFA sample will be presented in \citet{Huang11a} and
\citet{Huang11b}.

\subsection{ALFALFA Detections without Optical Counterparts}\label{sec:darkgals}

One of the scientific drivers behind blind HI surveys is the possibility of 
contributing gas rich but optically ``dark'' galaxies to the extragalactic census.
Previous analyses by e.g., \citet{Briggs90}, of the statistics of targeted HI line 
surveys have shown that such objects must be rare; otherwise there would have been
more sources detected serendipitously in the random off-source positions observed by the total
power position-switching observing mode used for most of those earlier surveys. Indeed,
perhaps the best example of an optically dark galaxy is the southwest component
of HI~1225+01 \citep{GH89, Chengalur95}, but it is not a purely isolated object,
being located on the outskirts of the Virgo cluster and 
part of a binary system with its dwarf galaxy companion to the northeast.
Of the 4315 HI sources reported in the HIPASS catalog, 84\% were identified
with one or more possible OCs \citep{Doyle05}. Most of the remainder are located
at low galactic latitude where Galactic extinction strongly inhibits the
hunt for the stellar counterpart. In fact, \citet{Doyle05} investigated
through followup observations the 13 HIPASS without OCs and 
with A$_V <$ 1 mag and concluded that not a single one could be claimed as an isolated
dark galaxy. Some might be intergalactic in the sense of being associated
with tidal debris fields or fragments of very extended HI disks, but always
there were nearby, visible (stellar) objects at the same redshift.

Because of ALFA's superior angular resolution at L-band in comparison with that of the
Parkes telescope (4\arcmin ~vs 15.5\arcmin), we are able to centroid the position of
the ALFALFA HI sources to better than 20\arcsec ~on average and to identify their
OCs likewise with better surety. Only \ndtot\ of the \ntot\
sources presented in Table \ref{tab:catalog} do not have assigned OCs. 
Of those, \nhvc\ blank field objects have observed
velocities which fall within the range characteristic of emission associated
with some Milky Way population. All of these are assigned an HI source category code of 9 in 
Column 12 of Table
\ref{tab:catalog}. They are likely to be HVCs, although a few
isolated objects with narrow velocity widths and small angular sizes (either
barely resolved or unresolved) are candidate low mass extragalactic halos \citep{Gio10}. 
Their distribution and nature will be discussed elsewhere.

Of the remaining \ndxgal\ HI sources ($< 2$\% of the total extragalactic population)
whose velocities suggest that they are  truly extragalactic, we have individually 
examined closely the SDSS and/or DSS2(B) fields to look for OCs; 
comments derived from that examination
are included in Table \ref{tab:comments}.
Some of these objects do not lie in the region covered by SDSS, making the
identification of OCs more difficult, but by design, only a few lie in regions
of significant optical extinction. 

Roughly 3/4 of the ``dark'' HI sources are located in fields where
objects of similar redshift are found, albeit beyond the reasonable limits
of coincidence given by the ALFALFA pointing accuracy. A number can be linked
to previously known extended HI distributions such as the Leo Ring \citep{Stierwalt09},
the tail of NGC~4254 \citep{Haynes07,Kent07}, the extended tail of NGC~4532/DDO~137
\citep{Koop08} or the intergroup gas found
in the NGC~7448/7463/7464/7465 group \citep{Haynes81}. Among the blank field
HI detections with SDSS data (including DR8) and not contaminated by the
presence of bright foreground stars, only about 50 remain as candidates to
be isolated ``dark'' objects. These objects are the targets of
a followup program that will confirm their reality as HI sources with
the Arecibo single pixel L-band receiver,
localize the HI emission via HI synthesis observations and search for 
associated low surface brightness stellar emission via optical imaging.

\subsection{OH Megamaser Candidates}\label{sec:ohms}

OH megamasers (OHMs) are powerful line sources associated
with the starburst nuclei in merging galaxy systems. 
\citet{Briggs98} has pointed out that OH megamasers at $z \sim 0.17$ may contaminate 
a blind extragalactic survey such as ALFALFA. 
An extremely rare phenomenon in the local universe, about 100 OHMs are known out 
to a redshift of 0.265 \citep{Darling02}. The main 18 cm OH
lines occur at rest frequencies of 1665 and 1667 MHz respectively. In OHMs, the
emission at 1667.359 MHz dominates; that line is redshifted
in the ALFALFA observing band for sources with $0.16 < z < 0.25$.
Using the large targeted survey for OHMs by \citep{Darling02} as a baseline for the
expected flux density and spectral characteristics of OHMs, it is probable that a few 
of the ALFALFA sources without OCs may in fact be OHMs with $0.16 < z < 0.25$. Confirmation
that an ALFALFA source is in fact an appropriately redshifted OHM and not
an optically dark HI galaxy will require follow-up HI synthesis observations to
localize the line emitting region and optical/IR spectroscopy to confirm
the redshift. 

\begin{figure}
\begin{center}
\epsscale{0.9}
\plotone{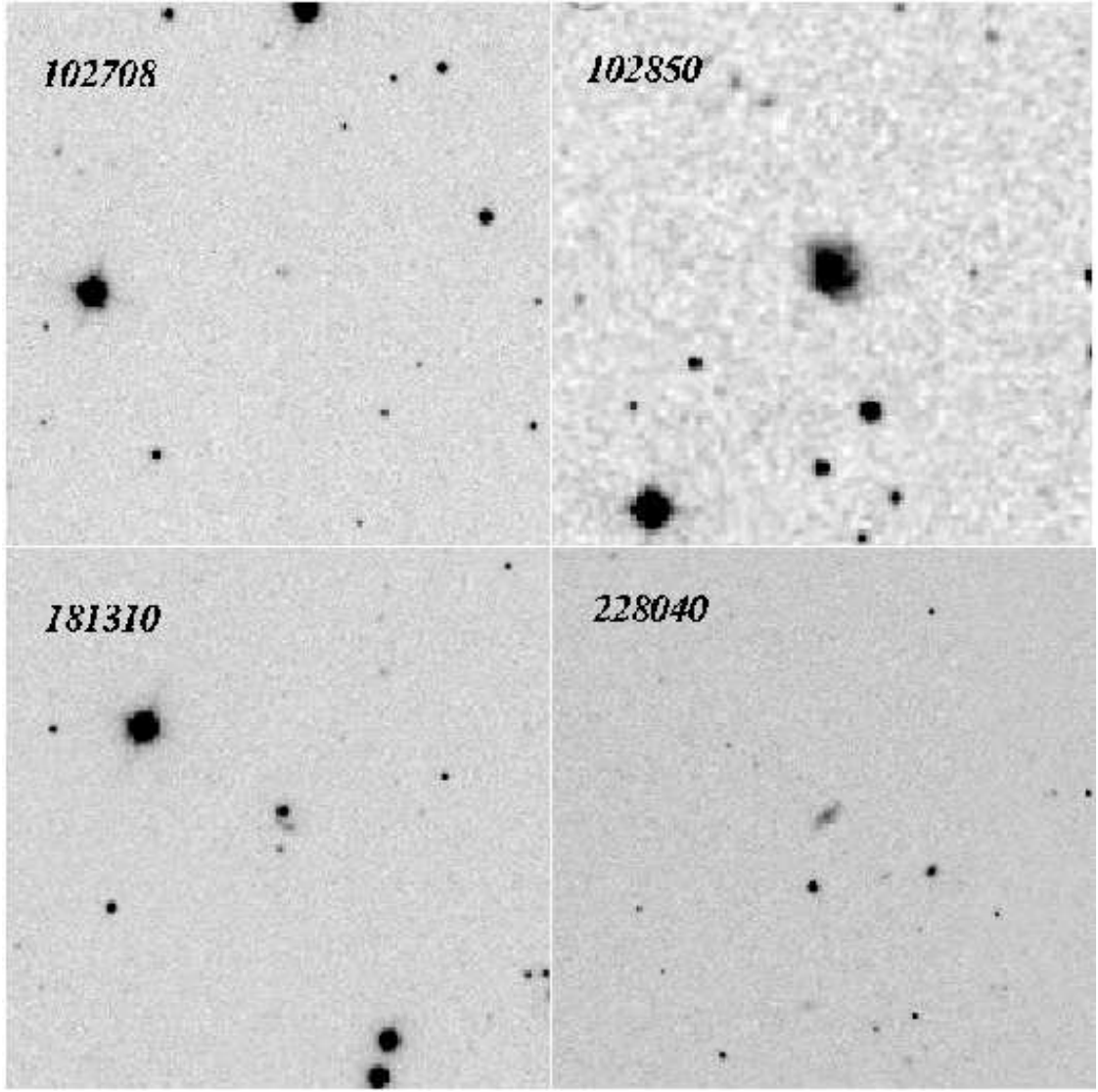}
\caption{Optical images of the four best OHM candidates listed in Table \ref{tab:ohmtab}.
The image of AGC 102850 comes from the DSS2(B) while the others are
SDSS-g; each image is 3\arcmin~ on a side.}
\label{fig:ohms}
\end{center}
\end{figure}

Already, however, there are four OHM
candidates which can be identified as such because the line emission occurs
at frequencies higher than 1422 MHz. Hence, under the
assumption that the emission arises from the HI 21cm line,
the observed $cz$ is too largely blueshifted for plausible interpretation
as an extragalactic or Galactic HI source. The properties of these four objects are given in 
Table \ref{tab:ohmtab} and optical images obtained from either the SDSS or DSS2(B) are 
shown in Figure \ref{fig:ohms}. The entries in Table \ref{tab:ohmtab} are as follows:

\begin{itemize}

\item Col 1: Entry number in the AGC

\item Col 2: Centroid (J2000) of the emission line source, in hhmmss.sSddmmss, as in Col. 3
      of Table \ref{tab:catalog}. The designation of the candidate then adopts the identifier
     ``OHMcand'' plus this centroid position.

\item Col 3: Position of the identified optical counterpart, in hhmmss.sSddmmss.

\item Col 4: $z_{opt}$, redshift of the optical counterpart, where known.

\item Col 5: $z_{OH}$, redshift of the candidate OHM assuming its emission is dominated by
   the OH line at 1667.359 MHz.

\item Col 6: $cz_{21}$, heliocentric velocity if the emission were associated with the HI line,
   in \kms.

\item Col 7: $F_{OH}$, OH line flux density, in Jy~\kms.

\item Col 8: S/N of the OH line emission, defined as in Col. 8 of Table \ref{tab:catalog}.

\item Col 9: RMS noise in the vicinity of the line emission, defined as in Col. 9  of Table \ref{tab:catalog}.

\end{itemize}

In all four instances, there is a small object visible
in public imaging databases which can be identified as the likely optical counterpart:

\begin{itemize}

\item AGC 102708 = OHMcand000337.0+253215
 is likely associated with SDSS J000336.02+253204.0,
a very tiny object also evident in DSS2(B). There is no NED entry or redshift measurement. 

\item AGC 102850 = OHMcand002958.8+305739 is likely associated with 2MASX J00295817+3058322,
a well-formed spiral galaxy. There is no confirming redshift measurement. 

\item AGC 181310 = OHMcand082311.7+275157 is likely associated with SDSS J082312.61+275139.8,
also known as IRAS 08201+2801 and 5C 07.206, a known ULIRG. 
For this single object, the optical redshift $cz$ = 50314 \kms, $z$ =  0.167830 $\pm$ 
0.000041 from the SDSS confirms the identification as an OHM; its OHM emission was
previously discovered by \citep{Darling01}.

\item AGC 228040 = OHMcand124540.5+070337 is likely associated with SDSS J124545.66+070347.3,
a spiral galaxy viewed at high inclination as evident in Figure \ref{fig:ohms}. No confirming
redshift measurement is available.
\end{itemize}

OHMs may also be identified in the subset of low S/N sources not included in the current
catalog because they do not meet the criteria of Codes 1 and 2. 

By the simplest argument based on the fraction of the usable ALFALFA bandwidth above 1422 MHz and 
assuming that these four candidates are, in fact, OHMs, it is possible that half of the
``dark galaxy'' candidates discussed in \S\ref{sec:darkgals} might be OHMs at 0.175 $< z <$ 0.245.
A similar estimate comes from considering the $\alpha.40$ volume and
the OHM luminosity function at low z \citep{Darling02}.
A more systematic approach to the identification of OHMs throughout the full bandpass
and using the 3-D ALFALFA dataset is 
currently being undertaken by members of the ALFALFA collaboration.

\section{Validation of ALFALFA HI Parameters}\label{sec:validate}

Most targeted extragalactic HI line flux densities are extracted from
spectra conducted using a total power position switching technique. As outlined
in \S\ref{sec:observations},
the ALFALFA dataset is generated using a very different approach whereby
ALFA drift scan data are obtained months and sometimes years apart and without
doppler tracking. The 2-D datasets (frequency versus time) for the two individual 
polarizations of each of the seven beams are bandpass subtracted and flagged 
for RFI. After the acquisition of all the drifts for a region of the sky,
the 3-D spectra grid is then generated. As with any new survey, it is
critical to verify that the spectral scales (velocity and flux density) at
each grid point are accurate. 

The Cornell Digital HI archive presents a large compilation of digital HI 
line spectra obtained using pointed observations of optically-selected targets
\citep{Springob05a} which have been digitally analyzed using similar
algorithms to those adopted for ALFALFA. Because those spectra were obtained
with a variety of single-dish telescopes and spectrometers, careful attention
was paid to correct for instrumental effects such as pointing errors,
source extent, instrumental broadening and spectral smoothing. Corrections for
the various effects were modeled and tested to produce a homogeneous catalog of 
extracted properties with their associated error estimates. Here, we present the 
validation of the ALFALFA velocity, velocity width and flux density scale by 
comparison of $\alpha.40$ catalog parameters with the previous targeted HI line 
observations of sources which have been re-detected by ALFALFA. Of the
2073 galaxies which are contained both in the \citet{Springob05a} 
and the $\alpha.40$ catalogs, 1887 are classified as ALFALFA Code 1 sources
and 186 are Code 2 detections.

\subsection{Validation of the ALFALFA Velocity Scale}\label{sec:vels}

The ALFALFA ``minimum-intrusion'' observing mode acquires data without 
doppler tracking, i.e., in topocentric mode. Heliocentric corrections are applied in 
the Fourier domain, whereby the appropriate velocity shift at each point (each spectrum 
associated with each one second record for each polarization of each beam) is
calculated, converted to a phase gradient across the bandpass and applied to the
Fourier transform of each spectrum. The inverse Fourier transform then gives
the spectrum in the heliocentric rest frame which is used thereafter to 
yield the systemic velocity $cz$ and the HI profile velocity width $W_{50}$.

\begin{figure}
\begin{center}
\epsscale{1.0}
\plotfiddle{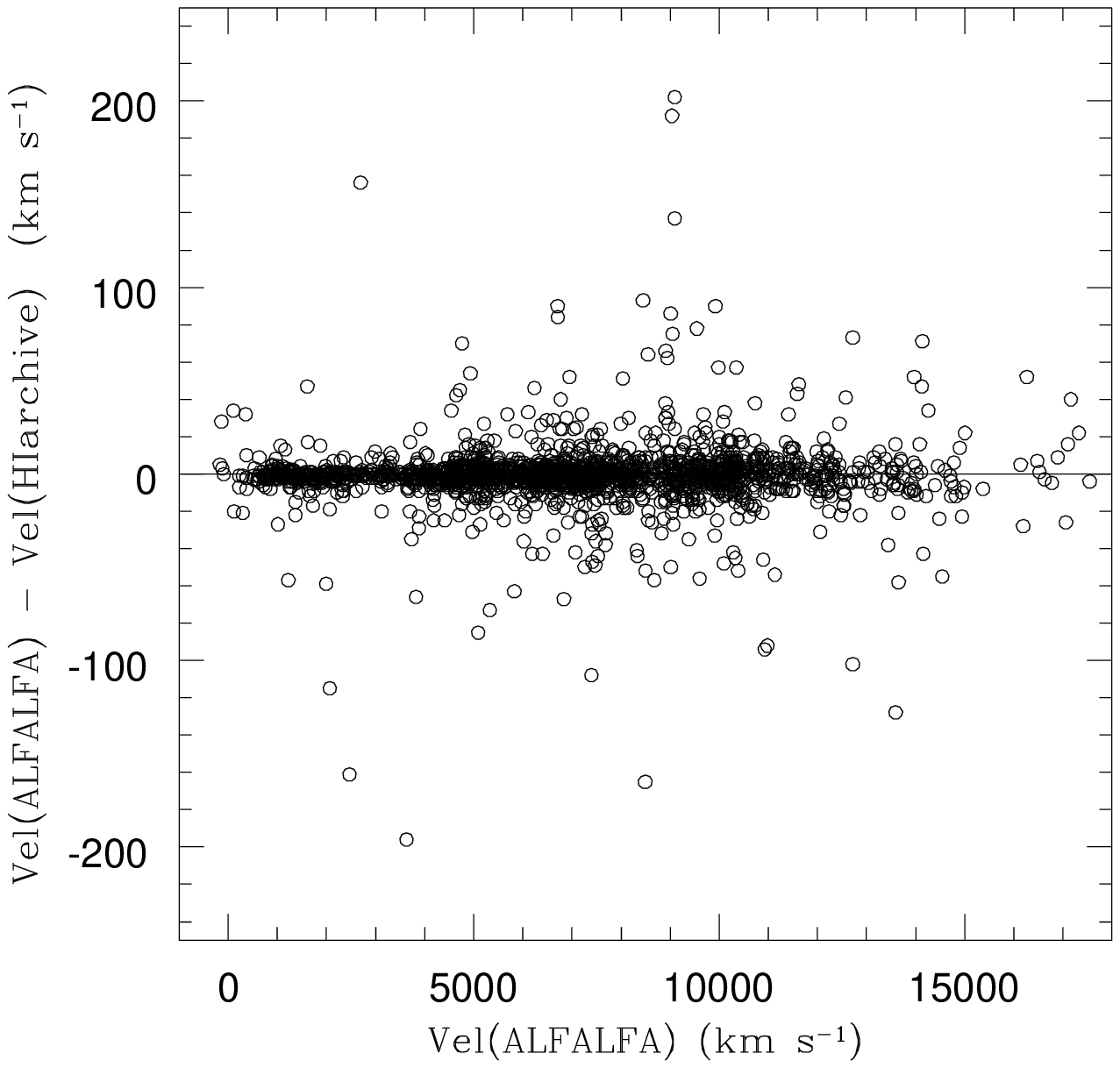}{2.0in}{0}{75}{75}{-470}{-280}
\plotfiddle{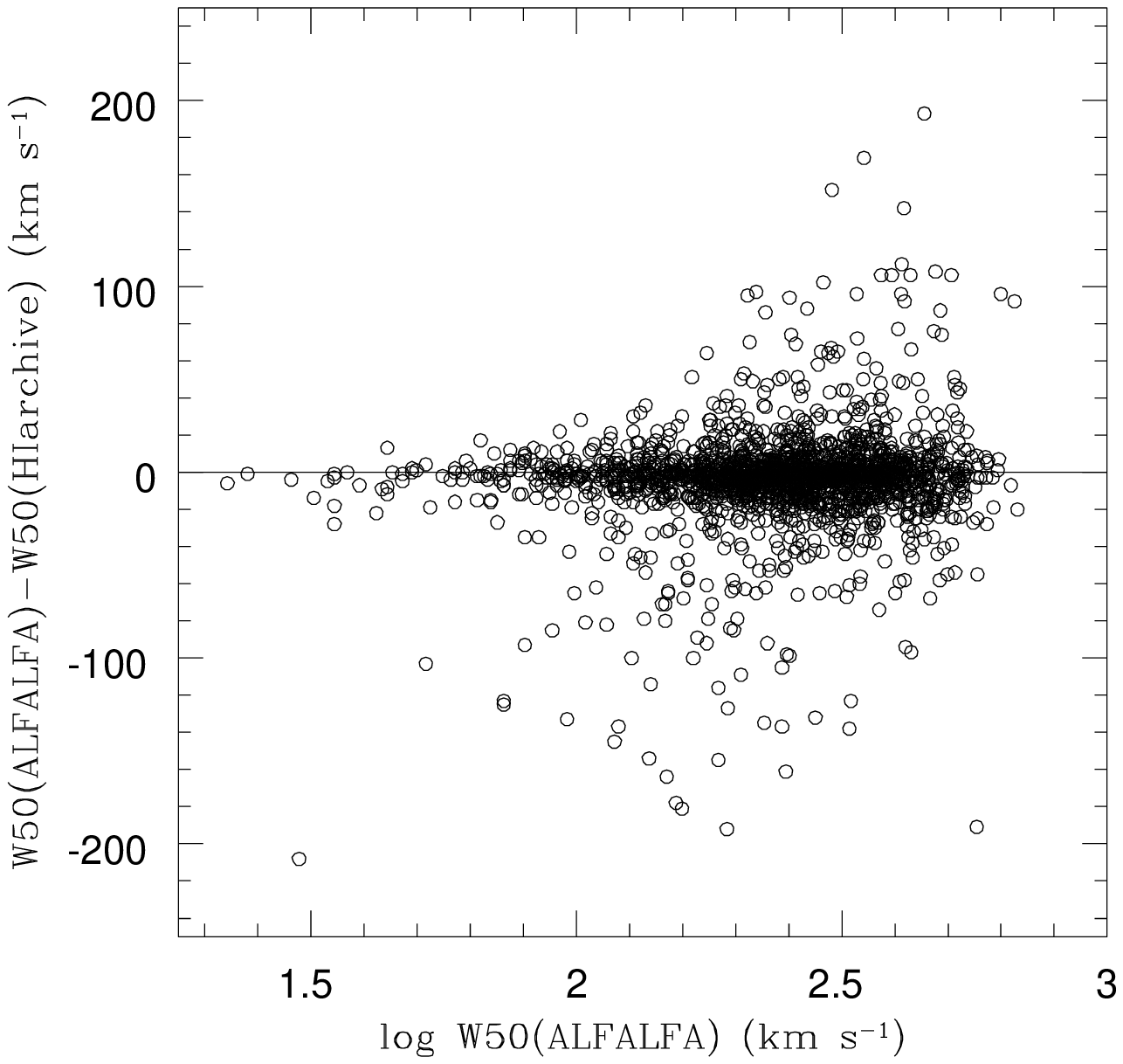}{2.0in}{0}{75}{75}{-470}{-410}
\vskip 8.0cm
\caption{Comparison of systemic velocity $cz_{\odot}$ (upper panel) 
and velocity width measurements $W_{50}$ (low panel) obtained by the ALFALFA
survey and values given in the Cornell Digital HI archive \citep{Springob05a}.
}\label{fig:velcompare}
\end{center}
\end{figure}

It is important to note that the specific definitions of HI systemic velocity 
and the global profile velocity width are not uniformly adopted in the literature. 
For ALFALFA, we adopt the same convention as that
used by \citet{Springob05a}, that is, polynomials are fit to each side of
the two-horned profile and then $cz$ and $W_{50}$ are measured at the level
of 50\% of the peak intensity on either horn: $cz$ is then the midpoint
and $W_{50}$ the full width at that level. Where appropriate (face-on galaxies;
dwarf systems), a single Gaussian provides the best fit and is similarly measured.
Figure \ref{fig:velcompare} illustrates the comparison of the two parameters
for the $\alpha.40$-\citet{Springob05a} HI archive overlap sample. In both panels,
the vertical axis shows the residual ALFALFA-HI archive. The occurrence of
outliers is expected because (a) ALFALFA spectra correspond only to 40 sec per
beam of integration time on source, whereas the targeted spectra are generally of much
longer integration; (b) targeted spectra are affected by pointing errors either in
the coordinates used to position the telescope or intrinsic telescope pointing inaccuracy;
and (c) blends with close companions where the pointed spectra
were taken with smaller single dish telescopes. A few cases of gross disagreement are explained
by errors in the velocity scales of very old HI data which were acquired in the days before
significant information was written into data headers, when the setup of the backend
electronics and spectrometer required physical cabling and hand dial-setting 
at the start of each observing 
run and when records of frequency offsets for different quadrants of the
spectrometer were kept only on hand-written index cards; 
these cases are noted in the comments in Table \ref{tab:comments}. 
The appearance in the lower panel of some outliers at relatively low $W_{50}$
reminds us that at low S/N or in the presence of residual baseline structure, 
broad widths may be underestimated. The dependence of the sensitivity on
line width is discussed in \S\ref{sec:completeness}.
As evident in Figure \ref{fig:velcompare}, the comparison of the velocity scales 
reveals no systematic offsets 
and agreement within the expected errors.

\subsection{Validation of the ALFALFA Flux Density Scale}\label{sec:fluxes}
 As discussed in \citet{vanZee97}, practical limitations and 
instrumental uncertainties restrict the accuracy with which HI line flux 
densities can be measured to not better than a few percent.
Despite regular calibration via the injection of a noise diode, 
drifts in the electronic gain, amplifier 
instabilities, sidelobe variations and standing waves (caused by multiple reflections
within the optical path of cosmic continuum sources or terrestrial RFI) induce
variations in the total power, while baseline irregularities and data loss due to 
RFI impact the measurement of flux density in noisy data. As discussed in 
\citet{Springob05a}, HI line flux densities derived from targeted (pointed) 
observations are typically accurate to not better than 15\%; older datasets
taken when amplifiers were substantially less stable than today are
probably accurate to not better than 25\%. 

Calibration of the ALFALFA dataset is performed in two separate stages. First, during
the course of an observing run, a noise diode, calibrated by the engineering
staff in the lab, is fired once every 600 seconds. The data stream then includes
a record with this additional power source (the ``cal-on'' record). All observing 
runs contain at least 9 such calibrations; longer ones may contain as many as 60. A 
polynomial fit is performed to the ratio of the total power with
the calibration diode on, versus when it is off, for the whole set 
for an observing block. This polynomial is then used to correct the individual
records of the drift scan data. The second method of calibration is performed
on the data after grid construction, making use of the radio continuum sources
which they contain. A comparison is made of the flux densities of the source contained
in the grids with published values in the NVSS \citep{Condon98}, and then
an average correction factor is applied to tie the ALFALFA flux scale to the
NVSS. Further details on calibration of the ALFALFA dataset are given in
\citet{Kent11}.

\begin{figure}
\begin{center}
\epsscale{1.0}
\plotfiddle{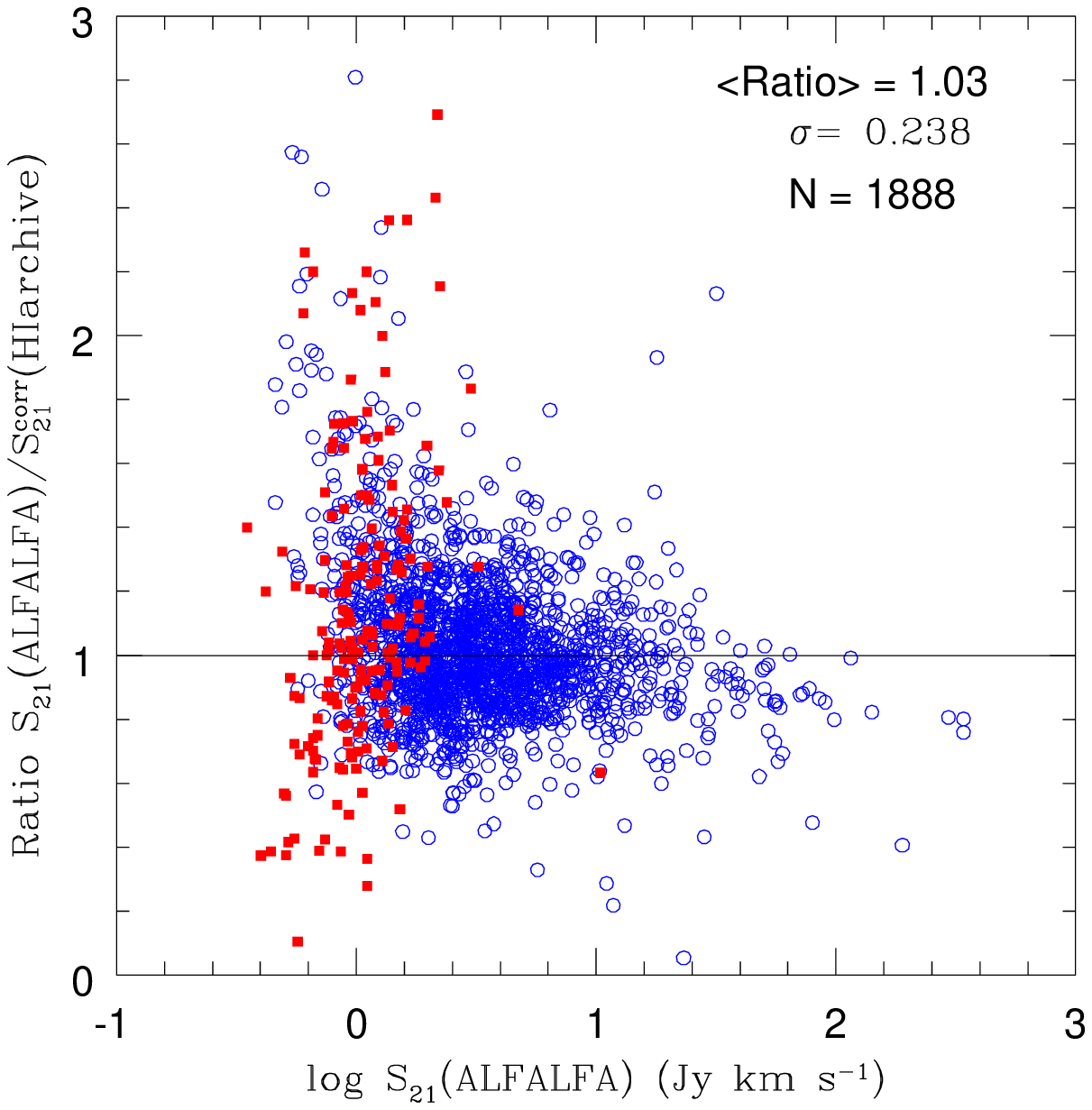}{2.0in}{0}{68}{68}{-460}{-260}
\plotfiddle{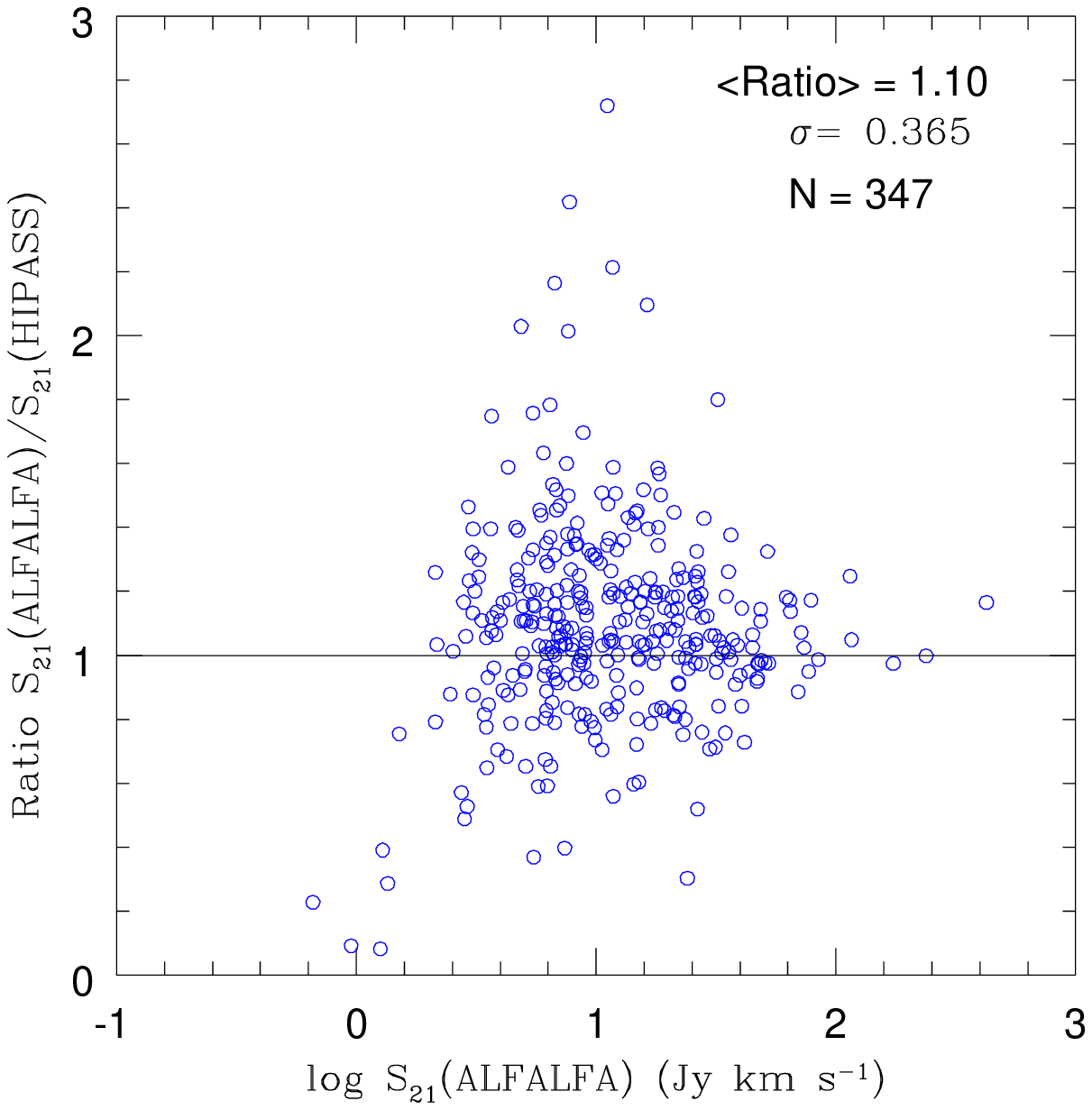}{2.0in}{0}{68}{68}{-460}{-360}
\vskip 7cm
\caption{Top: Comparison of HI line flux density measurements S$_{21}$ for 
the 1888 galaxies in
common between $\alpha.40$ and \citet{Springob05a}. 
The vertical axis displays the ratio of the HI line flux density detected
by ALFALFA to the correspnding value corrected for source extent and pointing
errors (but not internal HI absorption) reported by \citet{Springob05a}. 
ALFALFA Code 1 detections are plotted as blue open symbols, while Code 2 
(priors) detections are shown as red filled circles. The flaring of the
ratio at low fluxes is expected. Bottom: Similar comparison with 347 galaxies detected
by HIPASS. No Code 2 detections were detected by HIPASS.}
\label{fig:fluxes}
\end{center}
\end{figure}

Even when gain corrections
for frequency dependence and other effects are correctly calibrated out,
HI line flux densities observed with single point observations must be
corrected for beam dilution and, often, pointing errors. In addition to the
inaccuracy of telescope pointing, particularly important in early Arecibo
observations, the input positions used to point the telescope were accurate 
only at the level of 0.5--1\arcmin ~level for some of the oldest observations used
to acquire the archival data reanalyzed by \citet{Springob05a},

\citet{Springob05a} report both raw (as observed) and corrected values of the HI line flux
density for galaxies observed via pointed observations of optically-selected
targets. The true HI line flux density was derived by applying corrections
for telescope pointing errors, errors in the positions used to point the telescope 
(both of which apply more importantly to older datasets) and for partial resolution by
the telescope beam. The latter is derived by adopting a hybrid correction for 
source extent that is based on a modeled HI distribution scaled by the optical size 
and an average telescope beam power pattern. As a drift scan mapping survey, ALFALFA 
flux densities are not subject to such corrections.
Figure \ref{fig:fluxes} shows the comparison of the HI line flux densities
measured by ALFALFA with the values given in the \citet{Springob05a} HI archive.
The latter are corrected for pointing and position errors and for
source extent (but not for internal HI absorption). 
The vertical axis shows the ratio of the HI line flux densities reported in the 
two catalogs. ALFALFA detection Code 1 objects are shown as blue open circles;
the lower S/N Code 2 objects are shown in red filled circle. 
Since the error in the HI line flux density for both surveys depends
on the HI line flux density itself as well as the S/N of the spectrum and the
magnitude of the corrections applied to the pointed data, the increasing
scatter seen in the ratio at low HI line flux densities is as expected.
When corrections for source extent are applied to the pointed data, the 
flux density scales are coincident within the errors.

Among the sources with highest HI line flux density, Figure \ref{fig:fluxes}
suggests that the ALFALFA flux algorithm may be missing some flux.
The total flux should be recovered since ALFALFA is a mapping survey but the
HI line flux from very extended sources, especially those located towards the 
edges of the  constructed grids could be lost due to the finite grid size and
the bandpass subtraction and grid flattening schemes. For those, alternative
processing tools from the standard pipeline will be developed, after completion
of the main survey.

In order to assess the contribution of very diffuse, extended HI in 
vicinity of nearby, isolated galaxies, \citet{Haynes98} and \citet{Hogg07} observed a carefully-
selected sample of $\sim$100 galaxies with both the former 42~m and the Green
Bank Telescopes. As they note, the uncertainty in the HI line flux density for their
high signal-to-noise data is $<$ 1\%; on the other hand, the uncertainties contributed
by fitting the polynomial baseline and defining the boundaries of the emission 
profile are considerably larger. Because an unblocked aperture should deliver reduced
standing waves and minimal stray radiation, flux densities measured with the GBT
should be more accurate than ones measured with a complex instrument like Arecibo.
At this point, there are only 12 galaxies in common with the sample
observed very accurately with the former 42~m telescope by Haynes \etal ~(1998),
not enough for conclusive results. These issues
will be explored in a future work. 

Although it might be expected to serve as the better dataset to use in examining
systematic uncertainty and testing for missing flux from extended/bright sources, the northern
HIPASS survey \citep{Wong06} does not in practice provide an adequate comparison sample for several
reasons. First, as mentioned above, flux calibration uncertainties do not dominate most HI flux density
errors; baseline uncertainties, noise and beam effects do. A drawback of the northern HIPASS
catalog is that some of it suffers from residual baseline ripple, particularly when observations
were made during the daytime \citep{Wong06}. Secondly, the sensitivity difference between Arecibo and
Parkes means that the S/N of most ALFALFA/\citet{Springob05a} detections is typically much higher
than that of HIPASS. This fact alone, on top of the baseline issues, give the \citet{Springob05a} 
spectra a significant advantage over HIPASS in terms of parameter accuracy.
Thirdly, although there are 1000 galaxies in the northern HIPASS dataset \citep{Wong06} at Dec. $> 2$, only
~$\sim$350 lie in the overlap with the present $\alpha.40$ catalog. Lastly, there are 
no Code 2 sources detected by HIPASS (which is not surprising, given its much poorer sensitivity) 
so we cannot make the comparison between Codes 1 and 2. However, for the record, we include
in the lower panel of Figure \ref{fig:fluxes} the similar comparison of flux densities from ALFALFA
and HIPASS; clear cases of confusion within the Parkes beam are not included in this analysis.
Curiously, ALFALFA detects {\it more} flux density than HIPASS in some cases; examination
of those reveals that they are mainly ALFALFA detections with broad $W_{50}$ for which
HIPASS detects, at much lower S/N, a lower HI line flux density and a narrower $W_{50}$, clearly
missing some of the HI line emission detected by ALFALFA.

\section{ALFALFA Source Completeness and Reliability}\label{sec:completeness}

The practical exploitation of any survey requires an understanding
of its source sensitivity, completeness and reliability. In comparison with previous 
blind HI surveys, ALFALFA offers a much richer dataset which itself can be used to 
probe the robustness of its source catalog.

Source extraction and parameter measurement for ALFALFA is performed in a two-step 
process, which includes automated as well as interactive procedures. Initial 
source extraction is performed by a fully automatic matched-filtering method 
\citep{Saintonge07a, Saintonge07b}. The algorithm uses templates which vary in shape as a 
function of profile width (gaussian for narrow profiles, Hermite functions for broad 
profiles), and outperforms algorithms based on smoothing followed by peakfinding 
(see Figure 3 in \citet{Saintonge07a}). The reliability (i.e. the fraction of detections 
that correspond to real sources) of this automated method is estimated to be 
$\approx 95$\% for sources with $S/N_{extractor} > 6.5$, a value that was determined 
by performing source extraction on regions of the ALFALFA datacubes expected to
be devoid of cosmic signals 
(corresponding to the velocity range $-2000 < cz < -500$ \kms, see \S5.4 and Figure 8 
in \citet{Saintonge07a}). Source candidates are then visually inspected 
and source parameters are interactively measured and catalogued. It should
be noted that the parameters of the final catalogued sources (e.g. 
$S_{21}$, $S/N$, $W_{50}$, etc.) do not generally coincide with the values 
determined by the automatic signal extractor, because the two procedures use different 
definitions and calculation methodologies for the parameters \citep{Gio07}, and
the human intervention is designed to optimize the measurement accuracy 
and further improve the reliability of the catalog by rejecting spurious detections 
that correspond to low-level RFI, poorly sampled data and residual baseline 
fluctuations. We therefore expect the final reliability of ALFALFA Code 1
detections to be very close to 100\%.

However, as discussed by \citet{Saintonge07a}, the reliability of ALFALFA sources 
extracted by the matched-filtering algorithm drops precipitously
below a S/N of 6.5. The Code 2 sources included in Table \ref{tab:catalog} 
fall below this nominal ALFALFA S/N detection threshold, but are included in the catalog 
because they coincide with an optical galaxy of known (prior) and coincident redshift. 
Although these sources should not be included in statistical studies which require 
careful consideration of survey completeness and sensitivity limits, the vast majority 
are likely real HI line sources,
and the gas in them will also contribute to the overall HI density in the 
local universe. Hence, we include them in the following analysis of ALFALFA completeness
and their impact on measurements of cosmological parameters.

The two step process used to identify, extract and measure the ALFALFA detections
presented in Table \ref{tab:catalog}
results in a catalog of reliable detections that is
dependent on \emph{both} the integrated HI line flux density of a 
given source and its global HI line profile width, W$_{50}$.
Like all fixed integration-time spectroscopic surveys, ALFALFA is more sensitive 
to narrow HI profiles than to broader ones of the same integrated line flux. 
Based on the demonstrated performance during the single-pass precursor 
observations of the observing equipment and the signal extraction pipeline, 
\citet{Gio05b} predicted the specific relationship between the integrated 
flux density detection threshold (S$_{21, th}$ in Jy \kms) and the profile width 
(W$_{50}$, in \kms) of a source in terms of the S/N required 
for inclusion in the catalog:

\begin{equation}
   S_{21, th} = \left\{
     \begin{array}{lr}
       0.15 \, S/N \, (W_{50}/200)^{1/2}, \,\,\, &   W_{50} < 200 \\
       0.15 \, S/N \, (W_{50}/200), \,\,\, &   W_{50} \geq 200
     \end{array}
   \right.
\label{eq:eqnthresh}
\end{equation}
Note that the above expression differs from that given in Equation 2 of \citet{Gio05b}
(numerical factor of 0.15 vs. 0.22) because that work adopts the rms appropriate to 
the single-drift maps used in the precursor observations. One of the principal conclusions of
\citet{Gio05b} was that the two-pass strategy adopted for ALFALFA would improve
on that employed by the precursor program by a factor of 1.5.

``Sensitivity'' is a qualitative term that can be defined in terms of the survey 
``completeness''. We refer to the completeness of the ALFALFA survey as
that fraction of cosmic sources of a given 
integrated flux density and within the survey solid angle 
that are detected by ALFALFA and included in the 
$\alpha$.40 catalog. Other blind HI surveys (e.g. ADBS, HIPASS) have estimated 
the completeness of their catalogs as a function of profile width ($W_{50}$)
by examining their ability to recover synthetic sources of known characteristics 
(peak flux, $S/N$, $W_{50}$, etc.) injected into the spectral cubes. One of the 
motivations for such an approach is to assess the reliability of sources in the 
presence of non-Gaussian noise. As noted by \citet{Saintonge07a} (see \S5.4 and 
Figure 8 of that paper), the impact of non-Gaussian noise on the automatic signal 
extractor developed for ALFALFA is generally minimal above S/N = 6.5. Its presence, 
principally in the 
form of the very broad spectral standing waves resulting from reflections
in the telescope focal structure
\citep[e.g.,][]{Briggs97}, is responsible for the upturn at large 
$W_{50}$ in Equation \ref{eq:eqnthresh}. At the narrower widths, there is no evidence
that a Gaussian assumption is unfair. 

\begin{figure}
\begin{center}
\epsscale{0.45}
\plotone{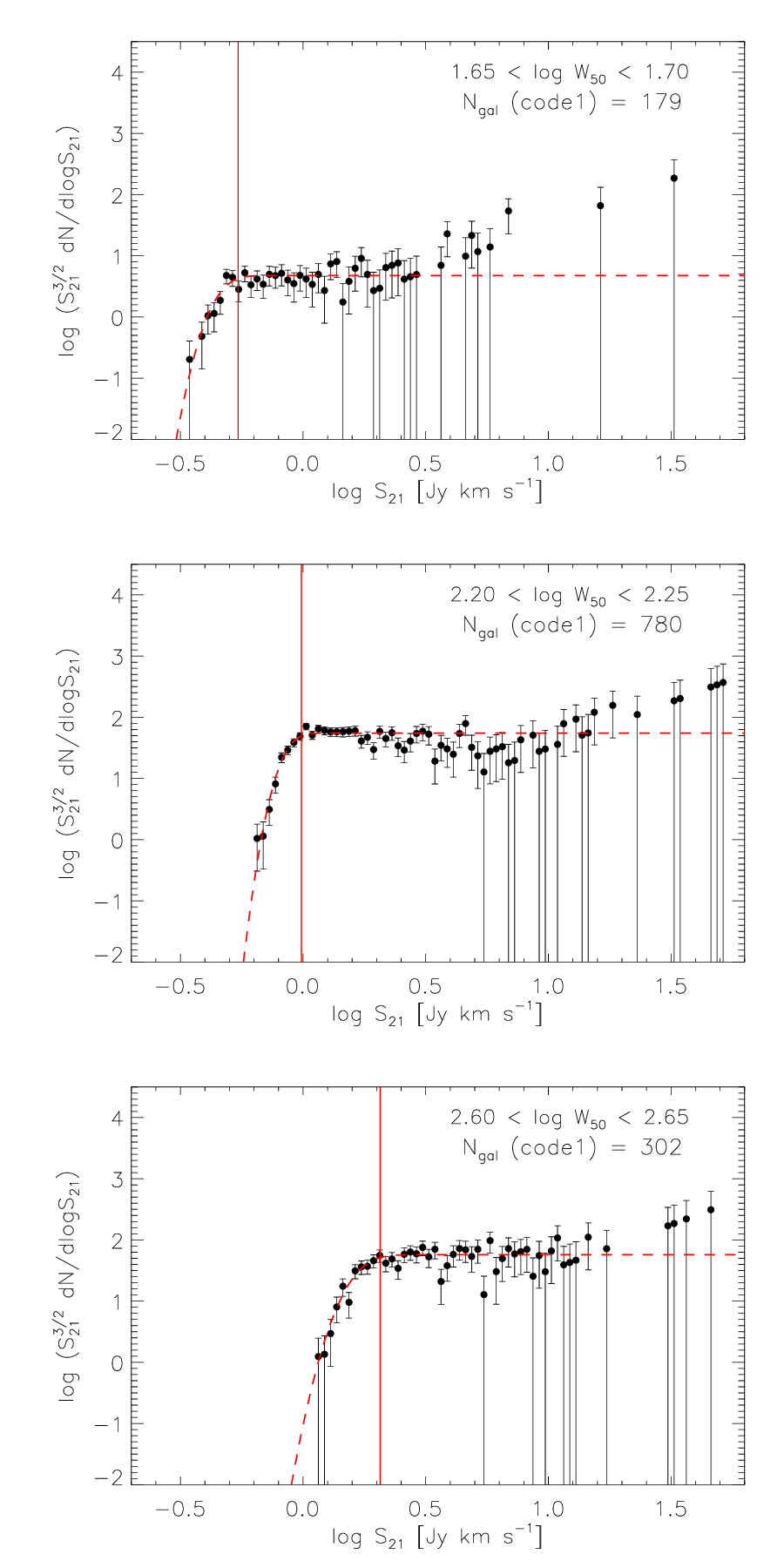}
\caption{Three representative examples of the 
$S_{21}$ - $S_{21}^{3/2} \: dN/d\log S_{21}$ 
distribution, used to evaluate completeness. 
Datapoints with errorbars ($1\sigma$ Poisson) represent the distribution of 
Code 1 sources in a low (upper panel), intermediate (middle panel) 
and high (bottom panel) profile width bin. The downturn of the 
distributions at low $S_{21}$ marks the limit where the survey completeness 
falls below unity. The red dashed 
line corresponds to an error function fit to the data, while the vertical 
red solid line represents the flux where the survey completeness is 90\% 
according to the fit, $S_{21,90\%,Code1}$. Values of $S_{21,90\%,Code1}$ for each 
width bin ($W_{50}$) are used to derive the 90\% completeness line of 
the survey presented in Equation \ref{eq:eqnthresh90c1}. A similar analysis has been
used for the combined catalog of Code 1 and 2 sources.}
\label{fig:logNlogS}
\end{center}
\end{figure}

\begin{figure}
\begin{center}
\epsscale{0.7}
\plotone{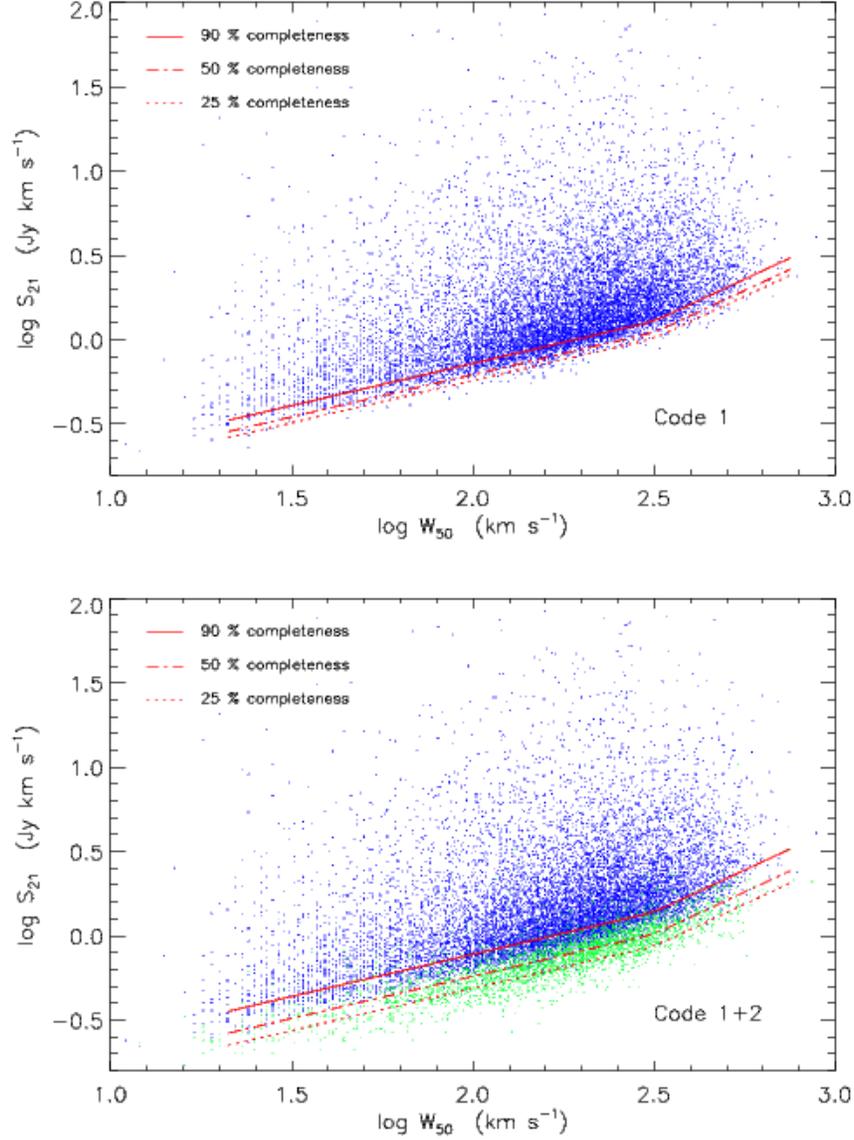}
\caption{Distribution of $\alpha.40$ extragalactic sources in the profile 
width versus integrated flux density $(\log W_{50} - \log S_{21})$ plane. 
The upper panel shows the distribution of Code 1 detections only, while
the lower panel shows the same for the whole $\alpha$.40 catalog, including 
Code 1 (blue symbols) and Code 2 (green symbols) detections. In both panels,
the solid red line corresponds to the 90\% completeness limit, while the red 
dash-dotted line corresponds to the 50\% (``sensitivity limit'') and the 
red dotted line to the 25\% (``detection limit'') completeness limits. 
See \S\ref{sec:completeness} for the analytical 
expressions for the plotted limits, as well as for an explanation of the 
derivation method.
}
\label{fig:fluxvsW}
\end{center}
\end{figure}

Now that a significant ALFALFA dataset exists, the data themselves can be used 
to derive the true sensitivity limits. The analysis of the real survey data is 
motivated both by a desire to use the real observables 
rather than predictions of the performance of the observing equipment 
and signal extraction pipeline, and especially by the fact that the ALFALFA survey has 
actually outperformed its predictions, as discussed in Appendix A of \citet{Martin10}.
Hence, we follow a different method to determine the ALFALFA completeness
that makes no use of ``fake sources'', but relies instead
on the $\alpha$.40 catalog itself. For a flux-limited sample from a uniformly distributed population,
number counts will follow a power-law with an exponent of $-3/2$.  We then can determine
the onset of incompleteness when our data deviates from this form.  Briefly, the details of this method
consist of the following steps:
\begin{enumerate}
 \item The Code 1 sources are divided into 32 equally spaced bins in $\log W_{50}$.
 \item  For each width bin, we count the number N of detected sources in logarithmic intervals of flux
        density to determine the $dN/d\log S_{21}$ histogram; apart from the impact 
        of large scale structure in the survey volume, number counts are expected 
        to follow a power-law with an exponent of $-3/2$.
  \item For each bin in $\log W_{50}$, we plot 
        $S_{21}^{3/2} \, dN / d\log S_{21}$ versus $S_{21}$; see Figure \ref{fig:logNlogS} for 
        three representative width bins.  This distribution should be flat if all 
        sources are accounted for. A downturn at low $S_{21}$ thus marks the onset
        of incompleteness.
  \item We fit an error function to each histogram (red dashed lines) and
        assume completeness over the well sampled range of $S_{21}$ over which the 
        distribution shows a flat plateau.
  \item We calculate the integrated flux density where the ALFALFA completeness crosses
        $90$\%, $50$\%, and $25$\% (vertical red lines mark the $90$\% 
        completeness in each bin).  In practice, the distributions drop off in the same
        in the same way, such that the 50\% and 25\% limits occur at a constant offset in $\log S_{21}$ 
        from the 90\% value across all bins.
 \item  The values
of  $S_{21,90\%}$ for each W$_{50}$ bin are then fit with the combinaton of two
straight lines, similar to Equation \ref{eq:eqnthresh90c1},  
with a break at $W_{50}=300$ km s$^{-1}$. 
\end{enumerate}
The resulting $90$\% completeness limit 
(red solid line in upper panel of Figure \ref{fig:fluxvsW}) for Code 1 sources 
can be expressed as:  
\begin{equation}
     \log S_{21,90\%,Code1} = \left\{
     \begin{array}{lr}
       0.5 \log W_{50} - 1.14, \,\,\, &  \log W_{50} < 2.5 \\
           \log W_{50} - 2.39, \,\,\, & \log W_{50} \geq 2.5  
     \end{array}
   \right.
\label{eq:eqnthresh90c1}
\end{equation}

\noindent
where $S_{21}$ is in Jy km s$^{-1}$ and $W_{50}$ is in km s$^{-1}$.
As mentioned before, the 50\% and 25\% completeness limits occur at a constant offset
from the 90\% value.  The derived offsets for the Code 1 sources only are:
\begin{equation}
  \begin{array}{l}
    \log S_{21,50\%,Code1} = \log S_{21,90\%,Code1} - 0.067 \\ 
    \log S_{21,25\%,Code1} = \log S_{21,90\%,Code1} - 0.102. 
  \end{array}  
\end{equation}

Of the \ngal ~extragalactic objects in the $\alpha.40$ sample, \nctwo ~are categorized
as Code 2 detections (low signal-to-noise detections with prior optical detection). The 
lower panel of Figure \ref{fig:fluxvsW} shows the corresponding plot of the distribution 
of sources in the $\log W_{50} - \log S_{21}$ plane
for the $\alpha.40$ extragalactic catalog, including the Code 2 detections
which are shown as green symbols. These additional HI sources are expected to have a lower 
detection threshold, clearly evident in the lower panel of Figure \ref{fig:fluxvsW}. 
An analysis identical to the above can be performed 
including both the Code 1 and 2 sources, yielding a relation for the 
combined catalog (red solid, dash-dotted and dotted lines in the bottom panel of Figure
\ref{fig:fluxvsW}):

\begin{equation}
\log S_{21,90\%} =  \left\{
     \begin{array}{lr}
       0.5 \log W_{50} - 1.11, \,\,\,  &\log W_{50} < 2.5 \\
            \log W_{50} - 2.36, \,\,\, &\log W_{50} \geq 2.5  
     \end{array}
   \right.
\label{eq:eqnthresh90c12}
\end{equation}
 \noindent
and

\begin{equation}
  \begin{array}{l}
   \log S_{21,50\%} = \log S_{21,90\%} - 0.130 \\ 
   \log S_{21,25\%} = \log S_{21,90\%} - 0.202.  
  \end{array}
\end{equation}

Excluding the Code 2 sources from the HIMF analysis as did \citet{Martin10} guarantees that more
confident detections with well-understood selection criteria are used. It could be argued
that the use of sources of Code 2 in the analysis could provide value added to
the determination of the HIMF. This is discussed further in 
\S\ref{sec:limits}. In practice, statistical studies requiring stringent requirements
on sensitivity limits should use only Code 1 sources and Equation \ref{eq:eqnthresh90c1}. 
With the proper caution associated with the incomplete nature of Code 2 sources, the
combination of Code 1 and Code 2 sources and Equation \ref{eq:eqnthresh90c12} can be
used in studies which can benefit from a larger sample.

In both cases, the 50\% completeness limit can be considered the ``sensitivity limit'' 
of the survey, since it is the most relevant completeness limit for the 
derivation of galaxy statistical distributions, such as the HIMF and the 
HI width function. \citet{RS02} have shown that adopting a step function cut at the 
50\% completeness limit of a survey produces approximately the same 
statistical results as adopting the survey's full completeness function.
The 25\% completeness limit can be identified with the ``detection limit'' 
of the survey, that is the integrated flux density level below which a source 
has only a small chance of being detected and cataloged. 

We remind the reader that the quoted limits given here refer to the full 
$\alpha$.40 catalog, and hence are representative of the average ALFALFA datacube 
noise properties. However, because of variations
in noise among and within grids and because some localized regions are entirely
contaminated by RFI, limits on the HI flux density at arbitrary positions
(e.g. upper limits for non-detections) must be computed individually, 
by specific inspection of the spectrum noise properties of the data cubes
and their associated ``weights grid'' and the continuum maps. It is the
availability of such ancillary information which enables the use of the
full ALFALFA dataset for stacking \citep{Fabello11} to probe statistical
ensembles more deeply.

\begin{figure}
\begin{center}
\epsscale{1.0}
\plotone{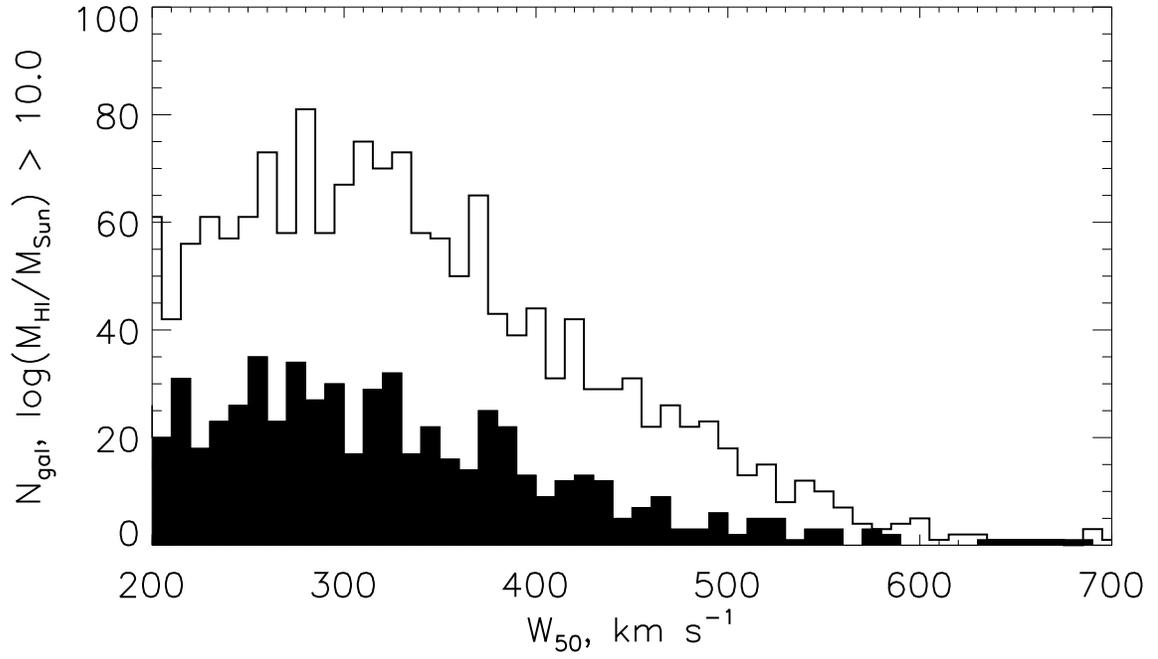}
\caption{The distribution of profile widths in $\alpha.40$ (open histogram) 
and HIPASS (filled histogram) for objects with log $M_{HI}/M_{\odot} > 10.0$.}
\label{fig:histbigW}
\end{center}
\end{figure}

As the previous generation blind HI survey, HIPASS \citep{Meyer04} set the standard
for survey completeness; by design, ALFALFA was intended to surpass and supercede HIPASS.
A reasonable comparison of the impact of the different source detection 
schemes (including the absolute level of flux density sensitivity) may be made by 
comparing the distribution of the highest-mass galaxies in HIPASS and 
ALFALFA. For example, one might have anticipated that the original HIPASS peak-flux 
density detection scheme \citep{Meyer04} could bias the catalog against edge-on 
(extremely wide) profiles at the highest masses, and such a bias could explain 
the finding of \citet{Martin10} that the HIPASS HIMF underestimated 
the number density of the highest mass galaxies. Figure \ref{fig:histbigW}
shows a comparison of the distribution of profile widths in $\alpha.40$ 
(open histogram) and HIPASS (filled histogram), for objects with 
log $M_{HI}/M_{\odot} > 10.0$.
No obvious difference which would explain a lack of high-mass sources in the HIPASS 
catalog is apparent. While the peak-flux density threshold detection 
could introduce such a bias, it is apparent that the matched filtering technique 
subsequently applied to the HIPASS dataset recovers high-width objects as 
does the technique used in ALFALFA. Instead, we attribute the lack of extremely 
high-mass sources in the HIPASS catalog to that survey's limited redshift 
extent and its lowered sensitivity near its bandpass redshift limit, both
of which resulted in inadequate sampled volume and thus an undercounting of the
rare, highest mass HI disks. 

\begin{figure}
\begin{center}
\plotone{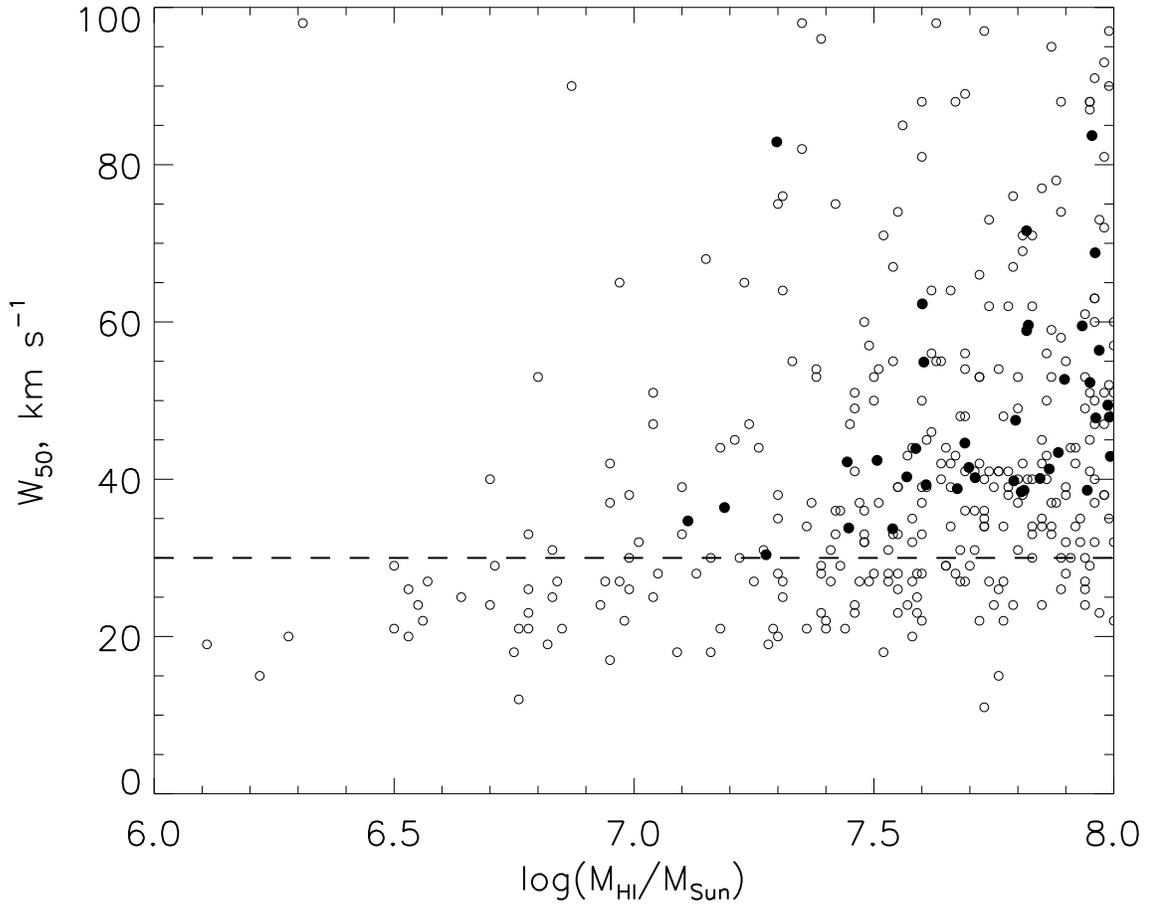}
\caption{The distribution of profile 
widths W$_{50}$ in ALFALFA (open circles) and HIPASS (filled circles, 
enlarged for visual clarity), for objects with log $M_{HI}/M{_\odot} < 8.0$. 
The overplotted horizontal dashed line shows the profile width cutoff at 30 \kms, 
the limit for inclusion in the HIPASS catalog.}
\label{fig:lowW}
\end{center}
\end{figure}

Furthermore, because of its lower sensitivity, poorer angular and spectral resolution and
source detection scheme, HIPASS was limited in its ability to probe the very low-mass 
and narrow-width HI sources. The spectrometer setup employed by HIPASS 
yielded a raw resolution of 13.2 \kms and of 18 \kms \ after Hanning smoothing; the 
narrowest objects included in the HIPASS catalog have W$_{50}$ = 30 \kms. 
In contrast, ALFALFA's velocity resolution is 11 \kms~ after smoothing is 
applied, and the $\alpha.40$ catalog thus includes sources with extremely narrow 
velocity widths. Although the signal extraction algorithm adopted by \citet{Saintonge07a}
applied a minimum template width of 30 \kms, the refined final process of parameter
extraction based on individual examination of the emission region permits finer
width estimation. In fact, 289 of the extragalactic objects included in Table
\ref{tab:catalog} have $W_{50} < 30$ \kms. Figure \ref{fig:lowW} examines
the distributions of low HI mass systems and their profile velocity 
widths in the two surveys; ALFALFA is clearly superior in its ability
to probe the lowest mass systems. 
This increased sensitivity to very narrow HI line
emission enhances ALFALFA's ability to probe the lowest HI masses, 
which in turn robustly constrains the faint-end slope of the HI mass 
function, $\alpha$. In fact, at the lowest HI masses, 
$\log M_{HI}/M_{\odot} <$ 8.0, the HIPASS catalog includes only 40 objects
while the $\alpha.40$ catalog contains 339. The ability of ALFALFA to sample
narrower HI line sources is also critical for the derivation of the
HI width function and its relation to the halo mass function \citep{Papastergis11}.

\section{The Impact of ALFALFA Survey Characteristics on Derivation of
the HIMF}\label{sec:impact}

In drawing conclusions from blind HI surveys about the HI-selected population in the 
local universe,
it is critical to understand the biases in the survey due to its sensitivity limits,
uncertainties in the HI line flux densities and distances leading to uncertainties in 
the derived HI masses, and the impact of large-scale structure in the survey volume.
\citet{Toribio11b} use a subsample of ALFALFA HI sources identified in low density
environments to establish a standard of normal HI content and performed an analysis
of the completeness of the particular version of the ALFALFA catalog they used.
\citet{Martin10} (see also Martin 2011) provided an overview of important effects
that impact the derivation of the HIMF by two different methods commonly used
to derive mass and luminosity functions, namely the 1/V$_{max}$ method and the
two-dimensional stepwise maximum likelihood (2DSWML) method. In the context of
applications such as the derivation of the HIMF by those two methods, we discuss here
in greater detail the magnitude and character of $\alpha.40$ survey properties, 
its limitations and biases. It is particularly important to understand 
these effects now because we anticipate the `100\% ALFALFA survey' to be available 
in the near future. The large increase in the number of galaxies available for that 
analysis will decrease the statistical uncertainties on the measurements, thus 
amplifying the relative impact of systematics and biases. Additionally, at that stage 
it will be less practical to create thousands of realizations to help understand 
the various effects. The results presented in this Section will provide a baseline 
and dictate procedure for the final measurement of the HI mass function from 
the completed ALFALFA survey.

\subsection{The Limits of ALFALFA: Code 2 ``Prior'' Sources and the RFI-imposed Redshift Cutoff}
\label{sec:limits}
Selection effects related to the Code 2 sources in $\alpha.40$ are poorly determined.
Because they require redshift information derived from other sources, 
they are subject to the limitations of the availability of such confirming data.
Additionally, ALFALFA's sensitivity as a function of distance is strongly affected 
by RFI especially in the frequency range contaminated by the aviation radar
at the San Juan airport (1350 MHz, corresponding to $cz \sim 15600$ \kms). 
For these reasons, \citet{Martin10} 
included only objects with Code 1, detected within 15000 \kms. Yet it may be 
argued that the additional information contained in Code 2 sources, dipping to 
lower flux limits, could provide additional insight. A first evaluation of the
value added to the HIMF by Code 2 sources relates to the observation that
most Code 2 sources fall near $M^*_{HI}$, a region of the HIMF well sampled 
by Code 1 sources: the value added is likely thus to be negligible. We explore
numerically this expectation.

\subsubsection{The Code 2 Sources}\label{sec:code2}

Because of the requirement that Code 2 sources be identified with an OC of known
(prior) redshift, most often contributed by optical/IR surveys like SDSS, those sources
may be biased toward overdensities, toward those regions of the local volume that 
have been included in specific targeted or wide-area redshift surveys, such 
as the Virgo Cluster, and in particular toward those regions of the sky that have 
been covered  in the spectroscopic catalogs of the SDSS. 

Does the inferred HIMF change if Code 2 sources are included in its derivation?
We account for HI mass and flux density errors by creating 500 
realizations of an HIMF that includes sources of both Code 1 and Code 2, 
and compare those to 100 realizations of the fiducial HIMF published in 
\citet{Martin10} which contained only the Code 1 objects. We use the 
2DSWML method, but do not jackknife resample. As did \citet{Martin10},
we restrict the analysis to the contiguous areas contained in $\alpha.40$
and limited to c$z < 15000$ \kms.
Over the same volume, the inclusion of Code 2 sources increases the 
sample size used for this analysis from the 10,021 included in \citet{Martin10}
to 11,177. 

\begin{figure}
\begin{center}
\plotone{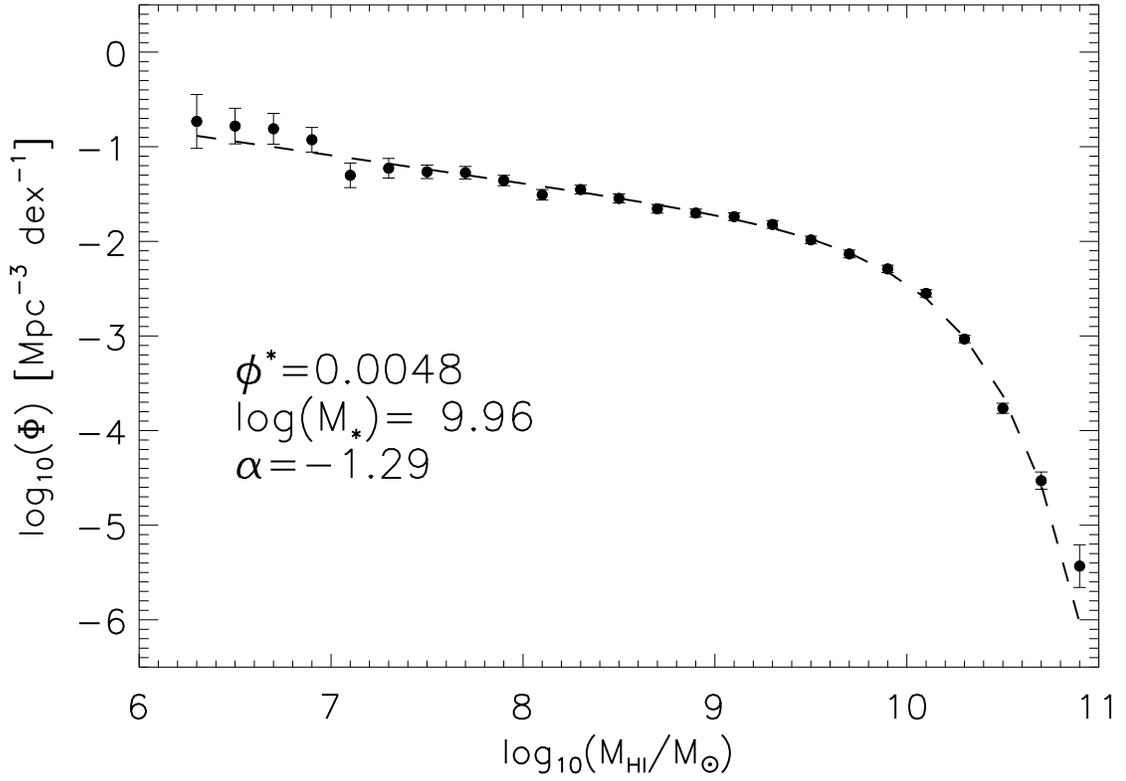}
\caption{The HIMF found via the 2DSWML method (without jackknife resampling) when 
Code 2 sources are included. The best-fit Schechter function is overplotted 
as a dashed line, with the best-fit parameters displayed. While $\Omega_{HI}$ 
and the overall Schechter function shape are not changed, the inclusion 
of the additional sources does slightly flatten the faint-end slope compared
to results obtained using only Code 1 sources (Table \ref{tab:himf}).}
\label{fig:HIMFc2}
\end{center}
\end{figure}

Figure \ref{fig:HIMFc2} displays the HI mass function found when Code 2 
sources are included in the analysis. The parameters of the function are 
not strongly affected by the inclusion of these sources. We find 
$\phi_{*}$ (h$_{70}^3$ Mpc$^{-3}$dex$^{-1}$) = 4.8 $\pm$ .3 $\times$ 10$^{-3}$, 
$\log $(M$_{*}/$M$_{\odot})$ + 2 $\log$ h$_{70}$  = 9.96 $\pm$ 0.02 and 
$\alpha$ = -1.29 $\pm$ 0.02. These correspond to 
$\Omega_{HI} =$ 4.1 $\pm$ 0.3 $\times 10^{-4}$ h$^{-1}_{70}$ found 
by integrating the Schechter function, or 
$\Omega_{HI} =$ 4.2 $\pm$ 0.1 $\times 10^{-4}$ h$^{-1}_{70}$ when summing 
the binned measurements directly. The fiducial HIMF which includes only
Code 1 objects as reported by \citet{Martin10} finds 
$\phi_{*}$ = 4.8 $\pm$ 0.3, $\log $(M$_{*}/$M$_{\odot})$ = 9.96 $\pm$ 0.02, 
$\alpha$ = -1.33 $\pm$ 0.02, $\Omega_{HI}$ (analytical) = 4.3 $\pm$ 0.3, 
and $\Omega_{HI}$ (summed) = 4.4 $\pm$ 0.1, all in the same units as expressed 
for the results with both Code 1 and 2 sources.

Encouragingly, these results indicate that the ALFALFA survey's detection coding
scheme does not systematically exclude significant sources of HI gas energy 
density in the local universe. Rather, the agreement between the Code 1 and 
the Code 1+2 HIMFs suggest that our robust understanding of the survey's 
sensitivity extends to those weaker sources identified as Code 2 objects. 

\begin{figure}
\begin{center}
\plotone{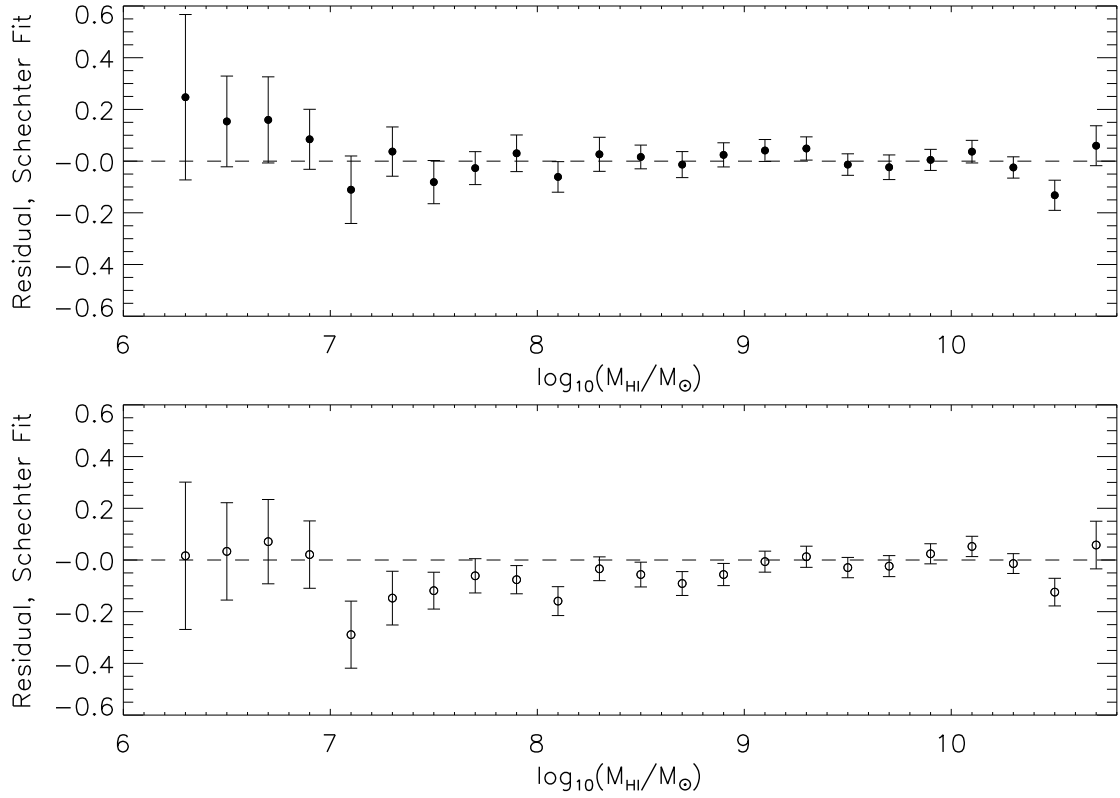}
\caption{Residuals (best-fit Schechter 
model subtracted from binned data) of HI mass functions calculated using only 
Code 1 sources (top) and both Code 1 and 2 sources (bottom). In both cases, 
the comparison model is the fiducial, Code 1-only Schechter function given
by \citet{Martin10}. The zero-residual reference line is overplotted 
as a dashed line.}
\label{fig:HIMFres}
\end{center}
\end{figure}

The only potentially significant impact is on the faint-end of the HIMF, 
influencing both the slope and the points there. The slope parameter $\alpha$ 
is flattened in the Code 1+2 case, though the two values are just barely 
within 1$\sigma$ of each other. In Figure \ref{fig:HIMFres}, we compare the 
residuals (the best-fit, fiducial HIMF Schechter model, subtracted from the 
binned data) for the case where we consider only Code 1 objects (top panel) 
and the Code 1+2 case (bottom panel). The figure clearly demonstrates 
that the Code 1+2 HIMF measured fewer low-mass objects per unit volume, thereby
yielding a flatter slope. This is unsurprising for, in comparison to HI surveys,
optical surveys undersample dwarf, low surface brightness galaxies. The typical
Code 2 detection is a galaxy near $L^*$, and the redshift distribution of Code 2
sources lacks the smattering of low redshift objects present in a HI-selected
sample. As a result, Code 2s add very few additional 
sources at the lowest redshifts, at which low HI masses are detectable. This is
an example of the fact that adding Code 2 sources to the sample is more likely
to subtract than to add value to the result: ``more is less''.

We note that in the process of source extraction, a second set of marginal HI line
detections has been identified which coincide with possible OCs for which no redshift 
measurement is available. Because the probability of these objects is yet too uncertain,
they are not included in the current $\alpha.40$ catalog.
Future followup observations to be made after the main survey is completed
will be undertaken to confirm the reality of the HI
line detections. This program will contribute additional low HI line flux density 
sources to the final ALFALFA catalog in this region of the sky.

\subsubsection{The Full Redshift Extent of the ALFALFA Survey \label{sec:redshift}}
(Unfortunately) we live on a planet occupied by technologically-active humans.
Figure 6 of \citet{Martin10} illustrates the relative spectral weight within the 
40\% ALFALFA survey volume as a function of observed heliocentric velocity. 
A relative weight of 1.0 indicates that the entire surveyed volume was accessible 
for source extraction and produced high-quality data. As also evident in the
deficit of sources near a distance of $\sim$225 Mpc seen in Figure \ref{fig:spanplot}, 
the FAA radar at the San Juan Airport contaminates the frequencies corresponding
to source at $cz$ between 15000 and 16000 \kms, rendering the detection of
sources in this range impossible when the transmitter is on. Beyond 16,000 \kms, 
ALFALFA's sensitivity recovers, but at the corresponding distance, it is
sensitive only to the most massive of galaxies. As a result, this distant volume 
contributes only a small number of galaxies to the overall $\alpha.40$ sample.

For these reasons, the analysis of the HIMF in \citet{Martin10} neglected 
galaxies beyond 15000 \kms, so that the results would not be influenced by the 
large spectral weight gap. This exclusion was especially important in the case 
of the 2DSWML method, since the 1/V$_{max}$ method allows the inclusion of 
explicit corrections for known missing volumes. 2DSWML, by contrast, determines 
the shape of the HIMF by comparing counts in HI mass bins to a built-in 
description of ALFALFA's flux density sensitivity as a function of distance and width. 
The large gap, which is not anticipated by this approach, may have caused 
problems in the analysis were those objects included. \citet{Martin10}
felt it was safer to limit 
the first measurement of the HIMF to regions where the spectral weights 
are relatively smooth, that is, to galaxies within 15000 \kms. Here, we
revisit the issue and consider the influence, if any, of including the full 
redshift extent of the survey in the HIMF analysis. 

In particular, we would expect that the increased bin counts at the very highest 
HI masses may increase the statistical significance of the HIMF measurement there. 
Such a possibility is of interest because \citet{Martin10} determined that 
ALFALFA is more sensitive to high-mass galaxies than HIPASS was, with
HIMF results indicating that previous blind HI surveys have missed a 
significant percentage of the most massive HI disks. 
To test this possibility, we calculated the HIMF using both methods and using 
all of the Code 1 sources out to 18,000 \kms. For each method, we created 
250 realizations of the survey to account for flux density, distance, and mass errors.
The fit parameters and $\Omega_{HI}$ values are displayed in Table \ref{tab:himf} 
for both the 1/V$_{max}$ and 2DSWML methods. It is worth noting that the 
2DSWML result is distorted,  likely because of the influence of the inaccessible 
volume and the inability of this method to correct for it. In fact, the 2DSWML
result drastically underestimates $\Omega_{HI}$, shifts 
$\log $ (M$_{*}$/M$_{\odot})$ to a higher value, and flattens out the low-mass 
slope $\alpha$. On the other hand, the 1/V$_{max}$ method continues to 
function as expected and results in a reliable measurement. Both of the 
1/V$_{max}$ results are consistent with each other, including the 
measured values for $\Omega_{HI}$, but the 2DSWML method in the presence
of the redshift gap performs poorly. This result confirms the decision
by \citet{Martin10} to limit their analysis to the volume within $cz < 15000$ \kms.

This analysis provides further evidence of the relative strengths and weaknesses
of the two available methods for estimation of the HIMF. While the 2DSWML approach
provides a powerful statistical tool, it functions as a `black box' method 
that cannot be manipulated by additional knowledge of the survey. In some cases, 
this may be an advantage, but in the case of ALFALFA where we have detailed 
information about the survey volume, the survey sensitivity, and other factors 
contributing to the HIMF, the 1/V$_{max}$ method provides a clearer path 
and a more understandable answer.

\subsection{Uncertainties in the HI Mass}\label{sec:mass}

On the low HI mass end, uncertainties in the conversion from HI line flux density
to HI mass are the primary source of error on the HIMF.  Unlike the practice
in the derivation of most HIMF results in the literature, the error analysis
undertaken here and by \citet{Martin10} has taken this explicitly into account. 
Because the HIMF is based on binning galaxies by HI mass and then considering 
each bin as an independent data point, it is not straightforward to carry 
HI mass uncertainties through analytically. Instead, the ALFALFA HIMF's 
uncertainties due to mass errors are calculated through the creation 
of many hundreds of realizations, each with randomly assigned mass (i.e., 
distance and flux density) errors. Here, we elaborate further on the distance estimate 
scheme used in ALFALFA, the biases that would be introduced by using 
alternative schemes (i.e., a pure Hubble flow model) and the overall 
impact of distance and flux density errors on the HI mass estimates used to construct 
the HIMF.

\begin{figure}
\begin{center}
\plotone{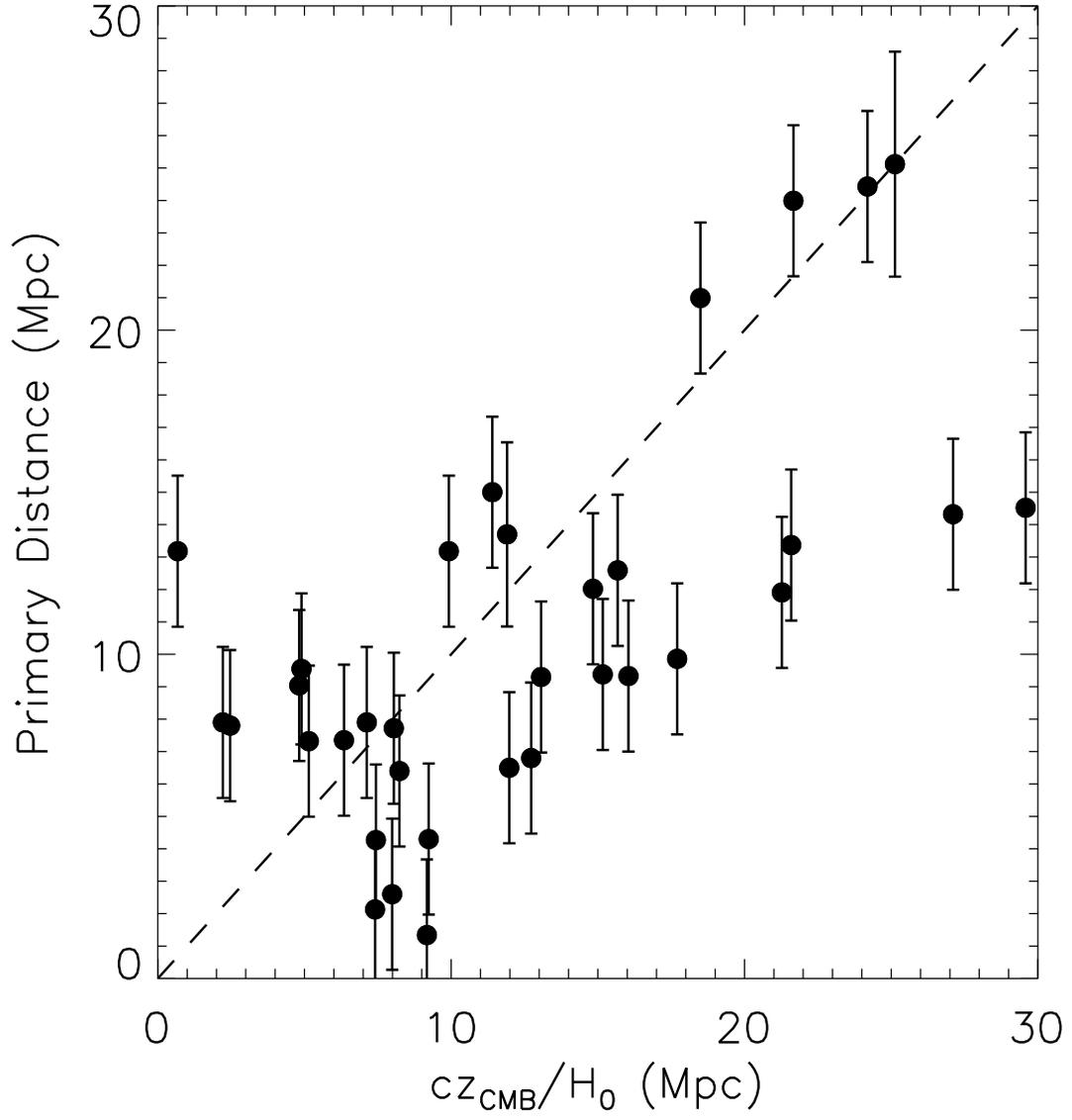}
\caption{Primary distances from the literature vs. estimates based 
only on pure Hubble flow, with the ALFALFA distance uncertainty 
estimates overplotted. The dashed line indicates a one-to-one correlation.}
\label{fig:dist}
\end{center}
\end{figure}

The distance estimation scheme adopted for ALFALFA was described by \citet{Martin10}
and is summarized briefly here. When distances are based on the adopted flow model, 
we employ the model's error estimates, constrained by the fit of the model 
to the observed velocity fit. When distances are estimated using pure Hubble 
flow, the error is estimated to be $\sim$10\%. We fix a minimum error of 
163 \kms, based on the local velocity dispersion measured by \citet{Masters05}. 
To demonstrate the importance of using the full suite of available 
information to estimate distances, Figure \ref{fig:dist} compares the primary 
distances (used in $\alpha.40$) to the values that would be obtained assuming 
pure Hubble flow. 

In their estimate of galaxy masses for the HIMF, the HIPASS team assumed Hubble flow. 
This is not a safe assumption, particularly in the regions of the sky 
surveyed in $\alpha.40$. The Virgo Cluster, in particular, represents a 
strong deviation from any assumed relationship between distance and recessional 
velocity. \citet{Masters04} showed the danger of assuming pure Hubble flow, 
especially because of the small redshifts accessible to blind HI surveys. 
Those authors concluded that the low-mass slope of the HIPASS HIMF was 
underestimated due to neglecting peculiar velocities, and predicted that a 
survey in the direction of Virgo could severely underestimate the low-mass slope.

Given the large-scale structure in the $\alpha.40$ volume, we would expect the 
HIMF to vary strongly if a poor choice of distance estimate were made. In 
order to test this, we have re-calculated the 2DSWML estimate of the HIMF using 
pure Hubble flow to estimate distances. That is, we converted the observed 
heliocentric velocities into the CMB rest frame, and then assumed 
D$_{Mpc}$ = $cz_{cmb}/H_0$, where we adopt the ALFALFA standard 
$H_0$ = 70 \kms Mpc$^{-1}$. In this case, we have no ideal estimate of the 
distance error, and therefore use 10\% of the Hubble flow distance or the 
local dispersion value 163 \kms, whichever is greater. As usual, flux density errors 
are also folded into the mass uncertainties. Once again, we create 250 
realizations to estimate uncertainties. 

\begin{figure}
\begin{center}
\plotone{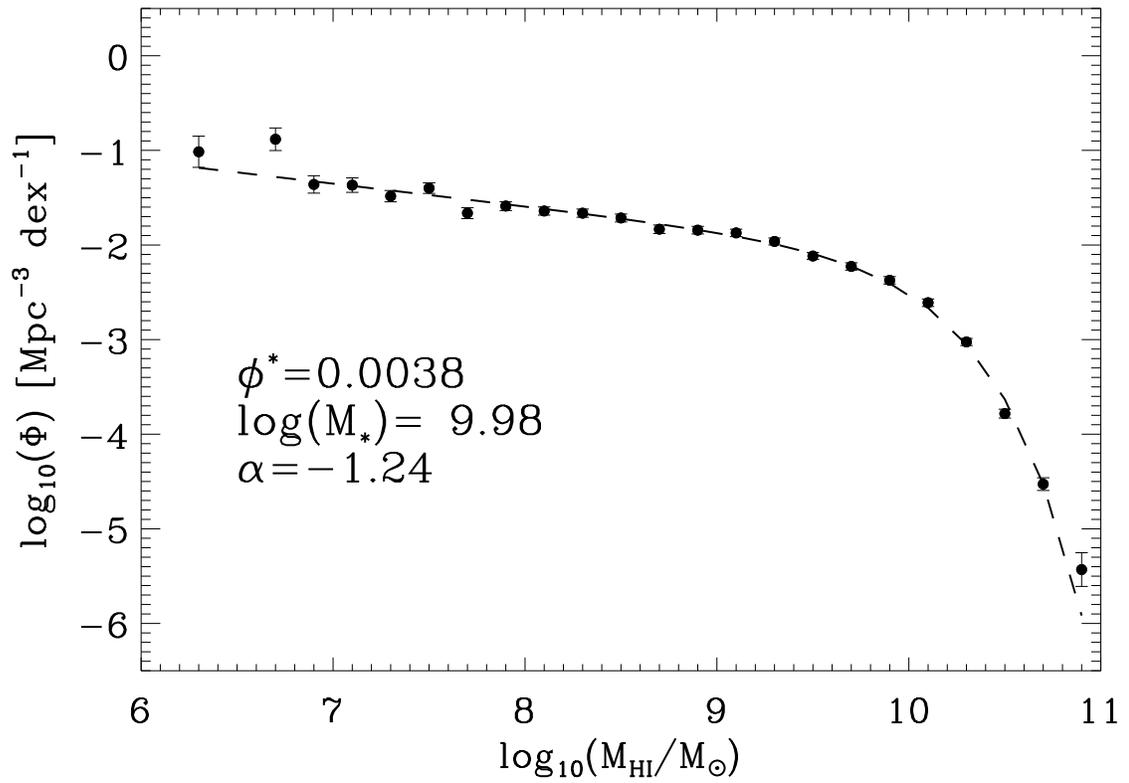}
\caption{The HI mass function obtained via the 2DSWML method when distances, and therefore 
masses, are obtained assuming pure Hubble flow with H$_0$ = 70 \kms Mpc$^{-1}$.
As anticipated by \citet{Masters05}, the adoption of pure Hubble flow yields an
underestimate of the low HI mass slope $\alpha$.}
\label{fig:himfhubble}
\end{center}
\end{figure}

The resulting HI mass function and Schechter fit parameters are displayed 
in Figure \ref{fig:himfhubble}. As anticipated \citep{Masters04}, the use of 
Hubble flow has caused a serious underestimate of the faint-end 
slope $\alpha$. ALFALFA's success at robustly measuring the HIMF 
depends not only on large sample size over a cosmologically significant 
volume, but also on the selection of a reasonable model for distance estimation.

\begin{figure}
\begin{center}
\plotone{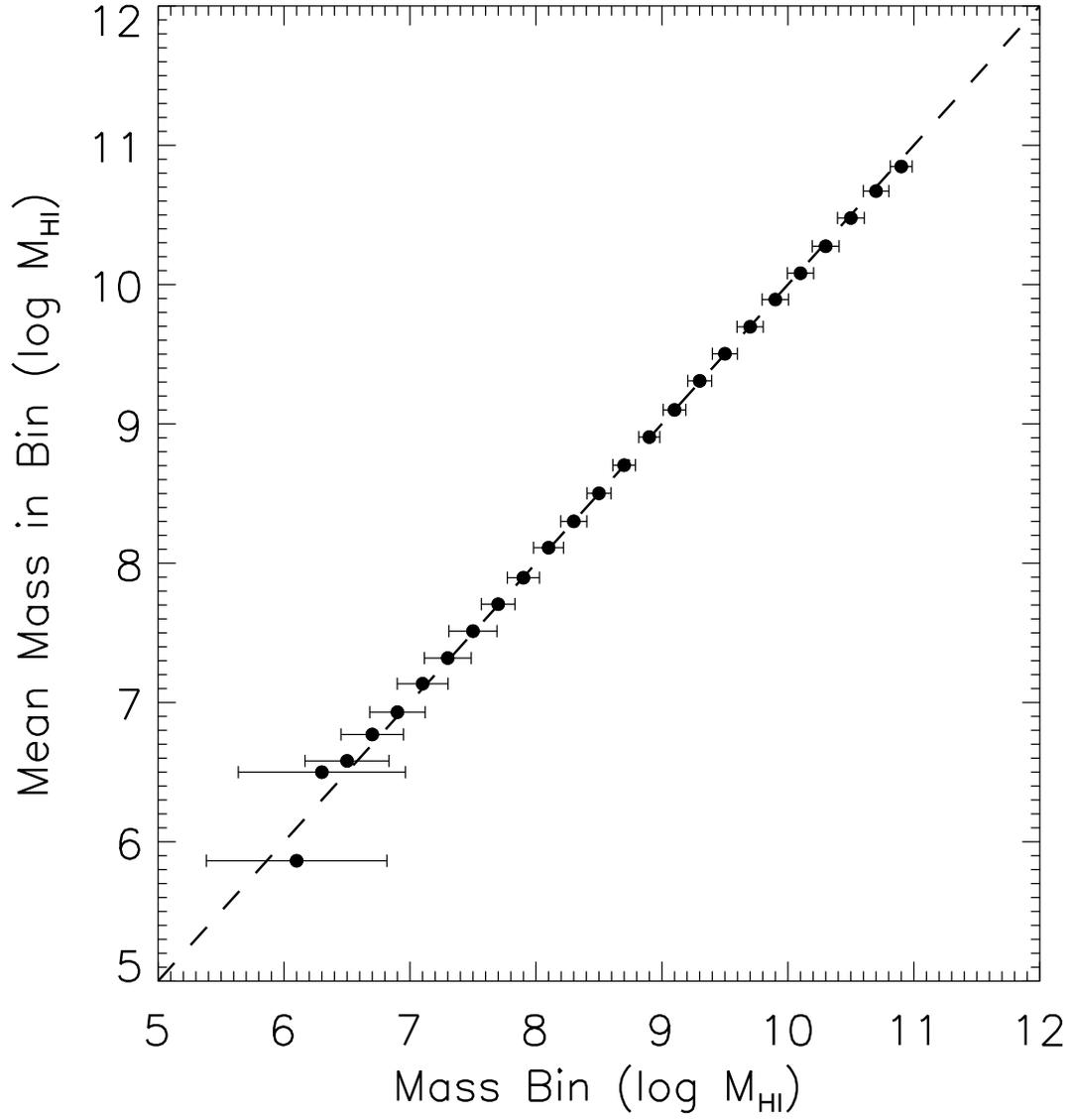}
\caption{The average (mean) mass falling into each HIMF bin. 
The estimated 1$\sigma$ uncertainty of a galaxy's HI mass is overplotted 
as error bars, along with a dotted line indicating a one-to-one relationship.}
\label{fig:massbin}
\end{center}
\end{figure}

Given the discussion of distance errors and their large impact on the HIMF 
and its uncertainties by \citet{Masters04}, it is reasonable to ask how large are the 
HI mass errors when both distance and flux density errors in the $\alpha.40$ 
sample are taken into account. To obtain robust estimates of HI mass errors, 
we created many thousand realizations of each galaxy in the $\alpha.40$ 
sample and applied distance and flux density errors. The result, displayed in Figure 
\ref{fig:massbin}, compares the HIMF mass bin galaxies would nominally fall 
into assuming a perfect measurement of distance and flux density (along the abscissa) 
to the mean mass of the galaxies assigned to that bin once realistic 
uncertainties are taken into account. The horizontal uncertainties 
indicate the 1$\sigma$ spread of potential `true' masses falling into 
nominally assigned mass bins. From the Figure, it is clear that ALFALFA's 
measurement of the HIMF and $\Omega_{HI}$ is not prone to large 
uncertainties above $10^{8.0} M_{\odot}$. In the mass range of interest 
to the missing satellites problem, dwarf galaxy studies, and the 
low HI mass slope of the HIMF, that is below $10^{8.0} M_{\odot}$,
 galaxies can easily be mis-assigned to bins, even when a realistic 
distance model is being used. Depending on the large-scale structure 
in the survey volume, this effect would lead to either an over- 
or under-estimate of $\alpha$. We therefore take great care to account, 
conservatively, for mass uncertainties.

\subsection{The Impact of Large Scale Structure in $\alpha.40$}\label{sec:lss}

Because blind HI surveys are relatively shallow, with ALFALFA probing the 
local universe only out to z $\sim$ 0.06, inhomogeneity in the survey volume 
has a strong impact on the derived HI mass function. This effect
is particularly true in the case of the 1/V$_{max}$ method, which is not 
as robust against large-scale structure, but the 2DSWML method is not
completely immune from these effects. To test the homogeneity of a sample, 
the usual statistical test applied is the V/V$_{max}$ test \citep{Schmidt68}. 
Much like the 1/V$_{max}$ method, this test considers the maximum 
volume out to which each source in a survey can be detected. By 
comparing the actual volume the source was detected in to the accessible 
volume, homogeneity in the sample can be evaluated; the expectation 
value $<V/V_{max}>$ is 0.5 in a homogeneous volume.

\begin{figure}
\begin{center}
\plotone{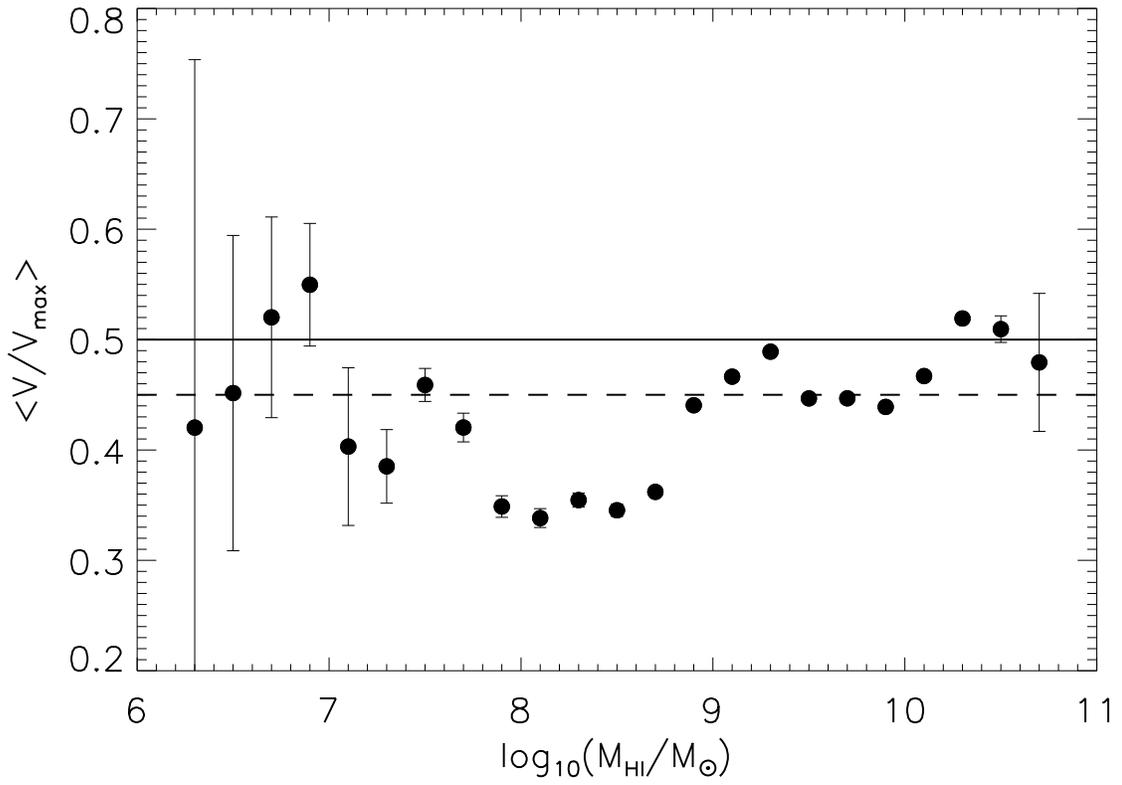}
\caption{The typical (mean) value of V/V$_{max}$, binned by HI mass. 
Error bars are Poisson counting uncertainties. The solid line 
indicates $<V/V_{max}>$ = 0.5, while the dashed line indicates 
$<V/V_{max}>$ = 0.45 for the $\alpha.40$ sample.} 
\label{fig:VVmax}
\end{center}
\end{figure}

In the case of the $\alpha.40$ volume, $<V/V_{max}>$ = 0.45. This indicates 
that, at 40\% completion, the survey does not yet contain enough volume to 
fully `smooth out' the effects of large-scale structure. This is reflected 
in Figure \ref{fig:VVmax} where $<V/V_{max}>$ is shown for each bin of
HI mass. The most obvious feature in this Figure, the dip near 
log (M$_{HI}$/M$_{\odot})$ $\sim 8.4$, is due to overdensities in the sample 
volume, primarily the Virgo cluster. Galaxies in those overdensities are 
found, preferentially, in those regions, rather than filling the full 
volume where ALFALFA's sensitivity could detect them, causing this dip. 

It is clear that $\alpha.40$ does not, yet, constitute a representative 
slice of the universe; as the survey progresses, we anticipate that the 
full sample will pass the $V/V_{max}$ test. Another, perhaps more intuitive, 
way to view the impact of voids and clusters in $\alpha.40$ is to 
compare the redshift distribution of cataloged galaxies to the prediction 
based on the survey's selection function (i.e., the percentage of galaxies
 at a given distance that are detectable in ALFALFA). The selection 
function is determined by the 2DSWML analysis of the HIMF, and when 
combined with the measurement of the HIMF, predicts the redshift 
distribution for a homogeneously distributed set of galaxies selected 
from the HIMF.

\begin{figure}
\begin{center}
\plotone{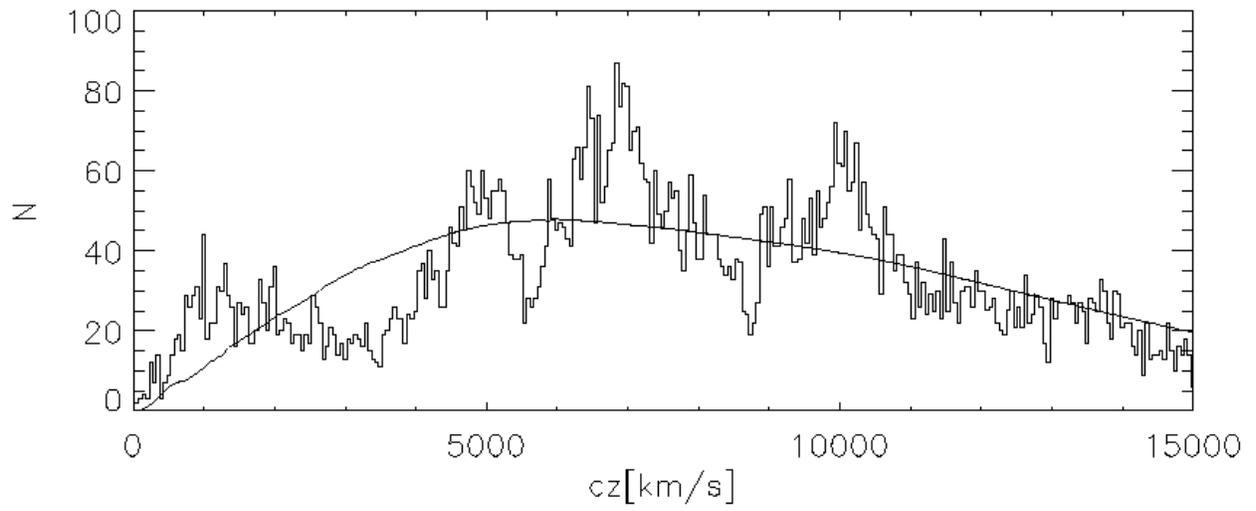}
\caption{The observed redshift distribution of $\alpha.40$ galaxies (histogram) 
compared to the expected distribution obtained via the survey's selection 
function.}
\label{fig:zdistr}
\end{center}
\end{figure}

Figure \ref{fig:zdistr} compares this expectation to the observations 
in $\alpha.40$. The bumps and dips in the histogram represent under- 
and overdensities, respectively, in the survey volume. For example, 
the Virgo Cluster explains the enhancement near 1,000 \kms. The 
Pisces-Perseus supercluster and its foreground void also make clear 
imprints on this figure.

\subsubsection{Subregions of the $\alpha.40$ catalog}\label{subregions}

If $\alpha.40$ does not represent a representative sampling 
of the universe, then statistical studies of the sample's characteristics, 
like the HIMF, may be subject to biases from large-scale structure. 
Because of its size, we can make an assessment by the impact of large
scale structure within separate subregions of the catalog.
The $\alpha.40$ sample is made up of 3 large, contiguous areas. 
In the Northern Galactic hemisphere, $\alpha.40$ covers 
07$^{h}$30$^{m}$ $< \alpha <$ 16$^{h}$30$^{m}$ in two separate blocks,  
4$^{\circ} < \delta < $16$^{\circ}$ and 24$^{\circ} < \delta 
< $ 28$^{\circ}$. We refer to these subregions as 
$\alpha.40.North1$ and $\alpha.40.North2$, respectively. 
In the Southern Galactic hemisphere, $\alpha.40$ covers 
22$^{h}$ $< \alpha <$ 03$^{h}$, 24$^{\circ} < \delta 
< $32$^{\circ}$, referred to as $\alpha.40.South$. The entire $\alpha.40$, 
combined together, covers enough cosmological volume for the effects of 
large-scale structure on the derivation of the HIMF to begin to become minimal, 
but reducing its coverage further leads to a situation in which the HIMFs 
derived for individual sub-regions are strongly affected by over- and 
under-densities within their volume.

\begin{figure}
\begin{center}
\plotfiddle{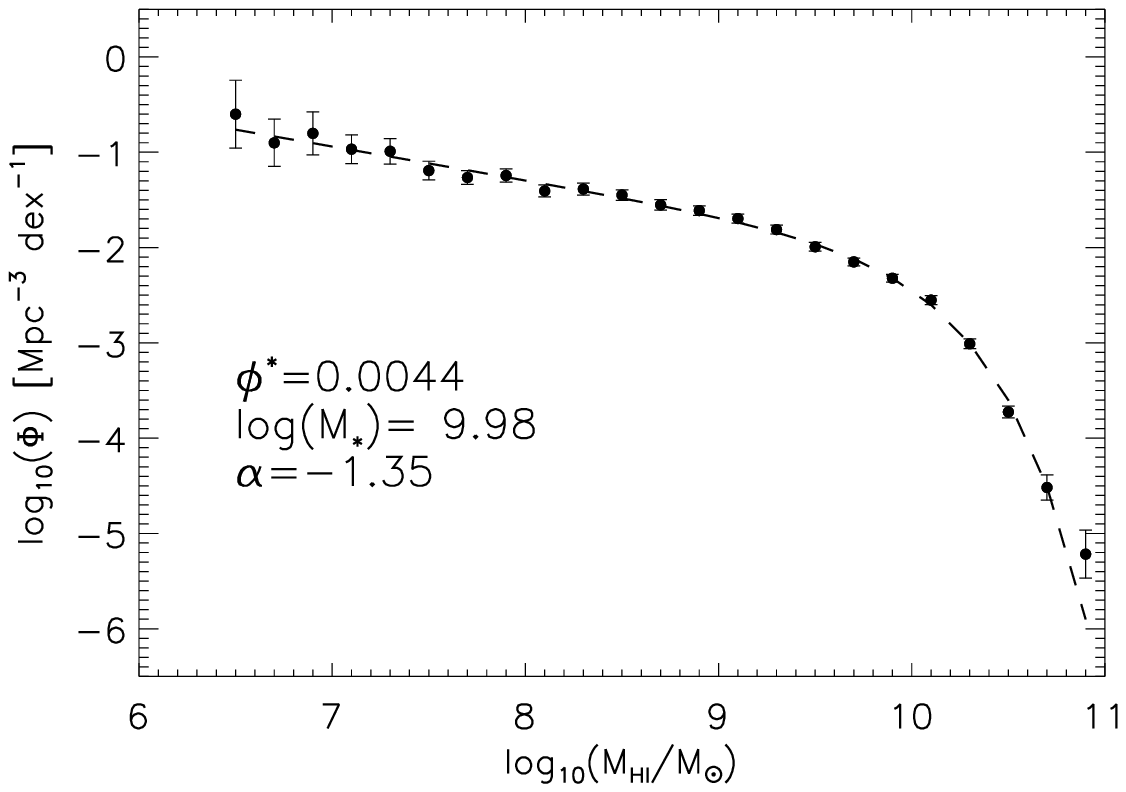}{2.0in}{0}{72}{72}{-420}{-260}
\plotfiddle{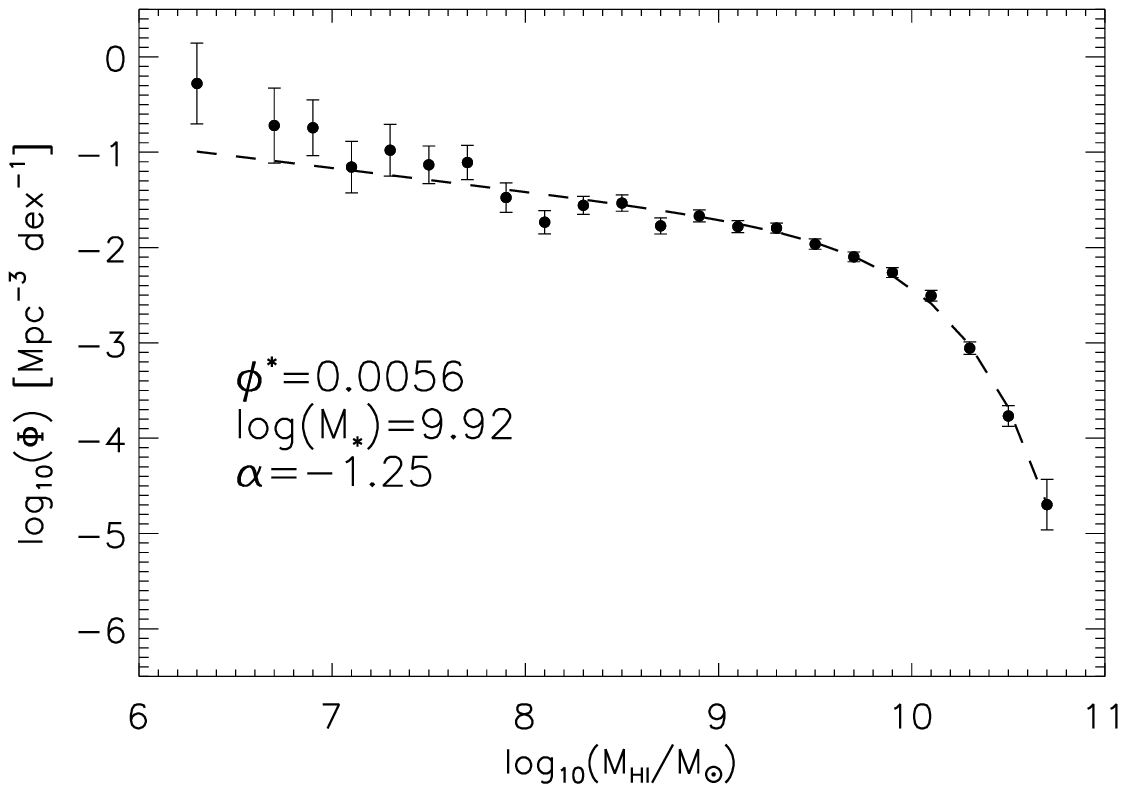}{2.0in}{0}{72}{72}{-420}{-290}
\plotfiddle{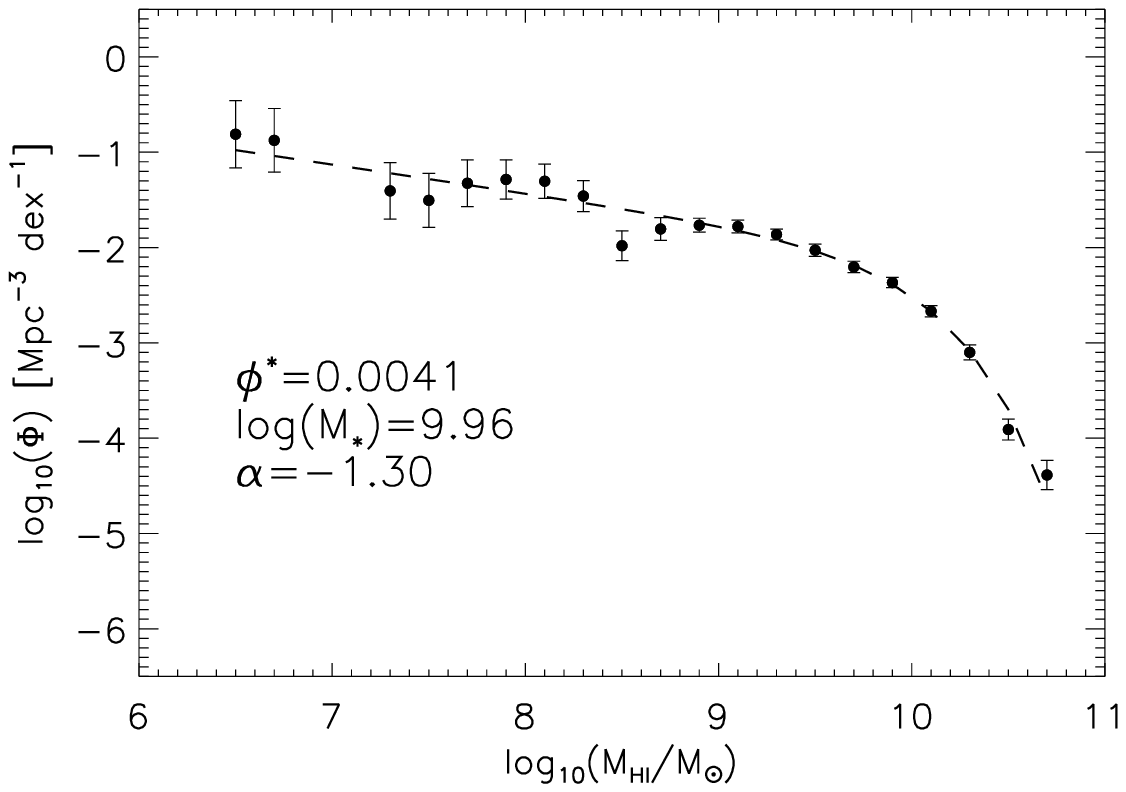}{2.0in}{0}{72}{72}{-420}{-320}
\vskip 1.5cm
\caption{The HIMF estimated for separate subregions of the
$\alpha.40$ catalog via the 2DSWML method with Schechter fit parameters.
Top panel: Results for the $\alpha.40.North1$ region. Middle panel: same, for the
$\alpha.40.North2$ region. Bottom panel: same, for the $\alpha.40.South$
sample. See Table \ref{tab:HIMFreg} for futher quantitative details.}
\label{fig:HIMFsub}
\end{center}
\end{figure}

Figure \ref{fig:HIMFsub} displays the the HIMFs for the three
subregions: $\alpha.40.North1$, $North2$, and $South$, from
top to bottom. The fit parameters and values of $\Omega_{HI}$ 
are given in Table \ref{tab:HIMFreg} along with the fiducial 
2DSWML HIMF for the entire $\alpha.40$ sample for comparison. 
The largest by a significant fraction is $\alpha.40.North1$, 
and it contributes over 50\% of the 10,000 galaxies in 
$\alpha.40$. As expected, the HIMF for this region, when 
isolated, follows the HIMF for the sample as a whole. 
Because of the large volume in this region, the HIMF 
displayed in the top panel of Figure \ref{fig:HIMFsub} is 
smooth and featureless.

In the case of the smaller samples in the middle and bottom panels
of Figure \ref{fig:HIMFsub}, features due to large-scale structure are 
clearly visible. Because of the inhomogeneity of the surveyed volume, 
the HIMFs do not follow the prescribed Schechter function. In the 
case of the $\alpha.40.North2$ subsample, the faint-end slope is 
better fit on its own, in which case it is measured to be 
$\alpha$ = -1.4 $\pm$ 0.1.

In every case, the `bumps' and wiggles in the sub-region HIMFs 
correspond to the cone diagram distributions in \citet{Martin10}. 
In essence, the combination of the ALFALFA survey's sensitivity and 
the scaling of survey volume with redshift leads to preferred distances 
for each of the HI mass bins in the HIMF (or preferred HI masses for 
every distance in the survey). A dip, for example, in the HIMF 
corresponds to an overdensity at the preferred distance for those 
HI mass scales. While the 2DSWML method has been designed to be 
less sensitive to large-scale structure, the volumes of these 
subregions are too small for these effects to average out. 

Such techniques can only work with the data they are given, but the 
1/V$_{max}$ approach allows for explicit correction for known structures. 
When these corrections are included in the 1/$V_{max}$ analysis of these 
subregions, the (unshown) results are very similar to those provided 
here. These corrections are based on the IRAS Point Source Catalog redshift
survey (PSCz; \citet{Branchini99}) density correction 
(see \S\ref{lsscorr}), but imperfections in this correction lead to the same 
bump and dip features. An additional weakness of the 1/V$_{max}$ 
density correction is that the counts can only be increased for 
galaxies that do end up in the sample, making the correction 
significantly less useful in voids. By contrast, 2DSWML essentially 
`self-corrects' for over- and underdense regions. Rather than 
looking at volumes and scaling counts by 1/V$_{max}$, 2DSWML 
constructs the relationship between bins by scaling the counts 
themselves and therefore automatically scales the HIMF downward 
for regions that are overdense and upward for regions that are underdense.

This consideration of subregions within $\alpha.40$ makes clear the 
impact that large-scale structure can have on blind HI surveys and 
the importance of cosmologically significant volumes before 
global conclusions can be drawn.

\subsubsection{Large-scale Structure Correction from Previous Surveys \label{lsscorr}}

As described in \citet{Martin10}, the 1/V$_{max}$ method can be corrected to account 
for large-scale structure in the survey volume. Essentially, overdense regions are 
considered to represent more effective volume ($\Sigma 1/V_{eff}$, rather than 
$\Sigma 1/V_{max}$) and vice versa for underdense regions, so that galaxies in 
various environments are weighted appropriately \citep{Springob05b,RS02}.

While this correction is successful, it does rely on datasets external to 
the ALFALFA survey. In \citet{Martin10}, the density map derived from 
the PSCz \citep{Branchini99}
was used to correct for large-scale structure. However, other options exist, 
in particular other PSCz maps (smoothed to different levels) and the density
reconstruction derived from the 2MASS Redshift Survey (2MRS; \citet{Erdogdu06}). 
The large-scale structure correction used has a large influence on the final 
HIMF estimate; a $\sim$ 20\% effect on the Schechter parameters, compared 
to neglecting the density correction, was reported in \citet{Martin10}. Given 
the magnitude of the effect, it is important to consider the impact 
that a different choice would make. In particular, since this portion of 
HIMF analysis is likely to always rely on external information, 
examining it here may be helpful in the future for the 100\% ALFALFA sample.

The parameter of interest reported by PSCz is the overdensity 
$\delta$, defined relative to the average number density of galaxies 
found in those surveys:

\begin{equation}
	\delta = \frac{n - \bar{n}}{\bar{n}}
\label{eq:eqndelta}
\end{equation}

\noindent
In the case of the ALFALFA survey and the HIMF, we are primarily interested 
in the average value of $\delta$ interior to each source's maximum 
detectable distance or, in other words, the average value of $\delta$ 
in the volume over which the source could have been detected. Both the 
2MRS and PSCz density maps report the value of $\delta$ in equal-volume 
cells throughout their survey volume. The PSCz map was chosen because of its greater
sensitivity in the nearby survey regions of $\alpha.40$, where the HIMF was especially
vulnerable to the impact of large-scale structure.

While PSCz was a good choice for the analysis of the 
$\alpha.40$ HIMF, there are actually several choices of maps available 
from \citet{Branchini99}, with the primary differences being the 
smoothing size of each volume cell and the maximum distance out 
to which the density fields were reconstructed. In \citet{Martin10}, 
the chosen map extended to 240 $h^{-1}$ Mpc and was smoothed 
with a Gaussian kernel of width 3.2 $h^{-1}$ Mpc. The alternative 
options include a map that extends to only 120 $h^{-1}$ Mpc 
with a 3.2 $h^{-1}$ Mpc kernel, and one that extends to 240 
$h^{-1}$ Mpc with a larger Gaussian kernel of 7.7 $h^{-1}$ Mpc. 
The smoothing scale of PSCz maps can lead to underestimation 
of density contrasts. Because the primary effect of the 
large-scale structure correction is on the lowest-mass bins of 
the HIMF, it is important to explore and understand the influence 
of this outside dataset.

Upon examination of the average interior overdensities determined for $\alpha.40$ in each map,
we find that the PSCz.240.G3.2 map used in \citet{Martin10} 
for the 1/V$_{max}$ analysis of the HIMF represents an extreme 
estimate of the impact of large-scale structure within the 
survey volume. It is the most conservative option, given 
that it attaches lower weight to those galaxies found in nearby 
overdensities, particularly the Virgo Cluster, to prevent them 
from artificially boosting the faint-end slope.

In order to quantify the effect of these options on the resulting 
HIMF, we use the PSCz.120.G3.2 and PSCz.240.G7.7 to re-analyze 
the $\alpha.40$ HIMF. Where the maps do not reach the full redshift extent of the $\alpha.40$ sample, 
we set the average interior $\delta$ to 0 for galaxies beyond 
the distance limit. In order to fit Schechter 
function parameters in each case, we use the same uncertainty 
estimates for each HI mass bin point as presented in \citet{Martin10}, 
as the PSCz map applied would only change 
the magnitude of each point and not its fluctuation due to HI mass 
uncertainties.

\begin{figure}
\begin{center}
\plotone{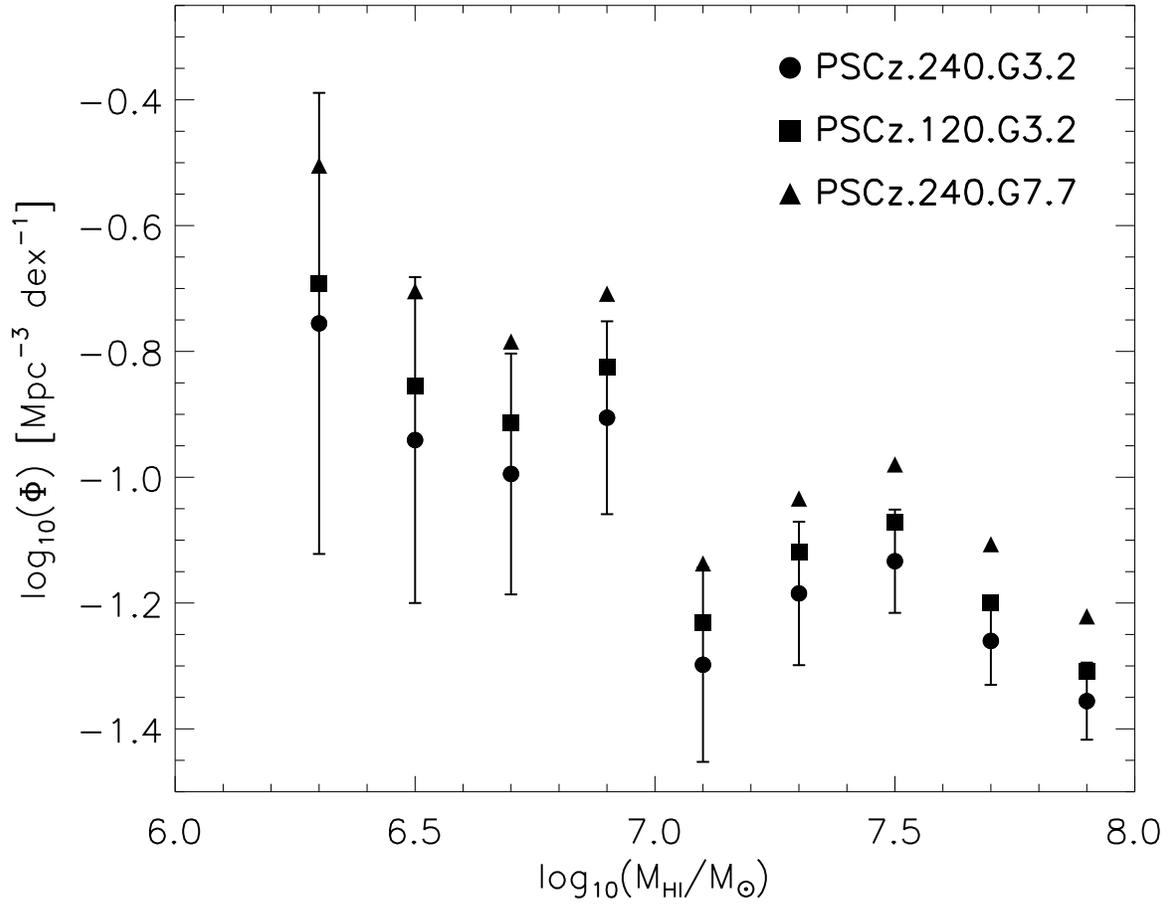}
\caption{The low-mass end of the HIMF, showing dependence on the chosen PSCz density 
reconstruction map. The fiducial 1/V$_{max}$ HIMF reported in \citet{Martin10}
is shown as a filled circle, with two other maps represented by squares and triangles.}
\label{fig:HIMFPSCz}
\end{center}
\end{figure}

Figure \ref{fig:HIMFPSCz} shows the results, focusing on the low-mass
end of the HIMF, since HI mass bins with $M_{HI} > 10^{8.0} M_{\odot}$ 
are not affected by the large-scale structure volume correction. 
The different large-scale structure corrections function effectively 
as a scaling in each bin, so that each option follows the fiducial 
case closely. Both PSCz.120.G3.2 and PSCz.240.G7.7 boost 
the faint-end slope, indicating that they are overcounting galaxies 
in the nearby overdensities, namely the Virgo Cluster. 
This analysis verifies that PSCz.240.G3.2 was the most conservative
 choice for correcting the 1/V$_{max}$ HIMF for the effects of large-scale 
structure. The changes to the low-mass slope $\alpha$ and the turnover 
mass M$_{*}$ are displayed in Table \ref{tab:HIMFPSCz}, along with 
the measured 2DSWML parameters for reference. It is clear that 
the PSCz map with the greatest extent and the smallest smoothing 
radius is most appropriate for estimating the $\alpha.40$ HI mass function.

\section{Summary}\label{sec:conclusion}

This paper presents the catalogued parameters for \ntot ~HI line detections
extracted from $\sim2800$ \sqd ~of high galactic latitude sky observed by
the ALFALFA survey. A (pleasant) surprise for us has been the higher than 
expected ALFALFA detection rate, 5.6 sources per \sqd, or, including only
the objects that are certainly extragalactic, 5.3 sources per \sqd. This
latter detection value is a factor of 29 times greater than the
rate of 0.18 sources per \sqd~ achieved by HIPASS. The characteristic resolution
of the ALFALFA spectral grids is about 4$\arcmin$; the positions of the HI sources
can be determined to an accuracy typically better than 20\arcsec. Using the
publicly available SDSS and DSS2 imaging datasets, we have assigned probable optical
counterparts to more than 98\% of the \ngal ~extragalactic detections and 
provide a cross-reference to the SDSS DR7 photometric and imaging databases.
An additional \nhvc~ HI line detections cannot be identified 
with stellar counterparts but lie within velocity ranges characteristic of the 
galactic/circumgalactic HVCs. Roughly 3/4 of the optically ``dark'' extragalactic 
HI sources are located in fields containing
galaxies of known optical redshift; many are likely to be associated with
tidal debris fields. We identify four objects as candidate OH megamasers
redshifted to $z \sim 0.17$; one of those is a rediscovery of a previously
recognized OHM and is
associated with a galaxy of the same optical redshift \citep{Darling01}.
Future works will explore more systematically the OHM candidates throughout
the ALFALFA bandpass and also will search for evidence of HI in absorption
\citep{Darling11}.
Unsurprisingly, a census of the HI bearing population of galaxies in the
local universe is strongly biased against galaxies on the red sequence, but some 
luminous, red galaxies are detected in the HI line. In particular, ALFALFA
provides a rich sampling of the low-to-moderate density universe at $z \sim 0$.

As a major ALFALFA data release, the $\alpha.40$ catalog presented here 
supercedes the datasets published by our team previously. In particular,
the HI line flux densities reported here are based on further improvements
in the software used for parameter extraction and increased knowledge of
the system performance. The ALFALFA reduction pipeline may 
miss flux for sources which are very large compared to the beam size
and offset from the center of the standard grids, but comparison with the
HI line flux densities derived from pointed single dish observations and
corrected for beam dilution and pointing errors with the ones reported
here shows no systematic offsets except for the very largest and very strongest
sources. The latter will need to be evaluated on a case-by-case basis
in grids produced and analyzed separately from the standard process and
where applicable, corrected for sidelobe contamination
\citep{Dowell10, Dowell11}.

The goals and expectations of the ALFALFA survey were outlined in
\citet{Gio05a} and survey source sensitivity and reliability was discussed
in \citet{Saintonge07a}. As discussed previously, the integrated
HI line flux density threshold of a blind HI survey like ALFALFA increases
with HI line profile width \citep{Martin10,Toribio11b}.  With the 
availability of the large $\alpha.40$ dataset, we test those expectations and 
give quantitative descriptions of the completeness and sensitivity of the ALFALFA survey
as functions of $\log W_{50}$. In addition to the highest quality, highly reliable
(Code 1) HI detections, the $\alpha.40$ catalog presented in Table \ref{tab:catalog}
includes also sources of lower S/N which coincide in position and redshift
with known optical galaxies (the ``priors''). Because the availability of
such prior information is highly dependent on the selection functions of other
surveys, these additional objects should not be used in studies which require 
stringent consideration of statistical completeness. However, the vast majority
are likely to be valid HI detections and hence they can be included
in studies where the number of sources is most critical (e.g., peculiar
velocity studies). Future work will
be undertaken to confirm these detections and an additional set of low S/N
possible detections which coincide with galaxies of unknown redshift. 

The sensitivity of ALFALFA and the thorough understanding of its
performance enable a robust measurement of the HIMF, and in 
particular, of its faint-end slope $\alpha$ and the energy density of neutral 
hydrogen $\Omega_{HI}$ at $z$ = 0.
On the low-mass end of the HIMF, ALFALFA improves on previous blind HI surveys 
in terms of sample size, angular and spectral resolution, sampling of cosmic
volume, and assumptions of pure Hubble flow. At the lowest HI masses, ALFALFA's
finer velocity resolution is an important factor in obtaining a full count
of the gas-rich dwarf population. 

On the high-mass end, previous HI surveys have 
overlooked the locally-rare population of very massive HI disks. We have 
evaluated the possible impact on the derived HIMF of missing sources at 
both the broad and narrow width ends, particularly in comparison with 
the HIPASS catalog \citep{Meyer04}. We conclude that
HIPASS did not recognize the richness of the very high HI mass population,
not because it failed to identify the systems with the broadest widths but 
because it did not have adequate sensitivity at large distances and 
was limited to only 64 MHz of bandpass.  It is ALFALFA's combination of 
sensitivity, spectral and angular resolution, frequency and sky coverage 
which yields a robust census of the HI bearing population at $z = 0$. 

With ALFALFA still only 40\% complete, we have shown
that the 2DSWML and 1/$V_{max}$ methods 
yield results on the HIMF in good agreement, but that the loss of significant volume in 
the ALFALFA survey beyond 15000 km s$^{-1}$ reduces the performance of the 
2DSWML approach if that region is included. 
A realistic treatment of distance and flux density uncertainties, 
translated into mass uncertainties, avoids the strong bias in $\alpha$ and the 
shape of the HIMF introduced by an assumption of Hubble flow in the local 
volume. While $\alpha.40$ does not yet provide a completely representative 
sampling of the local cosmological volume, our method for including the impact 
of large-scale structure is a conservative choice, 
and future data releases from ALFALFA will further improve both 
statistical and systematic uncertainties. We look forward to completing
the ALFALFA survey.

\acknowledgements
We thank the staff of the Arecibo Observatory, particularly Phil Perillat, 
Ganesh Rajagopalan, Arun Venkataraman, Hector Hernandez,
and the telescope operations staff, for their
invaluable help in support of the acquisition of the data used to produce this catalog,
and Tom Shannon and Adam Brazier for their critical support of hardware 
and database development at Cornell.
This work has been supported by the NSF grant AST-0607007 to RG and MPH
and by a Brinson Foundation grant. TB, DC, GLH, SJUH, DK, RK, JM, AO, RO, JR
and EW acknowledge support for the Undergraduate ALFALFA Team from NSF grants
AST-0724918, AST-0725267, AST-0725380, AST-0902211 and AST0903394. 
JLR acknowledges support from NSF AST-000167932. KS acknowledges support from the 
Natural Sciences and Engineering Research Council of Canada (NSERC).

This research has made use of data obtained from or software provided 
by the US National Virtual Observatory, which is sponsored by the National 
Science Foundation and of Montage, funded by the National Aeronautics 
and Space Administration's Earth Science Technology Office, 
Computation Technologies Project, under Cooperative Agreement Number 
NCC5-626 between NASA and the California Institute of Technology. 
Montage is maintained by the NASA/IPAC Infrared Science Archive.

We acknowledge the use of NASA's {\it SkyView} 
facility (http://skyview.gsfc.nasa.gov) located at NASA Goddard Space Flight Center
and the NASA/IPAC Extragalactic Database (NED) which is operated by the Jet 
Propulsion Laboratory, California Institute of Technology, under contract with 
the National Aeronautics and Space Administration. 
The Digitized Sky Surveys were produced at the Space Telescope Science Institute 
under U.S. Government grant NAG W-2166. The images of these surveys are based 
on photographic data obtained using the Oschin Schmidt Telescope on Palomar 
Mountain and the UK Schmidt Telescope. The plates were processed into the 
present compressed digital form with the permission of these institutions. 
The Second Palomar Observatory Sky Survey (POSS-II) was made by the California 
Institute of Technology with funds from the National Science Foundation, the 
National Geographic Society, the Sloan Foundation, the Samuel Oschin Foundation, 
and the Eastman Kodak Corporation. 

Funding for the SDSS and SDSS-II has been provided by the Alfred P. Sloan Foundation, 
the participating institutions, the National Science Foundation, the US Department 
of Energy, the NASA, the Japanese Monbukagakusho, the Max Planck Society and 
the Higher Education Funding Council for England. The SDSS Web Site is 
http://www.sdss.org/.
The SDSS is managed by the Astrophysical Research Consortium for the 
participating institutions. The participating institutions are the American 
Museum of Natural History, Astrophysical Institute Potsdam, University of Basel, 
University of Cambridge, Case Western Reserve University, University of Chicago, 
Drexel University, Fermilab, the Institute for Advanced Study, the Japan 
Participation Group, Johns Hopkins University, the Joint Institute for Nuclear 
Astrophysics, the Kavli Institute for Particle Astrophysics and Cosmology, the
 Korean Scientist Group, the Chinese Academy of Sciences (LAMOST), Los Alamos 
National Laboratory, the Max Planck Institute for Astronomy, the MPA, New Mexico 
State University, Ohio State University, University of Pittsburgh, University 
of Portsmouth, Princeton University, the United States Naval Observatory and 
the University of Washington.


\begin{centering}
\begin{deluxetable}{rlccrrrrrrrl}
\rotate
\tablecolumns{12}
\tablewidth{0pt}
\tabletypesize{\scriptsize}
\tablecaption{Properties of HI Detections\label{tab:catalog}}
\tablehead{
\colhead{~AGC}   & \colhead{Name} & \colhead{HI Coords} & \colhead{Opt. Coords} & \colhead{~~cz$_{\odot}$}  &
\colhead{$W_{50} ~(\epsilon_{w})$} & \colhead{~$S_{21}$} & \colhead{S/N} & \colhead{rms} & \colhead{~~Dist}  & 
\colhead{$\log M_{HI}$} & \colhead{Codes} \\
{\#~~~} &  & {J2000} &  {J2000} & {\kms} & {\kms} & {Jy\kms} & & {mJy} & Mpc & {$M_{\odot}$} & \\
(1)~~ & ~~~(2) & (3) & (4) & (5)~ & (6)~~~ & (7)~~~~ & (8) & (9) & (10) & (11) & (12)
}
\startdata
331061 & 456-013  & 000002.5+155220 & 000002.1+155254 &  6007 & 260(~45) &    1.13(0.09) &   6.5 &   2.40 &  85.2 &  9.29 & 1 I ~ \\
331405 &          & 000003.3+260059 & 000003.5+260050 & 10409 & 315(~~8) &    2.62(0.09) &  16.1 &   2.05 & 143.8 & 10.11 & 1 I ~ \\
102896 &          & 000006.8+281207 & 000006.0+281207 & 16254 & 406(~17) &    2.37(0.12) &  11.2 &   2.31 & 227.4 & 10.46 & 1 I * \\
102574 &          & 000009.1+280543 &                 &  -368 &  23(~~3) &    1.29(0.08) &  11.2 &   5.05 &       &       & 9 U * \\
102975 &          & 000012.3+290137 &                 &  -367 &  23(~~3) &    2.85(0.07) &  26.7 &   4.69 &       &       & 9 U * \\
102571 &          & 000017.2+272359 & 000017.3+272403 &  4654 & 104(~~3) &    2.00(0.06) &  19.0 &   2.29 &  65.9 &  9.31 & 1 I ~ \\
102976 &          & 000019.0+285931 &                 &  -365 &  26(~~2) &    2.53(0.11) &  18.3 &   5.76 &       &       & 9 U * \\
102728 &          & 000021.2+310038 & 000021.4+310119 &   566 &  21(~~6) &    0.31(0.03) &   7.5 &   1.92 &   9.1 &  6.78 & 1 I ~ \\
102575 &          & 000028.0+280845 &                 &  -371 &  33(~~7) &    0.47(0.03) &   8.6 &   2.11 &       &       & 9 U * \\
 12896 & 478-010  & 000030.1+261928 & 000031.4+261931 &  7653 & 170(~10) &    3.14(0.08) &  22.0 &   2.44 & 104.5 &  9.91 & 1 I * \\
102729 &          & 000032.1+305152 & 000032.0+305209 &  4618 &  53(~~6) &    0.70(0.04) &  10.5 &   2.02 &  65.4 &  8.85 & 1 I ~ \\
102576 &          & 000035.3+262712 &                 &  -430 &  21(~~2) &    0.60(0.04) &  11.7 &   2.50 &       &       & 9 U * \\
102730 &          & 000040.1+315610 & 000039.5+315618 & 12631 &  79(~23) &    0.66(0.05) &   7.3 &   2.25 & 175.8 &  9.68 & 1 I ~ \\
102578 &          & 000042.3+263311 &                 &  -429 &  22(~~3) &    0.67(0.04) &  12.8 &   2.44 &       &       & 9 U * \\
101866 &          & 000050.1+141612 & 000047.9+141639 & 10877 & 291(149) &    0.79(0.11) &   4.1 &   2.52 & 150.3 &  9.62 & 2 I * \\
 12901 & 499-035  & 000059.5+285431 & 000058.9+285441 &  6896 & 395(~~5) &    5.03(0.11) &  25.2 &   2.24 &  93.7 & 10.02 & 1 I * \\
102731 & FGC290A  & 000109.3+305221 & 000106.4+305247 &  7366 & 257(~~8) &    1.33(0.08) &   8.9 &   2.08 & 100.5 &  9.50 & 1 I ~ \\
102977 &          & 000108.7+284738 &                 &  -364 &  22(~~3) &    2.03(0.11) &  13.8 &   6.64 &       &       & 9 U * \\
102861 &          & 000110.1+320425 &                 &  -181 &  22(~~1) &    7.30(0.06) &  55.0 &   4.69 &       &       & 9 U * \\
102732 &          & 000114.8+312218 & 000115.0+312227 & 12532 & 292(~~5) &    1.54(0.09) &   9.1 &   2.20 & 174.3 & 10.04 & 1 I ~ \\
101869 &          & 000127.1+142431 & 000131.4+142427 & 12639 & 183(~16) &    1.00(0.09) &   6.4 &   2.57 & 175.4 &  9.86 & 1 I * \\
102733 &          & 000129.8+311418 & 000130.0+311403 & 12581 & 134(~12) &    1.03(0.08) &   8.6 &   2.29 & 175.0 &  9.87 & 1 I ~ \\
 12911 & N7806    & 000131.5+312629 & 000130.1+312631 &  4767 & 231(~23) &    1.40(0.08) &   9.4 &   2.19 &  67.5 &  9.18 & 1 I * \\
331082 & 433-016  & 000134.5+150448 & 000134.0+150454 &  6368 & 118(~~8) &    2.72(0.08) &  21.4 &   2.60 &  85.9 &  9.67 & 1 I ~ \\
748776 &          & 000142.4+135019 & 000141.3+135033 &  6337 &  53(~~5) &    0.65(0.05) &   8.7 &   2.27 &  89.9 &  9.09 & 1 I ~ \\
\hline
\enddata
\tablecomments{
Table \ref{tab:catalog} will be available as a datafile. A portion is shown here for guidance regarding its form and content.}
\end{deluxetable}
\end{centering}

\begin{centering}
\begin{deluxetable}{rccl}
\rotate
\tablecolumns{4}
\tablewidth{0pt}
\tabletypesize{\scriptsize}
\tablecaption{Comments on Individual Sources\label{tab:comments}}
\tablehead{
\colhead{~AGC} & \colhead{Cat.ID.} & \colhead{HI Code} & \colhead{Comment} 
}
\startdata
102896 &       & 1 & In region affected by RFI; parameters uncertain; near smaller AGC 102897 (000005.5+281129, unknown cz) at 0.7 arcmin\\
102574 &       & 9 & HVC; first of two knots near the top of the grid; see also AGC 102575 at 5.1 arcmin\\
102975 &       & 9 & HVC; part of filament that stretches through most of this grid\\
102976 &       & 9 & HVC; part of a filament that extends beyond this grid into 0004+29 \\
102575 &       & 9 & HVC; second of two knots near the top of the grid; see also AGC 102574 at 5.1 arcmin\\
 12896 &       & 1 & Near AGC 331800 (MCG+04-01-009, 0000316+261818, cz=7754) at 1.2 arcmin\\
102576 & 2-  4 & 9 & Compact HVC; one of two nearby knots (the other is AGC 102578\\
102578 & 2-  5 & 9 & Compact HVC; one of two nearby knots (the other is AGC 102576)\\
101866 &       & 2 & Ambiguous OC; several near including AGC 103024 (000049.5+141532, unknown cz) at 1.2 arcmin; others may be background\\
 12901 &       & 1 & Small companion at 0.4 arcmin AGC 103021 (000057.5+285427, unknown cz)\\
102977 &       & 9 & HVC; faint south end of a filament that stretches through most of this grid\\
102861 &       & 9 & HVC 110.7-29.6 part of nice arc\\
 12911 &       & 1 & Multiple system NGC 7805/6; UGC 12908 = NGC 7805 group; blend?\\
101869 &       & 1 & AGC 101869 (000149.5+142623, cz=12568) at 5.7 arcmin\\
102862 &       & 9 & HVC 110.5-31.0 part of nice arc\\
102978 &       & 9 & HVC; part of filament that stretches through most of this grid\\
102735 &       & 1 & Optical identification with bluer galaxy in pair; AGC 102831 (000250.0+281725, unknown cz) at 0.3 arcmin\\
102863 &       & 9 & HVC 110.8-30.0 part of nice arc\\
102979 &       & 9 & HVC; part of filament that stretches through most of this grid\\
749126 &       & 9 & HVC 1-6.04-45.19\\
102864 &       & 9 & HVC 110.7-30.7 part of nice arc\\
749127 &       & 9 & HVC 105.34-47.32\\
102981 &       & 1 & OC identified with larger of pair; second is AGC 103015 (000250.0+281725, unknown cz) at 1.6 arcmin\\
     7 &       & 1 & OC identified with larger of pair; second is AGC 100849 (000306.3+155834, unknown cz) at 1.2 arcmin\\
100011 &       & 2 & Poor spatial and spectral definition\\
\hline
\enddata
\tablecomments{
Table \ref{tab:comments} will be available as a datafile. A portion is shown here for guidance regarding its form and content.}
\end{deluxetable}
\end{centering}

\begin{centering}
\begin{deluxetable}{rccccrrcc}
\tablecolumns{7}
\tablewidth{0pt}
\tabletypesize{\scriptsize}
\tablecaption{The ALFALFA-SDSS DR7 Cross-reference\label{tab:sdsscross}}
\tablehead{
\colhead{~AGC} & \colhead{HI Code} & \colhead{SDSS} & \colhead{PhotoObjID)} & \colhead{SpectObjID}  &
\colhead{r$_{model}$} & \colhead{(u-r)} & \colhead{$z$} & \colhead{$\epsilon_z$}\\
(1)~~ & (2) & (3) & (4) & (5) & (6)~~ & (7) & (8) & (9)
}
\startdata
331061 & 1 & I & 587730775499407375 & 211330582074884096 & 14.77 &  1.59 & 0.02002 & 0.00010\\
331405 & 1 & I & 587740589481525478 &                    & 15.11 &  1.97 &         &        \\
102896 & 1 & I & 758874370996764887 &                    & 15.26 &  2.26 &         &        \\
102571 & 1 & I & 758874297994314032 &                    & 16.10 &  1.39 &         &        \\
102728 & 1 & I & 758874299066483769 &                    & 18.93 &  2.04 &         &        \\
 12896 & 1 & I & 758874370460680283 &                    & 13.98 &  1.29 &         &        \\
102729 & 1 & I & 758874299066548754 &                    & 18.32 &  1.43 &         &        \\
102730 & 1 & I & 758874299602960817 &                    & 16.94 &  1.55 &         &        \\
101866 & 2 & I & 587730773351989400 & 211330580741095424 & 15.15 &  2.48 & 0.03613 & 0.00010\\
 12901 & 1 & I & 758874371533308165 &                    & 13.69 &  2.64 &         &        \\
102731 & 1 & I & 758874372069392715 &                    & 16.01 &  1.70 &         &        \\
102732 & 1 & I & 758874299603223055 &                    & 14.91 &  1.93 &         &        \\
101869 & 1 & I & 587727221413707929 & 211330580573323264 & 15.82 &  1.85 & 0.04189 & 0.00010\\
102733 & 1 & I & 758874299603288292 &                    & 15.85 &  1.82 &         &        \\
 12911 & 1 & I & 758874299603222635 &                    & 13.25 &  3.00 &         &        \\
331082 & 1 & I & 587730774425796793 & 211330582490120192 & 14.87 &  1.46 & 0.02123 & 0.00007\\
748776 & 1 & I & 587730772815184088 &                    & 16.96 &  1.14 &         &        \\
102734 & 1 & I & 758874372605739306 &                    & 15.90 &  1.35 &         &        \\
101873 & 1 & I & 587727223561257129 & 211330582536257536 & 16.35 &  2.38 & 0.04254 & 0.00009\\
102735 & 1 & I & 758874299603419848 &                    & 18.38 &  0.67 &         &        \\
101877 & 1 & I & 587727221413773686 & 211330580648820736 & 16.70 &  1.40 & 0.01734 & 0.00033\\
102980 & 1 & I & 758874371533635834 &                    & 15.87 &  1.76 &         &        \\
 12920 & 1 & I & 758874298531316055 &                    & 15.13 &  2.04 &         &        \\
100006 & 1 & I & 758874372606001325 &                    & 14.28 &  2.70 &         &        \\
100008 & 1 & I & 758874372069982254 &                    & 16.35 &  1.48 &         &        \\
\hline
\enddata
\tablecomments{
Table \ref{tab:sdsscross} will be available as a datafile. A portion is shown here for guidance regarding its form and content.}
\end{deluxetable}
\end{centering}

\begin{deluxetable}{rccrrrrrr}
\tablecolumns{11}
\tablewidth{0pt}
\tabletypesize{\scriptsize}
\tablecaption{OH Megamaser candidates\label{tab:ohmtab}}
\tablehead{
\colhead{~AGC} & \colhead{OHM Coords (2000)} & \colhead{Opt. Coords (J2000)} & \colhead{z$_{opt}$}
& \colhead{z$_{OH}$} & \colhead{$cz_{21}$} & 
\colhead{~~$F_{OH}$} & \colhead{S/N} & \colhead{rms}\\
{\#~~~} & {hh~mm~ss.s+dd~mm~ss} &  {hh~mm~ss.s+dd~mm~ss} &  &  & {~~\kms} & {Jy~km~s$^{-1}$} & & {mJy}\\
(1) & (2) & (3) & (4)~~ & (5) & (6) & (7) & (8) & (9)
}
\startdata
102708 & 000337.0+253215 & 000336.1+253204 &         & 0.169 & -1335 & 0.91 &  5.7 & 2.33 \\
102850 & 002958.8+305739 & 002958.2+305832 &         & 0.172 &  -596 & 0.46 &  6.7 & 2.09 \\
181310 & 082311.7+275157 & 082312.7+275138 & 0.16783 & 0.168 & -1551 & 2.17 & 15.9 & 2.18 \\
228040 & 124540.5+070337 & 124545.7+070347 &         & 0.172 &  -624 & 0.33 &  5.1 & 2.11 \\      
\hline
\enddata
\end{deluxetable}

\begin{deluxetable}{cccccc}
\tablewidth{0pt}
\tabletypesize{\scriptsize}
\tablecaption{HI Mass Function Fit Parameters by Redshift Extent\label{tab:himf}}
\tablehead{ \colhead{Sample and} & \colhead{$\alpha$} & \colhead{$\phi_{*}$} & 
\colhead{$\log $ (M$_{*}$/M$_{\odot}$)} & \colhead{$\Omega_{HI}$, fit} & \colhead{$\Omega_{HI}$, points} \\
Fitting Function& & (10$^{-3}$ h$_{70}^{3}$ Mpc$^{-3}$ dex$^{-1}$) & + 2 $\log$ h$_{70}$ & 
($\times$ 10$^{-4}$ h$^{-1}_{70}$) & ( $\times$ 10$^{-4}$ h$^{-1}_{70}$)
}
\startdata
1/V$_{max}$, 15,000 \kms\tablenotemark{a} & -1.33 (0.04) & 3.1 (0.6)& 9.95 (0.05) & & 4.4 (0.1) \\
\\
1/V$_{max}$, 18,000 \kms\tablenotemark{a} & -1.34 (0.03) & 3.8 (0.6)& 9.92 (0.04) & & 4.3 (0.1) \\
\\
2DSWML, 15,000 \kms&-1.34 (0.02)& 4.7 (0.3) & 9.96 (0.01) & 4.3 (0.3) & 4.4 (0.1) \\
\\
2DSWML, 18,000 \kms&-1.26 (0.02)& 3.4 (0.2) & 10.00 (0.01) & 3.0 (0.2) & 3.1 (0.1) \\
\hline
\enddata
\tablenotetext{a}{In the 1/V$_{max}$ case, pure Schechter functions provide a poor fit to the faint-end
 slope $\alpha$, and the sum of a Schechter and a Gaussian function are used to complete the fit. The 
Gaussian component parameters are not shown in the table, given that they are not expected to be physical.}
\end{deluxetable}

\begin{deluxetable}{cccccc}
\tablewidth{0pt}
\tabletypesize{\scriptsize}
\tablecaption{2DSWML HIMF Schechter Parameters by Region\label{tab:HIMFreg}}
\tablehead{ \colhead{Sample and} & \colhead{$\alpha$} & \colhead{$\phi_{*}$} & 
\colhead{$\log $ (M$_{*}$/M$_{\odot}$)} & \colhead{$\Omega_{HI}$, fit} & \colhead{$\Omega_{HI}$, points} \\
Fitting Function& & (10$^{-3}$ h$_{70}^{3}$ Mpc$^{-3}$ dex$^{-1}$) & + 2 $\log$ h$_{70}$ & 
($\times$ 10$^{-4}$ h$^{-1}_{70}$) & ( $\times$ 10$^{-4}$ h$^{-1}_{70}$)
}
\startdata
North1&-1.35 (0.02)& 4.4 (0.3) & 9.98 (0.02) & 4.3 (0.4) & 4.4 (0.1) \\
\\
North2&-1.25 (0.04)& 5.6 (0.6) & 9.92 (0.02) & 4.2 (0.5) & 4.3 (0.2) \\
\\
South&-1.30 (0.04)& 4.1 (0.5) & 9.96 (0.3) & 3.6 (0.5) & 3.5 (0.2) \\
\\
Whole $\alpha.40$&-1.34 (0.02)& 4.7 (0.3) & 9.96 (0.01) & 4.3 (0.3) & 4.4 (0.1) \\
\hline
\enddata
\end{deluxetable}

\begin{deluxetable}{ccc}
\tablewidth{0pt}
\tabletypesize{\scriptsize}
\tablecaption{1/V$_{max}$ HIMF Schechter Parameters by PSCz Map\label{tab:HIMFPSCz}}
\tablehead{ \colhead{PSCz Map} & \colhead{$\alpha$} &  \colhead{$\log $ (M$_{*}$/M$_{\odot}$)} }
\startdata
2DSWML Result & -1.33 (0.02)& 9.96 (0.02)\\
\\
PSCz.240.G3.2& -1.33 (0.03)&  9.95 (0.04) \\
\\
PSCz.120.G3.2& -1.39 (0.03)&  9.96 (0.05) \\
\\
PSCz.240.G7.7& -1.44 (0.04)&  9.98 (0.06) \\
\\
\hline
\enddata
\end{deluxetable}

\clearpage

\clearpage

\clearpage 

\end{document}